\documentclass[10pt,onecolumn,preprintnumbers,amsmath,amssymb]{revtex4}

\usepackage{graphicx}
\usepackage{dcolumn}
\usepackage{bm}
\usepackage{graphicx,epstopdf}
\usepackage{xcolor}
\usepackage{textcomp}
\usepackage{soul}
\usepackage{csquotes}
\usepackage{mathrsfs} 
\usepackage{amsmath}
\usepackage{bigints} 
\usepackage{amsthm,amssymb} 
\usepackage[utf8]{inputenc}
\usepackage[english]{babel}
\usepackage[shortlabels]{enumitem}
\usepackage{mathrsfs,bigints,mathtools,dsfont}
\usepackage[colorlinks=true,linkcolor=blue,citecolor=blue]{hyperref}%

\newcommand\correspondingauthor{\thanks{Corresponding author.}}

\begin{document}


\title{Dynamical robustness of network of oscillators}
\author{Soumen Majhi$^a$}
\affiliation{Physics Department, University of Rome Tor Vergata, Via della Ricerca Scientifica 1, 00133 Rome, Italy\\
Physics and Applied Mathematics Unit, Indian Statistical Institute, 203 B. T. Road, Kolkata 700108, India }
\author{Biswambhar Rakshit$^a$}
\affiliation{Department of Mathematics, Amrita School of Physical Sciences, Coimbatore, Amrita Vishwa Vidyapeetham,  India}
\author{Amit Sharma$^a$}
\affiliation{Department of Physics, University Institute of Sciences, Chandigarh University, Mohali 140413, India}
\author{J\"{u}rgen Kurths}
\affiliation{Potsdam Institute for Climate Impact Research - Telegraphenberg A 31, Potsdam, 14473, Germany\\
Humboldt University Berlin, Department of Physics, Berlin, 12489, Germany}
\author{Dibakar Ghosh}
\correspondingauthor
\email{dibakar@isical.ac.in}
\affiliation{Physics and Applied Mathematics Unit, Indian Statistical Institute, 203 B. T. Road, Kolkata 700108, India}


\begin{abstract}
\par Most complex systems are nonlinear, relying on emergent behavior resulting from many interacting subsystems, which are often characterized by oscillatory dynamics. Having collective oscillatory behavior is an essential requirement for an appropriate functioning of various real-world systems. Complex networks have proven to be exceptionally efficient in elucidating the topological structures of both natural and artificial systems, as well as describing diverse processes taking place over them. Remarkable advancements have been achieved in recent years in comprehending the emergent dynamics atop complex networks. Specifically, among other processes, a large body of works intend to explore the dynamical robustness of complex networks, which is the networks' ability to withstand dynamical degradation in the network constituents  while maintaining collective oscillatory dynamics. Indeed, various physical and biological systems are recognized to undergo a decline in their dynamic activities, whether occurring naturally or influenced by environmental factors. The impact of such damages on network performance can be significant, and the system's robustness is indicative of its capability to maintain fundamental functionality in the face of dynamic deteriorations, often called aging. This review offers a comprehensive excerpt of notable research endeavors that scrutinize how networks sustain global oscillation under a growing number of inactive dynamical units. We present the contemporary research dedicated to the theoretical understanding and the enhancement mechanisms of the dynamical robustness in complex networks. Our emphasis lies on various network topologies and coupling functions, elucidating the persistence of networked systems. We cover variants of system characteristics from heterogeneity in network connectivity to heterogeneity in the dynamical units. Finally we discuss challenges ahead in this potential field and open areas for future studies. \\

$^a$ These authors equally contributed to the manuscript

\end{abstract}


\maketitle

\newpage
\tableofcontents

\section{Introduction}

Many natural systems are nonlinear and rarely isolated, and hence understanding of such complex systems requires system-level interpretation. The complexity of many social, biological, and physical systems emanates from the intricacy in the patterns of interaction among their constituents. Within this paradigm, the field of network science has been developed as the ideal platform for providing tools for modeling and analysis of complex systems~\cite{barabasi2016network,strogatz2001exploring,newman2003structure,boccaletti2006complex,albert2002statistical,watts1998collective,barabasi1999emergence}. These networked systems often function through the emergent behavior of many interacting units, where each unit exhibits oscillatory dynamics. Consequently, interacting oscillatory systems constitute an efficient framework to model many complex systems. In such a  framework, the inherent dynamics of an individual node can be modeled as a system of nonlinear differential equations, and different coupling functions can describe the interactions among the nodes~\cite{stankovski2017coupling}. Even simple nonlinear systems, when connected with each other through rather simple coupling functions, can generate complex collective dynamics.  With various types of coupling schemes and network structures, a wide variety of collective dynamics have been explored.  For example, synchronization processes are ubiquitous in nature and has been studied extensively in populations of locally interacting elements in the context of physical, biological, chemical,  and technological  systems~\cite{boccaletti2002synchronization,pecora1990synchronization,mirollo1990synchronization,arenas2008synchronization,ghosh2022synchronized}. Among various types of partial synchronization, the coexistence of coherent and incoherent patterns, commonly referred to as the chimera state~\cite{panaggio2015chimera,majhi2019chimera,abrams2004chimera,bera2017chimera,montbrio2004synchronization}, has been in the focus over the last decade across a diverse array of systems. Coupled oscillators, therefore, play a key role in many areas of science and technology, and their dynamics help us to explore various emergent behaviors of living and non-living systems.
 
\par Among all the research perspectives concerning dynamics of networks, studies focused on network robustness, which refers to the ability to withstand even strong perturbations, holds considerable significance from various aspects.
This phenomenon of network robustness can be conceptualized in two distinct ways: structural and dynamical robustness. i) Structural robustness addresses the endurance of network activities when faced with structural perturbations, which could involve the removal of links (bond percolation) or nodes (site percolation) within a network~\cite{cohen2010complex,albert2000error,callaway2000network,cohen2000resilience,dorogovtsev2008critical,buldyrev2010catastrophic,vespignani2010fragility,gao2016universal}. ii) On the other hand, dynamical robustness of complex networks, in general, is defined as the network’s capacity to sustain its dynamic activity despite local disturbances. Specifically, throughout this review, we uncover dynamical robustness of networks of coupled oscillators which refers to the network’s ability to sustain its global dynamical activity even when a portion of its dynamical components are functionally degraded. In the literature, this is, popularly presented in terms of \textit{aging transition}. Aging transition is an emergent collective phenomenon in networked systems comprising of self-oscillatory and non self-oscillatory nodes. This transition occurs when the network shifts from a globally oscillatory state to a oscillation quenched state as the proportion of inactive (aged) oscillators exceeds a critical threshold. 

\par Possessing collective oscillatory behavior is an essential prerequisite for the regular functioning of many complex systems. Examples include circadian rhythms~\cite{bernard2007synchronization}, biological pacemaker cells~\cite{miake2002biological}, cardiac and respiratory systems~\cite{Jalife_1998_Chaos} etc.  Oscillation plays important roles in other several dynamic processes within both single cell and multicellular processes~\cite{kruse2005oscillations,chen2010modeling}. Additionally, coupled oscillator models are applicable to electric power-grid networks, where components such as power sources need to be synchronized to the same frequency~\cite{Menck_2014_Nat.Com,filatrella2008analysis}. Robust oscillatory dynamics is thus a fundamental characteristic of such systems~\cite{winfree1980geometry,strogatz2001exploring}. Despite being regularly subjected to internal and external disturbances, these systems can maintain their rhythmic activities to a certain degree. If a limited number of units below a specific threshold fail to generate oscillatory behavior, the remaining units can compensate, enabling the entire system to resiliently preserve its proper functioning. However, if a substantial number of units transit to an inactive quenched state, it can significantly impede their functions, potentially resulting in a partial collapse or even complete failure of the system in question.

\par Keeping this fundamental and inevitable context in mind, in the last two decades, researchers from the nonlinear dynamics community have been working on the dynamical robustness of a network of coupled oscillators 
~\cite{Daido_2004_PRL,Daido_2007_PRE,tanaka2012dynamical}. It is defined as the ability of a network to sustain its collective macroscopic oscillation when a few of its nodes fail to produce rhythmic dynamics due to local degradation 
~\cite{tanaka2012dynamical}. Daido and Nakanishi~\cite{Daido_2004_PRL} in their pioneering work laid the mathematical framework for studying dynamical robustness. They examined a situation in which oscillating nodes gradually transform into stable points. If the quantity of nodes transitioning to stable points exceeds a critical threshold, it could disrupt the typical oscillatory patterns in these systems, leading to an abrupt phase shift toward a universally non-oscillatory state.
They explored a scenario, where oscillatory nodes progressively transition to fixed points. If the number of nodes that shift to fixed points surpasses a critical threshold, the usual oscillatory behavior of these systems could be disrupted, resulting in a sudden phase transition towards a globally non-oscillatory state. 
This phenomenon, characterized by such abrupt and catastrophic emergence, is termed an \textit{aging transition}. The authors demonstrated that in a global network of diffusively coupled oscillators, an aging transition can be characterized by a universal scaling law of an order parameter involving inactivation fraction and the strength of coupling. After their initial work a series of research articles have been published. Owing  to its  widespread relevance, aging transition has been studied in diverse models with different coupling functions and network structures. Among the many significant attempts made along this topic, Paz\'o et al.~\cite{pazo2006universal} studied aging transition in an ensemble of globally coupled Morris-Lecar model which exhibits a saddle-node bifurcation on an invariant circle. A similar scaling law is established for an ensemble of excitable and oscillatory units. Tanaka et al.~\cite{tanaka2012dynamical} have explored the aging transition in a complex network and have shown that scale-free  networks are highly resilient  to random inactivation but extremely vulnerable to targeted inactivation of low-degree oscillators with respect to the dynamical robustness. Their finding is not in agreement with the structural robustness of a scale-free network where high degree nodes play a crucial role. Dynamical robustness has been studied in the context of metapopulation dynamics by  Kundu et al.~\cite{Rakshit_2017_PRE}. Their results reveal how the network  topology plays a crucial role in metapopulation survivavbility. Thakur et al.~\cite{thakur2014time} studied the  influence of time-delayed coupling on the nature of the aging transition in globally coupled Stuart-Landau oscillators. Their findings divulge that time delay in the coupling does not favor dynamical robustness. In Ref.~\cite{morino2011robustness} the authors demonstrated aging transition in a multi-layer network of active and inactive units, while contemplating with various interlayer coupling functions.
 
\par As mentioned above, such global dynamical degradation in the form of aging transition can have outright impacts on the substantive system performance. Thus for many of real-world and man-made systems, continued dynamic oscillatory activity of the components are often extremely crucial for maintaining proper functioning. For instance, in case of neuronal activity~\cite{lisman2008neural}, from physiological processes like cell necrosis within organs~\cite{gurtner2007progress} to cardiac and respiratory systems~\cite{jalife1998self}, robust oscillatory dynamics is quite necessary. Due to such high practical importance many researchers  proposed remedial measures to enhance the dynamical robustness against aging or deterioration of individual units. Among the notable attempts in this regard, Liu et al.~\cite{liu2016enhancing} proposed an efficient method to enhance dynamical persistence by introducing an additional parameter that controls the diffusion rate. In Ref.~\cite{kundu2018resumption} authors established that by adding a linear positive mean-field feedback term, network's dynamical robustness can be improved substantially. Effectiveness of self-feedback delay to increase  the dynamical resilience has been studied by Sharma and Rakshit~\cite{sharma2021enhancement}. The increasing body of vast literature on aging transition and dynamical robustness is itself a testimony of its relevance in various fields of science and engineering.  In this report, we intend  to provide an exhaustive overview on aging transition by integrating prevailing knowledge achieved in the last two decades. Thus the relevant results and methodology related to dynamical robustness  will be more generally accessible for researchers in diverse  communities of science and technology. There are also several open challenging problems emphasized here. 



\section{Dynamical robustness for different network structures}

Based on the network topologies, we discuss the dynamical robustness by taking different types of networks. Most of the results are explored for globally connected network and complex networks. We also discuss the results on dynamical robustness for multiplex, time-varying and long-ranged networks.


\subsection{Dynamical robustness of globally coupled networks} \label{analytical}

Initially, we address the outcomes regarding dynamical robustness concerning instantaneous diffusive coupling, followed by an exploration of time-delay diffusive coupling topology. Subsequently, we introduce findings related to weighted conjugate coupling and interactions characterized as attractive-repulsive. Finally, we delve into dynamical robustness in scenarios where inactive oscillators are absent.


\subsubsection{Diffusive coupling}\label{Diffusive instantaneous coupling}

\begin{figure}[t]
\centering
\includegraphics[width=0.40\linewidth]{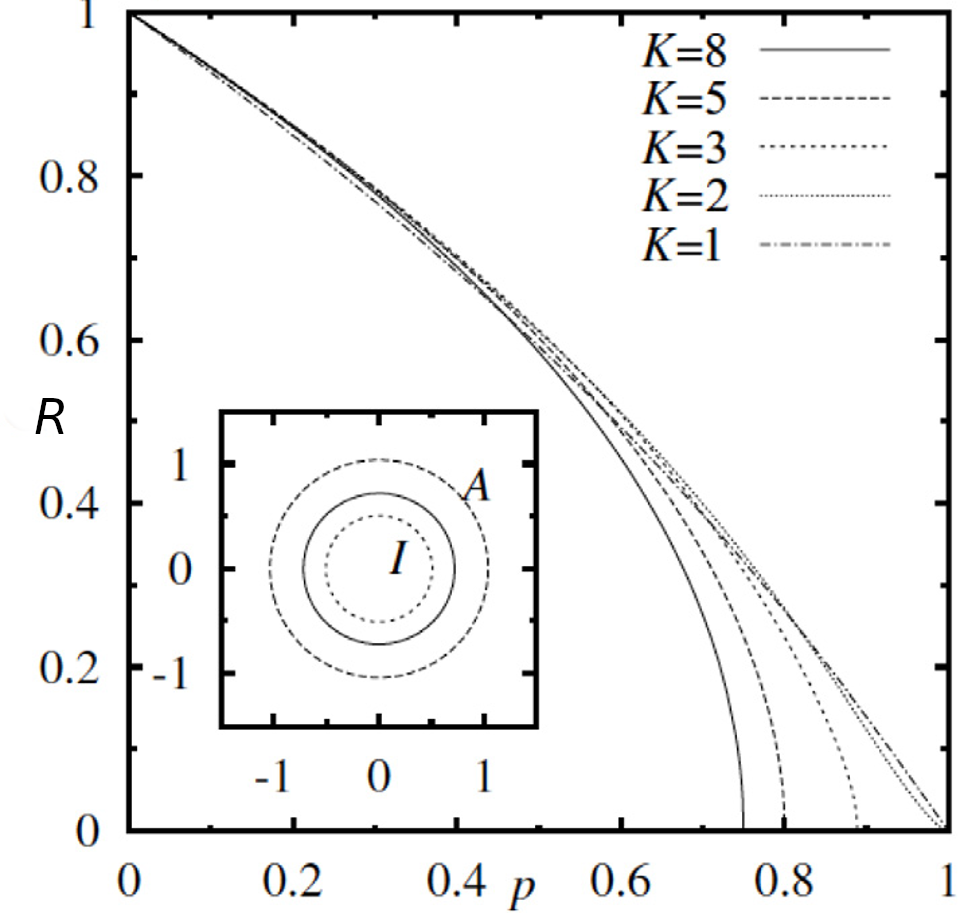}
\caption{The normalized order parameter $R$ ~(Eq.~\eqref{order-para}) is plotted as a function of $p$ to illustrate the aging transition in the globally coupled Stuart–Landau limit-cycle system~(Eq.~\eqref{eq-1br}). Different (increasing) values of the coupling strength $K$ are chosen ranging from $K=1$ to $K=8$, with $a=2$ and $b=1$.
Reprinted figure with permission from Ref.\cite{Daido_2004_PRL}}
\label{f-1br}
\end{figure}
Diffusive coupling represents the predominant form of coupling observed in numerous real-world systems\cite{rosenblum2003synchronization}, consequently garnering significant attention when investigating aging transitions. We first discuss the aging transition phenomena in a system of $N$ globally coupled Stuart–Landau oscillators \cite{Daido_2004_PRL}. The mathematical form of all-to-all diffusively coupled network is,

\begin{equation}
\dot{z}_j(t)=(\alpha_j+i\omega-|z_j(t)|^2)z_j(t) +\frac{K}{N}\sum_{l=1}^N(z_l-z_j),
\label{eq-1br}
\end{equation}
\noindent where $j = 1,2, . . . ,N$ and $z=x+iy$ is a complex variable. Here $\omega$ signifies the internal frequency of each oscillator, and $\alpha_j$ is the bifurcation parameter of the $j$-th oscillator, denoting its proximity to a Hopf bifurcation point. When $\alpha_j>0$, each isolated Stuart-Landau oscillator exhibits a stable sinusoidal oscillation, while it converges to the stable trivial fixed point $z_j = 0$ for $\alpha_j<0$. The second term on the right-hand side of Eq.(\ref{eq-1br}) denotes the presence of diffusive coupling with a strength represented by $K$.

The framework introduced by Daido and Nakanishi \cite{Daido_2004_PRL} facilitates the study of robustness in a manner where an active oscillator with $\alpha_j=a>0$ transitions to an inactive state with $\alpha_j=-b<0$, with a parameter $p$ representing the fraction of inactive oscillators. To simplify, we can designate the group of oscillators as active ones, denoted by $j = 1, 2, ..., N(1-p)$, and the remaining oscillators as inactive, denoted by $j = N(1-p)+1, N(1-p)+2, ..., N$. This implies that  $0 < p < 1$ signifies the fraction of inactive oscillators in the entire networked system. The degree of macroscopic oscillation of the whole network is then measured by the order parameter $|\Bar{Z}(p)|$,  
\begin{equation}
 \Bar{Z}(p)=\frac{1}{N}\sum_{j=1}^N z_j,  
 \label{eq-opr}
\end{equation}
and subsequently by the normalized order parameter $R$, which is defined as follows: 
\begin{equation}
    R=\frac{|\Bar{Z}(p)|}{|\Bar{Z}(0)|}.
    \label{order-para}
\end{equation} 
\noindent As one increases the inactivation parameter $p$, the order parameter gradually diminishes and at a critical value $p = p_c$, the loss of global oscillation takes place and leads to a transition  in the mean-field dynamics called \textit{aging transition}. Figure \ref{f-1br} portrays the normalized order parameter $R$ as a function of $p$. It depicts the gradual loss of global oscillation for  various strengths of coupling constant $K$. For decreasing $K$ the critical value of the inactivation parameter $p_c$  monotonically increases until it reaches unity at a threshold value of $K=K_c$, below which $p_c$ remains at unity.

\par Synchronized activities among the oscillators permit us to assume that within each group of active and inactive nodes oscillators  behave identically. With this presumption in mind, we assign $z_j$ as $A$ for the active ensemble and $z_j$ as $I$ for the inactive ensemble of oscillators. Consequently, Eq.~(\ref{eq-1br}) transforms into the subsequent interconnected systems: 

\begin{eqnarray}
\dot{A}(t)&=&(a-Kp+i\omega-|A(t)|^2)A(t)+KpI(t),\nonumber\\
\dot{I}(t)&=&(-b-Kq+i\omega-|I(t)|^2)I(t)+KqA(t),\nonumber\\
\label{eq2-br}
\end{eqnarray}
where $q=1-p.$
\noindent A linear stability analysis of the reduced system leads to an analytical formula of the critical point $p_c$ as 
\begin{eqnarray}
p_c=\frac{a(K+b)}{(a+b)K}.
\label{eq-3br}
\end{eqnarray}

\noindent In the limiting case, $\lim_{K \to \infty} p_c=\frac{a}{a+b}$. Consequently, one can conclude that the dynamical robustness is stronger when the active oscillators have larger amplitude of oscillation.  We can derive the scaling property of the order parameter near the critical point $p_c$ \cite{Daido_2004_PRL}. The scaling law is represented as $|\Bar{Z}|\propto(p_c-p)^{\beta}$, where the critical exponent $\beta$ varies based on the coupling strength in the following manner, 

\begin{eqnarray}
\beta =
 \begin{cases} 
      1/2 & \text{for $K<K_c$}\\
      1 & \text{for $K=K_c$}\\
      3/2 & \text{for $K>K_c$}.\nonumber
    \end{cases} 
\end{eqnarray}
The critical exponent $\beta$ increases by increasing the coupling strength $K$.

\par Until recently, the investigation of dynamical robustness was primarily focused on coupled Stuart–Landau oscillators, which exhibit typical sinusoidal oscillations. However, many natural systems can be modeled by networks of non-sinusoidal oscillators, such as the Van der Pol oscillator. Rakshit et al.~\cite{rakshit2020abnormal} have studied the aging transition in a network of globally coupled Van der Pol oscillators and highlighted how it differs from the Stuart-Landau model. Their investigation uncovers distinct pathways to aging transition in networks of Van der Pol oscillators compared to typical sinusoidal oscillators like Stuart–Landau oscillators. Unlike sinusoidal oscillators, where the order parameter smoothly undergoes a second-order phase transition, they observed an unconventional phase transition characterized by the abrupt emergence of unbounded trajectories at a critical point. Through detailed bifurcation analysis, they elucidated this abnormal phase transition, demonstrating that it is driven by the boundary crisis of a limit-cycle oscillator, paving the way for an unconventional and discontinuous aging transition path.


\subsubsection{Time-delay diffusive coupling}

Dynamical systems characterized by temporal delays, are prevalent in nature. They manifest due to finite signal propagation time and memory effects in a diverse range of natural phenomena~\cite{sismondo1990synchronous,yeung1999time,boutle2007nino,dhamala2004enhancement,lakshmanan2011dynamics,reddy1998time}, spanning physical, chemical, engineering, economic, and biological domains, including their respective networks. Several dynamical systems can be portrayed by delay differential equations with single constant delay~\cite{mackey1977oscillation,ikeda1980optical,fort1999time}, discrete delays~\cite{saad1998comparative,asea1999time,kuang1993delay}, distributed delay~\cite{haldane1934contribution,baylor1974electrical}, state-dependent delay~\cite{insperger2008criticality,hartung2006functional}, and time-dependent delay~\cite{kye2004synchronization,senthilkumar2007delay,ghosh2010generalized,majhi2016synchronization,ghosh2010projective}. 

\par Let us now introduce a time delay in the signal transmission and observe how it affects the dynamical robustness of networked systems. For simplicity, we contemplate with globally (all-to-all) coupled systems of oscillators. The system of $N$ all-to-all coupled Stuart-Landau oscillators subject to linear time-delayed coupling can then be represented by the following set of equations,

\begin{equation}
\begin{array}{lcl}\label{dl1}
\dot {z_j}=(\alpha_j+ i\omega-|{z_j}|^2){z_j}+\dfrac{\kappa'}{N}\sum\limits_{k=1, k \neq j}^{N}[{z_k}(t-\tau)-{z_j}(t)];~~~j= 1,2,\cdots,N,
\end{array}
\end{equation}
with $\kappa'=2\kappa$ as the interaction strength and $\tau$ as the parameter describing the time delay in the signal transmission. 
\par
\begin{figure}[ht]
	\centerline{\includegraphics[scale=0.300]{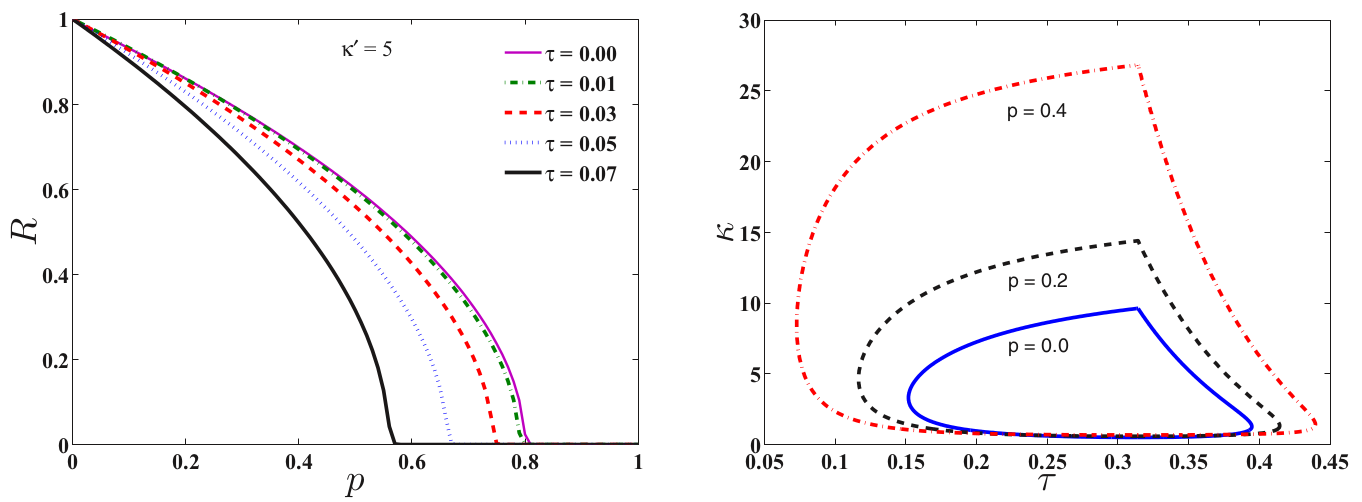}}
	\caption{(a) The normalized order parameter $R$ for the system~\eqref{dl1} as a function of the proportion $p$ of inactive oscillators for different values of the time-delay $\tau$ for $N=500$ dynamical units, whenever $\omega=10$ and $\kappa'=5$ with $a=2,b=1$. (b) Aging islands in the $(\kappa,\tau)$ parameter plane for various values of $p$. Reprinted figure with permission from Ref.\cite{thakur2014time}}
	\label{fig2}
\end{figure}

 Choosing a network of $N=500$ dynamical units, we depict in Fig.~\ref{fig2}(a) how the normalized order parameter $R$ (Eq.~\eqref{order-para}) changes with respect to increasing inactivation represented by the parameter $p$, for various values of $\tau$. We keep $\kappa'=5$ fixed and start with the non-delay case (i.e., $\tau=0$). Then we choose a higher time-delay with $\tau=0.01$ and observe a slightly faster aging transition in the system. More significant change takes place when we increase $\tau=0.03$ where the critical inactivation ratio $p_c$ evidently decreases. This scenario remains valid as we increase the delay to $\tau=0.05$ and $\tau=0.07$. In fact, with increasing $\tau$ this scenario becomes more pronounced and the aging transition occurs faster. So, for a fixed interaction strength, the critical inactivation ratio $p_c$ decreases with increasing $\tau$. Thus, the aging takes place faster and hence the robustness of the networked system decreases due to the introduction of time-delay. 
\par Going further, one can perform a linear stability analysis of this reduced system of equations around the origin (as in Ref.~\cite{thakur2014time}), ending up with a characteristic equation for the eigenvalues, from which the aging islands in the $(\kappa,~\tau)$ parameter plane can be determined. Subsequently, for a better perception of the robustness subject to time-delay, along with varying the time-delay, we consider simultaneous change in the interaction strength, and plot the phase diagram in the $(\kappa,~\tau)$ parameter plane for different values of the inactivation ratio $p$ (in Fig.~\ref{fig2}(b)). Specifically, $p=0.0,0.2,$ and $0.4$ are chosen and it is clear that the aging island expands for raising the ratio of the inactive oscillators. The aging islands expand in both directions of $\kappa$ and $\tau$ reflecting the fact that along with the interaction strength and the inactive elements, the time-delay also carries the capability of suppressing oscillations to the trivial fixed point and hence to decrease the robustness of the system.  

\par Further, the study by Rahman et al.~\cite{rahman2017aging} examines a globally connected network comprising active and inactive oscillators with distributed-delay coupling. It establishes conditions for aging transition, derived for both uniform and gamma delay distributions. The findings suggest that in the case of a uniform distribution, increasing the width of the delay distribution, while maintaining the same mean delay, enables an aging transition to occur with a smaller coupling strength and a lower proportion of inactive elements. Provided the coupling strength falls within a specific range and the mean time delay is substantial, it may be feasible to achieve an aging transition for any proportion of inactive oscillators for a gamma distribution.   


\subsubsection{Mean-field diffusive coupling}
\begin{figure}[t]
\centering
\includegraphics[width=0.4\textwidth]{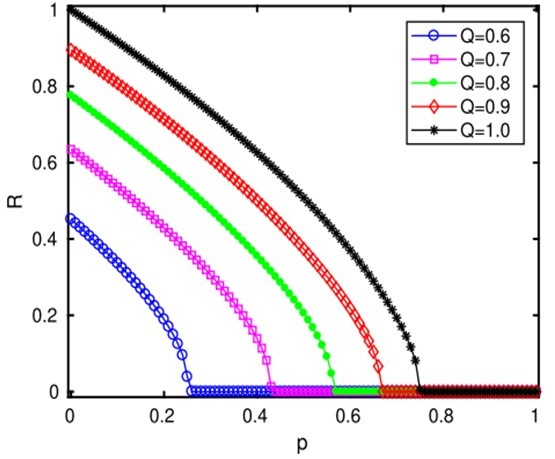}
\caption{The influences of the mean-field density parameter for the system~\eqref{eq-mff} on the aging transition scenarios has shown for a certain coupling value. The order parameter $R$ versus the fraction of inactivation ratio p corresponding to the various mean-field density parameter $Q = 0.6$, $0.7$, $0.8$, $0.9$ and $Q = 1.0$ with fixed coupling strength $\epsilon=2.0.$ Reprinted figure with permission from Ref.\cite{bera2020}.}
\label{mfc_f1}
\end{figure}

The study of mean-field coupling is extensively studied due to its prevalence in numerous natural occurrences within the realms of physics, biology and engineering. Effects of mean-field diffusion have previously been explored in synchronization~\cite{shiino1989synchronization}, multistable dynamics of synthetic genetic networks~\cite{garcia2004modeling,ullner2007multistability,koseska2007inherent,ullner2008multistability}, oscillation suppression processes~\cite{mirollo1990amplitude,de2003coherent,sharma2012amplitude,ghosh2015revival}, and also in the dynamics of chimera-death~\cite{banerjee2015mean}. Recently, the dynamical robustness in the coupled Stuart-Landau oscillators through mean-field diffusion has been reported in \cite{bera2020}. The mathematical model of the coupled oscillators is given by
\begin{eqnarray}
	\dot{z}_j &=& (\alpha_j + i \omega -  |z_j|^2)z_j+\epsilon(Q\bar{z} - z_j),
 \label{eq-mff}
\end{eqnarray}
\noindent where $\bar{z}$ is mean value which represent by $\frac{1}{N}\sum_{l=1}^Nz_l$.  In this coupling scheme, the control parameter $Q$ associated with the mean-field interaction describes the influx of the mean- field in the dynamical units. The parameter $Q$ essentially controls the rate of mean-field interaction in the diffusive coupling among the interacting oscillatory systems. When $Q=0$, there is no interaction between the oscillators and they behave like uncoupled ones subjected to self-feedback, while  $Q=1$  maximizes the interaction with the mean field \cite{PRE_2012_Amit}. Linear stability analysis around the origin enables us to analytically derive the critical value $p_c$ for Eq.~\eqref{eq-mff} as~\cite{bera2020},

\begin{eqnarray}
	p_c = (b+\epsilon)\left(\frac{1}{a+b}+\frac{a-\epsilon}{Q\epsilon(a+b)}\right),~~with~~Q > 1-\frac{a}{\epsilon}, 
	\label{eq-2_mfc}
\end{eqnarray}

\noindent where $b-a-Q\epsilon+2\epsilon>0$. The above expression for $p_c$ clearly indicates that lowering the mean-field parameter $Q$
has a negative effect on the dynamical robustness of the networked system. This result is in agreement with the previous study that the mean-field
control parameter plays an important role in the suppression of oscillations \cite{PRE_2012_Amit}. As $Q\rightarrow 0$, the effect of the mean-field interaction in coupling decreases, which in turn hinders the collective macroscopic oscillation. Figure \ref{mfc_f1} depicts the normalized  order parameter $R$~(Eq.~\eqref{order-para}) as a function of the inactivation parameter $p$ for various values of $Q$. It is discernible from the plot for the maximum mean-field scenario (with $Q=1$) that the order parameter sharply decreases with increasing $p$, and eventually vanishes at the critical value $p_c\sim 0.75$, indicating the occurrence of an aging transition. The value of $Q$ is decreased subsequently and the order parameter is plotted for $Q=0.9$, for which aging transition takes place at a lower value of $p$, and hence the dynamical robustness of the coupled system reduces. Similarly for a lower value of $Q=0.8$, the robustness decreases even more. This trend remains intact for even lower values of the mean-field parameter $Q=0.7$ and $Q=0.6$. These numerical results are in line with the analytically obtained critical values of the inactivation ratio as well. Thus, these plots demonstrate that a lower mean-field density forces the entire system to dynamically collapse for lower values of the critical inactivation ratio, and hence leads to the decrement of dynamical robustness. Similar results are observed for networks of delay-coupled systems as well~\cite{bera2020}. 


	
\subsubsection{Weighted conjugate coupling}
Apart from the traditional coupling via the similar variables among the dynamical systems, coupling through dissimilar or conjugate variables is also natural in a number of experimental scenarios in which the units are coupled by feeding the output of one into the other. Examples include the experiments by Kim et al.~\cite{kim2005scaling} of coupled semiconductor laser systems. Later, in studying the phenomenon of oscillation quenching~\cite{saxena2012amplitude,koseska2013oscillation}, this form of coupling is utilized in detail~\cite{karnatak2007amplitude,punetha2018dynamical}, as coupling through dissimilar variables naturally breaks the rotational symmetry of diffusively coupled systems. Conjugate coupling has further been shown to be effective in enhancing coherence~\cite{zhao2018enhancing} and inducing explosive death~\cite{zhao2018explosive}. 

\par Here we consider a globally coupled network of Stuart-Landau limit cycle oscillators with weighted conjugate coupling \cite{Ponrasu2019}, for which the governing equation can be represented as,  
\begin{eqnarray}
	\dot{z}_j &=& (\alpha_j+i\omega - |z_j|^2)z_j + \frac{\epsilon m_{(1,2)}}{N} \sum_{k=1}^N [(Img(z_k)-\beta Re(z_j)) + i(Re(z_k)-\beta Img(z_j))],
 \label{eq-wc}
\end{eqnarray}
\noindent where $\epsilon$ is the overall coupling strength, and $\beta$ is the feedback control parameter ($0 \leq \beta \leq 1$). Here $Re(z)$ and $Img(z)$ are the real and imaginary parts of $z$, respectively. $\alpha_j$ is the bifurcation parameter for Stuart-Landau oscillators, specifically we choose $\alpha_j = 2$ for the active set of oscillators, and $\alpha_j = -1$ for the inactive ones. The intra-group coupling strengths of active and inactive oscillators are $m_1$ and $m_2$.  When $m_1 = m_2$, the interaction is symmetric, while for $m_1 \neq m_2$ the interaction becomes asymmetric. 

\par For the stability analysis, the global network is divided into two groups of active (A) and inactive (I) oscillators. The $A = A_r+iA_{im}$, and $I = I_r+iI_{im}$, where $A_r, I_r$ and $A_{im}, I_{im}$ are the real and imaginary variables which satisfy the following equations, 
\begin{eqnarray}\label{eqw}
	\dot{A}_r &=& aA_r-A_{im}\omega-A_rA^2_{im}-A^3_r+m_1\epsilon[(1-p)A_{im}+pI_{im}-\beta A_r], \nonumber \\
	\dot{A}_{im} &=& aA_{im}+A_{r}\omega-A_{im}A^2_{r}-A^3_{im}+m_1\epsilon[(1-p)A_{r}+pI_{r}-\beta A_{im}], \nonumber \\
	\dot{I}_r &=& bI_r-I_{im}\omega-I_rI^2_{im}-I^3_r+m_2\epsilon[(1-p)A_{im}+pI_{im}-\beta I_r], \nonumber \\
	\dot{I}_{im} &=& bI_{im}+I_{r}\omega-I_{im}I^2_{r}-I^3_{im}+m_2\epsilon[(1-p)A_{r}+pI_{r}-\beta I_{im}].
\end{eqnarray}	

\begin{figure}[ht]
\centering
\includegraphics[width=0.70\textwidth]{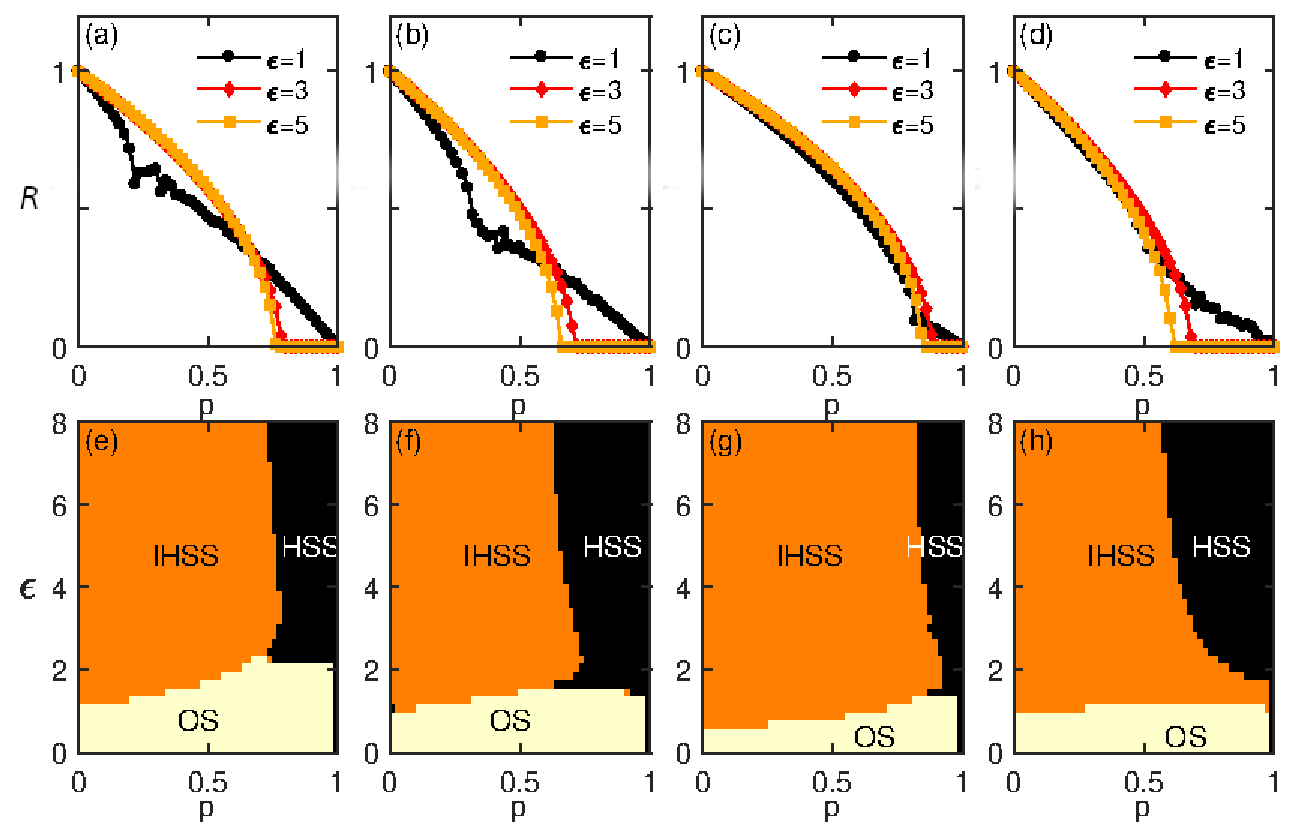}
\caption{(a-d) The order parameter $R$ for the system~\eqref{eq-wc} varies with the inactivation ratio $p$ across various coupling strengths, 
and (e-h) the phase diagram in ($\epsilon-p$) space for the natural frequency $w=1$ for $(m_1,m_2)= (1,1), (1,2), (2,1)$ and $(1,3)$ (left to right). Dynamics regions: HSS for homogenous steady state  corresponding to the aging transition (black), IHSS for inhomogeneous steady state (orange), and OS for oscillatory state (yellow).}
\label{wcc_f1}
\end{figure}
	
\noindent The determination of the critical value $p_c$ involves conducting a linear stability analysis on Eq.~\eqref{eqw} at the origin, denoted as $(\bf A, \bf I) = (\bf 0, \bf 0)$.
For this purpose, the 4th order characteristic equation is deduced from the Jacobian, and the Hopf and pitchfork bifurcation points using the Routh-Hurwitz stability criterion are computed (for more details see \cite{Ponrasu2019}). 

\par For the numerical analysis, first we consider the symmetric case ($m_1 = m_2 = 1$) in the conjugate coupled system of Stuart-Landau oscillators with the natural frequency $\omega=1$ and the control parameter $\beta=1$. In Figure~\ref{wcc_f1}(a), we illustrate the variation of the order parameter $R$ as a function of the inactivation ratio $p$ across various coupling strengths $\epsilon$. 
The critical value $p_c$ of aging transition to the homogeneous steady state (HSS) decreases with increasing the coupling strength $\epsilon$. To show the effect of $\epsilon$ on the aging transition, we portray the corresponding phase diagram in the $p-\epsilon$ parameter plane in Fig.~\ref{wcc_f1}(e), which confirms the fall of $p_c$ with coupling strength for HSS. To untangle the impact of asymmetry parameter on the aging transition, we assume three cases with the parameter sets $(m_1, m_2)= (1, 2), (2, 1)$ and $(1, 3)$. The aging transition in terms of $R$ and the corresponding phase diagrams for these three cases are respectively depicted in Figs.~\ref{wcc_f1}(b-d) and \ref{wcc_f1}(f-h). As the impact of inactive (active) oscillators on the active (inactive) oscillators intensifies with rising values of $m_2 (m_1)$, it either enhance or shrink the HSS region in the parameter plane. 
\begin{figure}[t]
\centering
\includegraphics[width=0.70\textwidth]{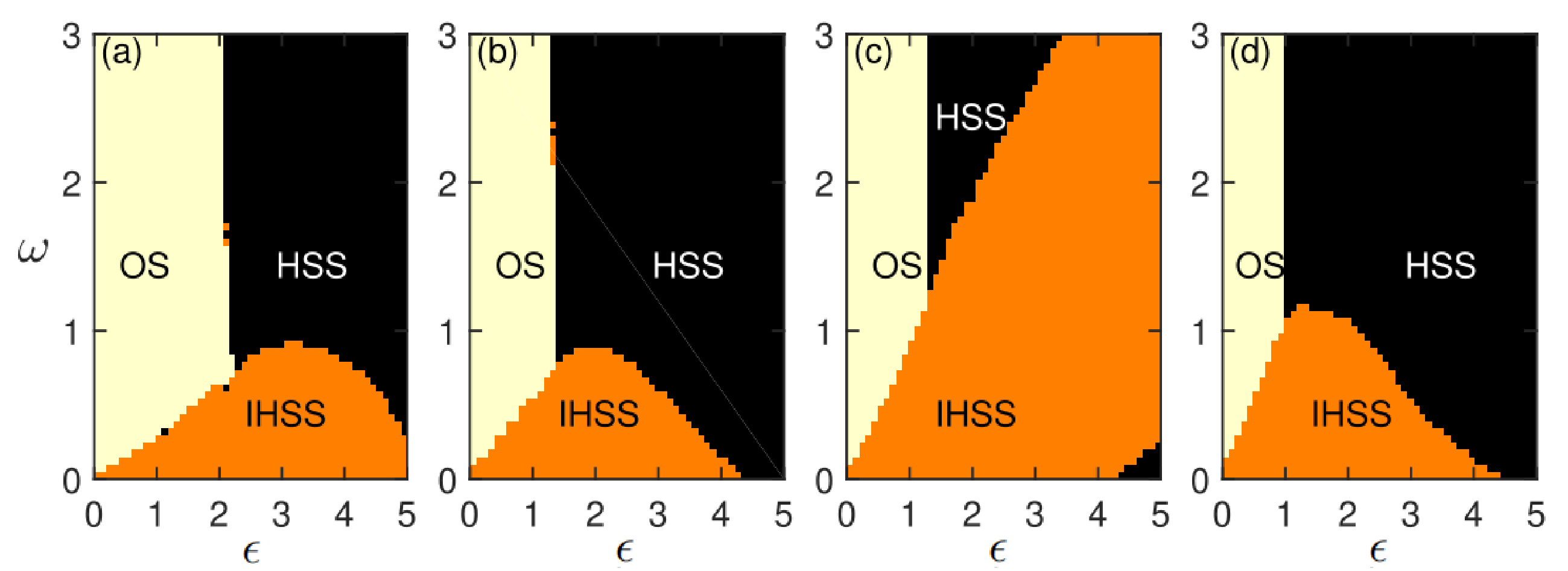}
\caption{Phase diagrams for the system~\eqref{eq-wc} in the $(\epsilon, \omega)$ space for $p=0.8$ for conjugate coupled Stuart-Landau oscillators for coupling strength for intra groups (a) $m_1=m_2=1$, (b) $m_1=1, m_2=2$, (c)$m_1=2, m_2=1$ and (d) $m_1=1,m_2=3$.}
\label{wcc_f2}
\end{figure}

\par The impact of the natural frequency on the aging transition is shown in the ($\epsilon-\omega$) parameter plane for different combinations of ($m_1, m_2$) at fixed inactivation ratio $p=0.8$ in Fig.~\ref{wcc_f2}(a-d). In the parameter plane, dynamical transitions occur through three bifurcations (similar to Fig.~\ref{wcc_f1}) among the oscillatory regime (OS), HSS and the inhomogeneous steady state (IHSS).  The variation in the spread of the aging transition island as a function of $\omega$ and $\epsilon$ for different combinations of ($m_1, m_2$) is depicted in Fig.~\ref{wcc_f2}. Increasing the coupling strength of inactive oscillators through $m_2$ enlarges the aging transition region (HSS), as a result of the influence of the inactive oscillators over the active group of oscillators. On the other hand, with the increase of $m_1$ the aging transition region, i.e., the HSS region shrinks in the parameter plane due to the influence of active oscillators over the inactive ones in the globally conjugate coupled Stuart-Landau oscillators in Figs.~\ref{wcc_f2}(a-d). \\



\subsubsection{Attractive and repulsive interaction}

Up to this point, we have explored the concept of aging transition, predominantly focusing on scenarios where interactions between dynamical nodes are attractive (positive). However, real-world systems often exhibit a more intricate nature, involving mixed coupling with both attractive (positive) and repulsive (negative) connections~\cite{majhi2020perspective}. It is important to emphasize that the coexistence of positive and negative couplings can simultaneously occur in diverse physical, ecological, and biological systems~\cite{giron2016synchronization,daido1987population,dixit2020static}. In recent years, numerous studies have delved into the emergent dynamics within an ensemble of oscillators featuring attractive-repulsive coupling. These investigations have showcased the emergence of manifold collective dynamics, including chimera states, solitary states, extreme events, amplitude (or oscillation) death, and anti-phase synchrony in a network of oscillatory nodes experiencing both attractive and repulsive interactions~\cite{majhi2020perspective,hens2013oscillation,dixit2020static,sharma2021explosive,chowdhury2021antiphase,chowdhury2020effect}. The interplay between attractive and repulsive couplings can lead to suppression of oscillatory activities among coupled oscillators. As a result,  the phenomenon of aging transition  exhibits qualitative differences in the presence of competitive attractive-repulsive interactions.  In this direction the very first work was carried out by Bera~\cite{bera2020additional}. The mathematical form of a  $N$ coupled network of Stuart-Landau oscillators having attracting and repulsive interactions is,

\begin{eqnarray}
	\dot{z}_j &=& (\alpha_j+i\omega-|z_j|^2)z_j+\frac{K}{N}\sum_{k=1}^{N}C_{jk}G(z_k,z_j)-\frac{\epsilon}{N}\sum_{k=1}^{N}B_{jk}H(z_k,z_j),
 \label{diff-repul}
\end{eqnarray}

where $j=1,2,\dots,N$. Here $C_{jk}$ and $B_{jk}$ represent the adjacency matrices of attractive and mean repulsive interactions, respectively. The functional forms of  attractive and repulsive interactions are defined  by  $G(z_k, z_j)=(z_k-z_j)$ and $H(z_k, z_j)=(z_k+z_j)$. In this context, the parameters $K$ and $\epsilon$ represent the strengths of attractive and repulsive interactions, respectively.

To determine the critical threshold $p_c$ of the aging parameter, we set $z_j = A$ for all active oscillators and $z_j = I$ for all inactive oscillators. Consequently for a globally connected network  Eq. \eqref{diff-repul} reduced to,

\begin{eqnarray}
	\dot{A} &=& (a+i\omega-|A|^2-2 \epsilon-Kp-p\epsilon)A+(K-\epsilon)pI, \nonumber \\
	\dot{I} &=& (-b+i\omega-|I|^2- \epsilon-Kq-p\epsilon)I+(K-\epsilon)qA, \nonumber \\	
\end{eqnarray}
where $q = 1-p$. Employing  a linear stability analysis of this reduced model around the origin, we derive the  critical value $p_c$ as 

\begin{eqnarray}
	p_c &=& \frac{(a-2\epsilon)(b+K+\epsilon)}{(a+b)(K-\epsilon)}, ~~~\text{for $\epsilon+K\ge a$}.
\end{eqnarray}

\noindent When $\epsilon + K \leq a$, the critical parameter $p_c$ remains at unity. In the limiting case as $\epsilon$ approaches 0, $\lim_{\epsilon \to 0} p_c = \frac{a(b+K)}{K(a+b)}$, aligning with the critical aging parameter for a diffusively coupled global network. Additionally, $\lim_{K \to \infty} p_c = \frac{a - 2\epsilon}{a+b}$. Consequently, it can be inferred that an increase in the repulsive interaction strength $\epsilon$ leads to a decrease in dynamical robustness.

\par
\begin{figure}[h]
\centering
\includegraphics[width=0.70\textwidth]{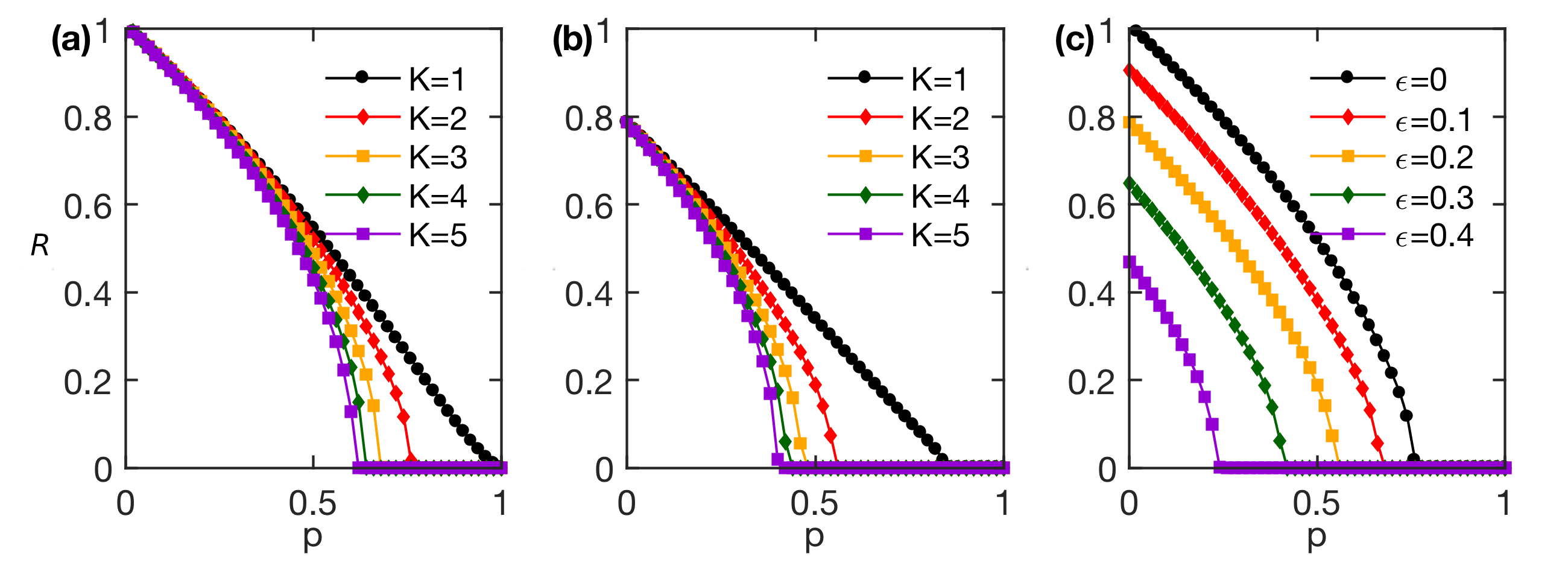}
\caption{The order parameter $R$ as a function of aging parameter $p$, for the system~\eqref{diff-repul}. (a)  In absence of repulsive interactions  ($\epsilon$=0 ) for different values of attractive coupling strengths $K$. (b) In presence of repulsive interactions ($\epsilon=0.2$) for different values of $K$. (c) For different values of repulsive coupling strength $\epsilon$ at a fixed value of $K=2$.}
\label{drl_f1}
\end{figure}

 We plot the order parameter $R$ as a function of aging parameter  $p$  for various scenarios. Figs.~\ref{drl_f1}(a) and~\ref{drl_f1}(b) depict the transition scenario for $\epsilon=0$ and $\epsilon=0.2$ respectively, considering a spectrum  of attractive coupling strengths. The general trend of aging transition remains same for both settings (absence and presence of repulsive interaction) for increasing attractive coupling strength $K$. But, for the latter case of $\epsilon=0.2$, the aging occurs for smaller values of the inactivation ratio $p$ compared to the former case of $\epsilon=0$, implying a decline of the robustness. This illustration demonstrates that the inclusion of repulsive interaction significantly reduces the dynamical robustness.  In Fig.~\ref{drl_f1}(c), we portray, how the transition takes place for increasing values of $\epsilon$, with a fixed attractive coupling strength $K=0.2$. The figure clearly shows that for increasing $\epsilon$, the aging transition keeps occurring for lower values of $p$. Thus,  Fig.~\ref{drl_f1}(c) further confirms that  repulsive interactions negatively impacts dynamical robustness.\\


\begin{figure}[ht]
\centering
\includegraphics[width=0.50\textwidth]{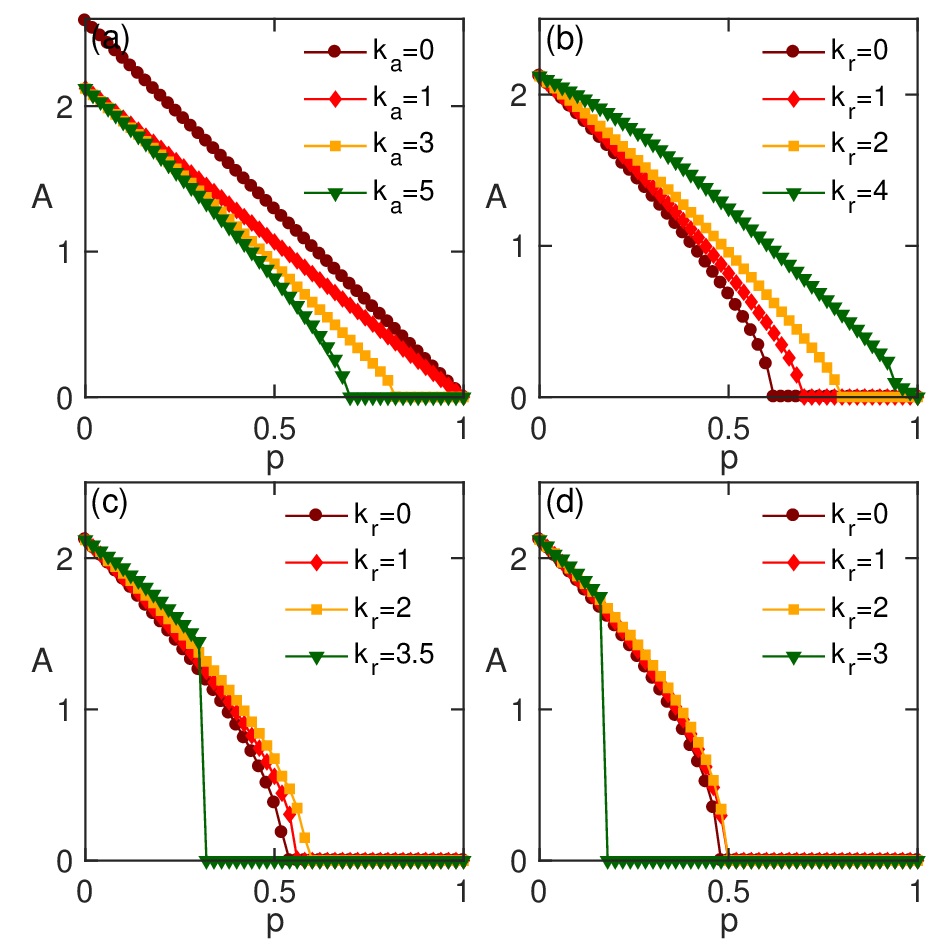}
\caption{The average amplitude (A) for the system~\eqref{e1} varying with the inactivation ratio $p$ for various values of (a) attractive coupling strength $k_a$ at fixed repulsive coupling $k_r=1$ and for different repulsive coupling strength $k_r$ at fixed attractive coupling strength (b) $k_a=5$ ,(c) $k_a=7$ and (d) $k_a=9$ for $N=500$ coupled Stuart-Landau oscillators with $\omega=5$. Reprinted figure with permission from Ref.\cite{sharma2022}.}
\label{arc_f1}
\end{figure}
------------------------------------------------------------------------------------------

In contrast to the above study based upon the assumption of attractive and repulsive interactions in both real and imaginary variables of Stuart-Landau oscillators, our next analysis delves into the robustness of an alternative form of attractive and repulsive interactions, as outlined in \cite{sharma2022}. Consider $N$ coupled  Stuart-Landau oscillators with attractive and repulsive interactions given by,

\begin{eqnarray}
\dot{z}_j(t) &=& (\alpha_j+i\omega-|z_j(t)|^2)z_j(t)+\frac{k_a}{\lambda_j}\sum_{l=1}^NB_{jl}(\mbox{Re}(z_l)-\mbox{Re}(z_j))-\frac{k_r}{\lambda_j}\sum_{l=1}^NB_{jl}(\mbox{Img}(z_l)-\mbox{Img}(z_j)),
\label{e1}
\end{eqnarray}

\begin{figure}[t]
\centering
\includegraphics[width=0.50\textwidth]{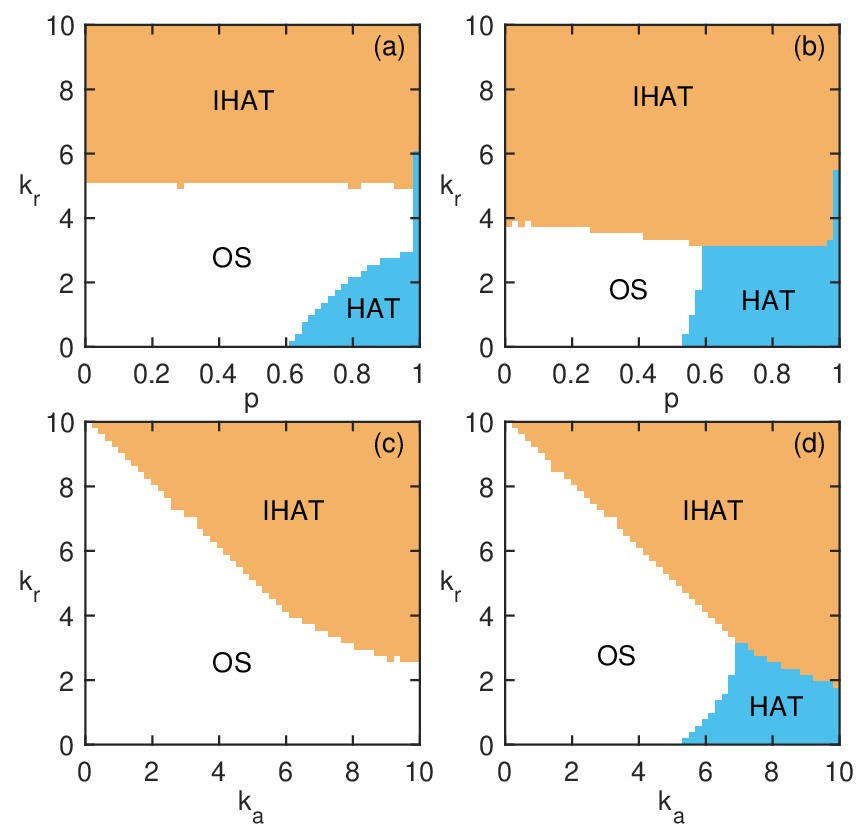}
\caption{The ($p, k_r$) parameter plane's phase diagram is depicted for a network of Stuart-Landau oscillators~\eqref{e1} that are coupled in an attractive-repulsive manner for (a) when the attractive coupling strength is set to $k_a=3$, and (b) when it is adjusted to $k_a=7$. The ($k_a, k_r$) parameter  plane exhibits a phase diagram for two  inactivation ratio (c) $p = 0.3$ and (d) $p = 0.6$. The regions corresponding to homogeneous aging transition (HAT), inhomogeneous aging transition (IHAT), and oscillatory state (OS) are indicated by light blue, orange, and white colors, respectively. Reprinted figure with permission from Ref.\cite{sharma2022}.}
\label{arc_f2}
\end{figure}

\begin{figure}[t]
\centering
\includegraphics[width=0.50\textwidth]{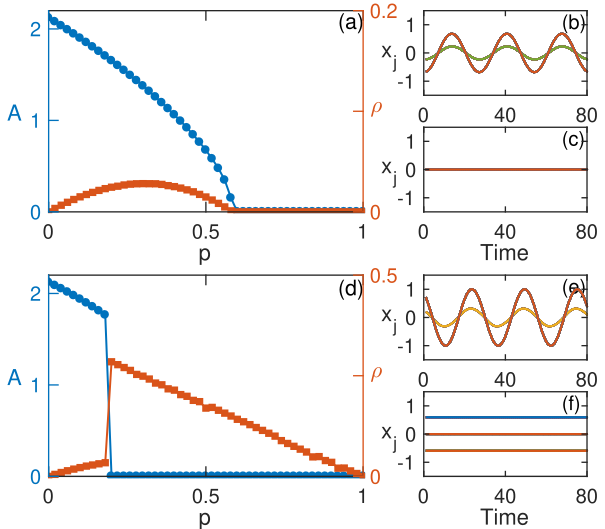}
\caption{Average amplitude ($A$) and variance ($\rho$) for the system~\eqref{e1} as a function of aging parameter $p$ for repulsive coupling (a) $k_r=2$ and (d) $k_r=3.6$ at  a fixed attractive coupling $k_a = 7$. The time series of the $x_j$ variable for the oscillator is depicted at various values of the inactivation ratio (b) $p=0.4$, (c) $p=0.6$ at fixed $k_r=2$, and (e) $p=0.1$, (f) $p=0.3$. Reprinted figure with permission from Ref.\cite{sharma2022}.}
\label{arc_f3}
\end{figure}

\par 
Here  $k_a$, and $k_r$  indicate the magnitude of attractive and repulsive interactions in coupled oscillators, respectively. $B_{jl}$ is the connection matrix, and $\lambda_j$ is the average degree of the $j$th node of the coupled oscillators.  Two order parameters, namely, the average amplitude  
 $A_{amp}= \frac{1}{N}\sum_{i=1}^{N}(\langle x_{i,max}\rangle_t-\langle x_{i,min}\rangle_t)$ and the variance $\rho=\sigma^2(\langle(x_i)\rangle_t)$  are used to identify the aging transition. These order parameters are employed to differentiate between homogeneous and inhomogeneous steady states. Specifically, when $A_{amp} = \rho = 0$, all oscillators converge to HSS, while $A_{amp} = 0$ and $\rho \neq 0$ indicate the presence of oscillators in IHSS.

\par In Fig.~\ref{arc_f1}, the average amplitude ($A_{amp}$) is plotted with respect to $p$ for different combinations of the attractive coupling $k_a$ and the repulsive coupling strength $k_r$. Figure~\ref{arc_f1}(a) suggests that as  we augment $k_a$, the critical inactivation ratio $p_c$ decreases, indicating a faster occurrence of the phase transition due to aging. This observation aligns with earlier studies suggesting that increased attractive couplings serve to dampen oscillations. Conversely, it is evident from Fig.~\ref{arc_f1}(b) that repulsive interactions enhance robustness to large extent. Yet, in Figs.~\ref{arc_f1}(c-d), it becomes evident that as the repulsive interaction strength $k_r$ reaches higher values, a sudden phase transition unfolds in the order parameter $A$. This transition is characterized by its catastrophic nature, with the magnitude of the order parameter providing no forewarning of the impending collapse of the system. Consequently, our observation suggests that depending on parameter values, repulsive interactions have the potential to bolster dynamical robustness in coupled oscillators, but concurrently, they might instigate an abrupt suppression of macroscopic oscillations within the network.

\par Subsequently, we delve into the mechanism behind this sudden catastrophic phase transition by examining the 2-dimensional parameter space. The phase diagram on the $k_r-p$ plane is presented in Figs.~\ref{arc_f2}(a-d) for various values of attractive coupling strength. This figure delineates three distinct regions: OS (oscillatory state), IHAT (inhomogeneous aging transition state), and HAT (homogeneous aging transition state). With an increase in $k_a$, we note an expansion in both the IHAT and the HAT regions. Additionally, it is observable that for higher values of $p$, the transition from homogeneous steady-state to an inhomogeneous one occurs with variations in $k_r$. This result is also apparent in Fig.~\ref{arc_f3}, where we have graphed $A_{amp}$ and $\rho$ as functions of $p$, alongside the corresponding time series near the aging transition point. In Fig.~\ref{arc_f3}(a), a typical aging transition is evident as the order parameters $A$ and $\rho$ exhibit smooth functions of $p$ and converge to zero at a critical value $p=p_c$. The corresponding time-series of the oscillators on both sides of the aging transition point $p_c$ is depicted in Figs.~\ref{arc_f3}(b-c). However, Figs.~\ref{arc_f3}(d-f) portray an aging transition that is qualitatively distinct from the previous one. In this case, a discontinuity in the order parameters is noticeable at the aging transition point $p_c$. Additionally, the corresponding time series indicates an aging transition through an inhomogeneous steady-state, where oscillators settle into three different steady states after the aging transition. Therefore, we can infer that the abrupt discontinuous jump in the order parameter is a consequence of a bifurcation from the oscillatory state to inhomogeneous steady states.


\subsubsection{Dynamical robustness in the absence of inactive oscillators}

Thus far, whatever we have discussed above is the scrutinization of the phenomenon of dynamical robustness in ensembles of dynamical systems achieved by increasing the proportion of inactive oscillatory systems. Now, we move aside from this specification of the dynamical units, and demonstrate that even an ensemble of co- and counter-rotating oscillatory systems~\cite{prasad2010universal,bhowmick2011synchronization} can exhibit similar aging transition upon raise in the fraction of counter-rotating oscillators. The impact of the mean-field feedback in the symmetry preserving as well as the symmetry breaking coupling is explored in Ref.~\cite{sathiyadevi2019aging}. The network becomes dynamically vulnerable completely and hence the regime of global oscillation transits to aging through a Hopf bifurcation, whereas the transition from aging to oscillation death occurs via a pitchfork bifurcation. 

\par The dynamical evolution of a network of $N$ globally interacting Stuart-Landau oscillators with symmetry preserving coupling is described as, 
\begin{equation}
	\begin{array}{lcl}\label{abs1}
		\dot {z_j}=(\lambda+ i\omega_j-|{z_j}|^2)z_j+\dfrac{K}{N}\sum\limits_{k=1}^{N}(\alpha z_k-z_j);~~~~~~j= 1,2,\cdots,N,
	\end{array}
\end{equation}
where $\alpha$ refers to the strength of the mean-field feedback. Moreover, $\omega_j$ is the frequency of the $j$-th dynamical unit. Specifically, for a frequency $+\omega$, the system rotates in a counter-clockwise direction, and in the clockwise direction whenever the frequency is $-\omega$. Analogous to the studies with inactive oscillators, we divide the entire system into two groups. To be precise, the frequency $\omega_j=\omega$ for $j \in 1,2,\dots, N-Np$ and frequency $\omega_j=-\omega$ for $j \in N-Np+1,\dots, N$, and consequently the parameter $p$ characterizes the proportion of the counter-rotating oscillators in the networked system. 

\begin{figure}
	\centerline{\includegraphics[scale=0.400]{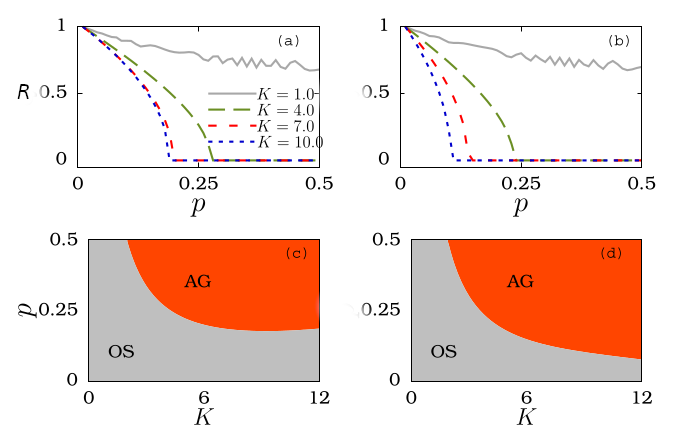}}
	\caption{ Normalized order parameter $R$, for the system~\eqref{abs1} with respect to the fraction $p$ of the counter-rotating oscillators for (a) $\alpha=1.0$ and (b) $\alpha=0.95$, with fixed $\omega=5$, $\lambda=1$ and $N=100$. Four different values of the interaction strength are chosen as $K=1,4,7$, and $10$. The corresponding phase diagrams in the $(K,p)$ parameter plane are plotted in (c) and (d), respectively. The oscillatory and aging regimes are respectively represented by OS and AG. Reprinted figure with permission from Ref.\cite{sathiyadevi2019aging}.}
	\label{fig12}
\end{figure}

\par Further choosing the network size $N=100$, frequency $\omega=5.0$, and defining the normalized order parameter $R$ similarly as before, we plot $R$ with respect to the fraction $p$ of counter-rotating oscillators in Fig.~\ref{fig12}, for two different values of the mean-field feedback strength $\alpha=1.0$ and $\alpha=0.95$. The dynamical outcomes being symmetric for both the ranges $p \in (0,0.5)$ and $p \in (0.5,1)$, we effectively confine ourselves to the range $p \in (0,0.5)$ in the plots. In Fig.~\ref{fig12}(a), we encounter that whenever $\alpha=1.0$, with $K=1$, $R$ remains non-zero finite for the entire range of $p$, depicting oscillatory regime of all the individual units and hence the whole network. However, with higher coupling strengths $K=4,7,$ and $10$, the order parameter $R$ transits from non-zero to zero value and hence aging transition takes place through Hopf bifurcations at $p_{HB}=0.28,0.20,$ and $0.19$, respectively. Thus, increasing coupling strength decreases the critical value of $p$ at which aging occurs via Hopf bifurcation. In order to perceive the impact of the mean-field feedback on the dynamical robustness, we depict similar transitions for a lower feedback strength $\alpha=0.95$, with the same set of interaction strengths (in Fig.~\ref{fig12}(b)). Similarly as before, the entire system shows oscillatory nature for $K=1$. For increasing $K=4,7,$ and $10$, the aging transition occurs at the Hopf bifurcation points $p_{HB}=0.24,0.15,$ and $0.10$ respectively. Thus, even for a tiny decrement in the feedback strength $\alpha$, the onset of aging transition takes place for smaller values of the fraction $p$. For a better understanding, we further portray this scenario through the cooresponding phase diagrams in the $(K,p)$ parameter planes, in the lower panel of Fig.~\ref{fig12}. As can be witnessed, for both values of the mean-field feedback strength ( Figs.~\ref{fig12}(c) and~\ref{fig12}(d)), the networked system displays oscillatory solutions for a small interaction strength and smaller proportion of the counter-rotating oscillators. However, as $K$ and $p$ surpass certain values, aging transitions take place. The regimes of oscillation and aging are classified as OS and AG, respectively. More importantly, for smaller feedback strength, the aging region expands in the phase diagram, as expected from earlier observations. \\
Our investigation thus demonstrates how networked dynamical systems comprising of co- and counter-rotating oscillators instead of active-inactive decomposition, can also give rise to the phenomenon of aging transition and, hence, a potential of the study of dynamical robustness of networked systems from a different perspective.   


\subsection{Dynamical robustness of complex networks}

In the above, we have examined the dynamical robustness of globally coupled networks under various coupling schemes. In this section, we turn our attention to aging transitions within different complex network topologies. In the past two decades, network science has experienced significant advancements due to the discovery of various topological structures in many real-world networks~\cite{barabasi2016network,newman2003structure,boccaletti2006complex,strogatz2001exploring,albert2002statistical,estrada2012structure,costa2011analyzing}. The complexity of a network structure can be characterized by the connectivity properties of interaction pathways (links) among network components (nodes).  In terms of the degree distribution (the probability distribution of node degrees across the network), complex networks mainly fall into two categories: homogeneous and heterogeneous networks. Homogeneous networks, exemplified by random graphs~\cite{erdHos1960evolution} and small-world models~\cite{watts1998collective}, exhibit a binomial or Poisson degree distribution where node degrees cluster around the mean degree. In contrast, heterogeneous networks like scale-free networks display a heavy-tailed degree distribution that approximately follows a power-law distribution~\cite{barabasi1999emergence}. We will first discuss the results obtained for homogeneous complex networks, followed by the study of heterogeneous networks. Then, we will present the findings for weighted complex networks, and finally, we will discuss the dynamical robustness within a correlated network topology.


\subsubsection{Homogeneously coupled complex networks}
 Tanaka et al. \cite{tanaka2012dynamical} extended the work of Daido and Nakanishi \cite{Daido_2004_PRL} to complex network structures. They considered a  homogeneously coupled network having a Poisson degree distribution. In particular they explored an Erdös-Rényi random graph \cite{erdHos1960evolution} by considering a network of diffusively coupled oscillators as follows:
\begin{equation}
\dot{z_j}(t)=(\alpha_j+i\omega-|z_j(t)|^2)z_j(t) +\frac{K}{N}\sum_{l=1}^N A_{jl}(z_l-z_j).
\label{sf0}
\end{equation}

Using the system reduction techniques proposed by Daido and Nakanishi \cite{Daido_2004_PRL}, the critical value $p_c$ of the inactivation parameter can be analytically calculated. For a homogeneous network, the degree of a node can be approximated by the average degree of the network. With this assumption, we can conclude that each active oscillator is connected to neighboring $(1-p)\langle k \rangle$ oscillators and each inactive oscillator is to $p\langle k \rangle$ neighboring oscillators. Here $\langle k \rangle$ represents the mean degree of the network. When we designate $z_j$ as 'A' for the active group and 'I' for the inactive group of oscillators, Eq.~(\ref{sf0}) in its original form transforms then into the following coupled system: 

\begin{eqnarray}
\dot{A}(t)&=&(a-Kpd+i\omega-|A(t)|^2)A(t)+KpdI(t),\nonumber\\
\dot{I}(t)&=&(-b-K(1-p)d+i\omega-|I(t)|^2)I(t)+K(1-p)dA(t).\nonumber\\
\label{eq5-br}
\end{eqnarray}
\noindent A linear stability analysis of the equilibrium point $(A,I)=(0,0)$ leads to 

\begin{eqnarray}
p_{c}^{hom}=\frac{a(Kd+b)}{(a+b)Kd} & \text{for $K>K_c^{hom}$},
\label{eq-6br}
\end{eqnarray}
where $d$ represents the link density, defined as $\langle k\rangle/N$, and $K_c^{hom}=a/d$ provides the critical coupling strength, below which $p_c^{hom}=1$. Here we consider $\alpha_j=a$ for an active oscillator and $\alpha_j=-b$ for an inactive oscillator. From the Eq.~(\ref{eq-6br}), we can conclude that for a fixed  $K$, the robustness increases with the decrease of link density $d$.


\subsubsection{Dynamical robustness of heterogeneous networks}\label{het}

To understand how robust a networked dynamical system is under increasing inactivation of its dynamical units whenever the degree distribution of the network is heterogeneous, we present a detailed comparative analysis of aging transition between homogeneous and heterogeneous networks~\cite{tanaka2012dynamical}, while considering a network of $N$ dynamical systems as described by the networked system Eq.~\eqref{sf0}. In order to derive the desired expression for the critical inactivation ratio $p_c$ for heterogeneous networks under random inactivation, we follow the degree-weighted mean field approximation~\cite{pastor2001epidemic}. Then the original system (Eq.~\eqref{sf0} can be approximated as
\begin{equation}
	\begin{array}{lcl} \label{sf3}
		\dot{z_j} = (\alpha + i\omega - |z_j|^2)z_j + \dfrac{K{k_j}}{N}\Big[(1-p)M_A(t)+pM_I(t)-z_j\Big],\\
	\end{array}
\end{equation}
where

\begin{equation}
	\begin{array}{lcl} \label{sf4}
		M_A(t) = \dfrac{\sum_{j\in S_A} {k_j}z_j(t)}{\sum_{j\in S_A} {k_j}},
        ~~M_I(t) = \dfrac{\sum_{j\in S_I} {k_j}z_j(t)}{\sum_{j\in S_I} {k_j}}
	\end{array}
\end{equation}
are the degree-weighted mean fields for the active and inactive sets of dynamical units,
with ${k_j}(j=1,2,...,N)$ being the degree of the $j$-th node in the network.\\
Based upon the fact that the oscillators display phase synchronization with frequency $\Omega$, let us now assume that the state variables can be written as $z_j(t)=r_j(t)e^{i(\Omega t+\theta)}$, $r_j$ being the amplitude and $\theta$ being the phase shift. On replacing this in Eq. \eqref{sf3}, we get
\begin{equation}
	\small{\begin{array}{lcl}  \label{sf6}
			\dot{r}_j = \Big(\alpha_j - \dfrac{Kk_j}{N}-r_j^2\Big)r_j + \dfrac{Kk_j}{N}\Big[(1-p)R_A(t)+pR_I(t)\Big],\\
	\end{array}}
\end{equation}
where
\begin{equation}
	\begin{array}{lcl} \label{sf7}
		R_A(t) = \dfrac{\sum_{j\in S_A} {k_j}r_j(t)}{\sum_{j\in S_A} {k_j}},
		~~R_I(t) = \dfrac{\sum_{j\in S_I} {k_j}r_j(t)}{\sum_{j\in S_I} {k_j}}.
	\end{array}
\end{equation}
Further supposing that in the stationary oscillatory regime, $R_A(t)$ and $R_I(t)$ are time-independent, the phase transition from the oscillatory ($R_A,R_I>0$) regime to the non-oscillatory ($R_A=R_I=0$) regime eventuates due to the change in the stability of the equilibrium point at the origin. The stability is governed by the following Jacobian matrix 
$$J_0=\begin{pmatrix}
	\dfrac{\partial G_A(R_A,R_I)}{\partial R_A} & \dfrac{\partial G_A(R_A,R_I)}{\partial R_I} \\
	\dfrac{\partial G_I(R_A,R_I)}{\partial R_A} & \dfrac{\partial G_I(R_A,R_I)}{\partial R_I}
\end{pmatrix}\Biggr\rvert_{R_A=R_I=0},$$
in which
\begin{equation}
	\begin{array}{lcl}  \label{sf8}
		G_A(R_A,R_I)=\dfrac{\sum_{j\in S_A} {k_j}r_j^*(R_A,R_I)}{\sum_{j\in S_A} {k_j}},\\\\
		G_I(R_A,R_I)=\dfrac{\sum_{j\in S_I} {k_j}r_j^*(R_A,R_I)}{\sum_{j\in S_I} {k_j}}.
	\end{array}
\end{equation} 
The stationary amplitude $r_j^*$ is a positive real solution of the following equation,
\begin{equation}
	\begin{array}{lcl} \label{sf9}
		r_j^3 - \Big(\alpha_j-\dfrac{Kk_j}{N}\Big)r_j - \dfrac{Kk_j}{N}\Big((1-p)R_A+pR_I\Big)=0.
	\end{array}
\end{equation}
Eq.~\eqref{sf9} has only one positive real root if 
\begin{equation}
	\begin{array}{lcl} \label{sf10}
		\alpha_j - \dfrac{Kk_j}{N}<0,~\forall  j \in S_A. 
	\end{array}
\end{equation}

Now we differentiate Eqns. \eqref{sf8} and \eqref{sf9} with respect to $R_A$ and we find the $(1,1)$-th entry of $J_0$ as,
\begin{equation}
	\begin{array}{lcl} \label{sf11}
		\dfrac{\partial G_A}{\partial R_A}\Bigr\rvert_{R_A=R_I=0} = \dfrac{\dfrac{(1-p)}{N}K}{\sum_{j \in S_A}{k_j}} \Big[\sum\limits_{j \in S_A}\dfrac{{k_j}^2}{(Kk_j/N)-\alpha_j}\Big] \\ \\
		~~~~~~~~~~~~~~~~~~~\simeq \dfrac{1}{d}\Big(\dfrac{1}{N} \sum\limits_{j \in S_A}\dfrac{{d_j}^2}{d_j-\alpha_j/K} \Big) ,
		
		\\ 
	\end{array}
\end{equation}
in which $d_j=k_j/N~(j=1,2,...,N)$ is the normalized degree of the $j$-th node, where the approximation $\sum_{j \in S_A}{k_j} \simeq (1-p)dN^2$ is used. Also, in the large $N$ limit, the link density is $d= \langle k \rangle /(N-1)$. \\

If we now define
\begin{equation}
	\begin{array}{lcl} \label{sf12}
		H(K,\alpha)=\dfrac{1}{N}\sum\limits_{j=1}^{N}\dfrac{{d_j}^2}{d_j-\alpha/K},
	\end{array}
\end{equation}
then we can write
\begin{equation}
	\begin{array}{lcl} \label{sf13}
		\dfrac{\partial G_A}{\partial R_A}\Bigr\rvert_{R_A=R_I=0}\simeq
		(1-p)H(K,a)/d. 
	\end{array}
\end{equation}
Similarly, we obtain
\begin{equation}
	\begin{array}{lcl} \label{sf14}
		\dfrac{\partial G_A}{\partial R_I}\Bigr\rvert_{R_A=R_I=0}\simeq
		pH(K,a)/d,\\
		\dfrac{\partial G_I}{\partial R_A}\Bigr\rvert_{R_A=R_I=0}\simeq
		(1-p)H(K,-b)/d,\\
		\dfrac{\partial G_I}{\partial R_I}\Bigr\rvert_{R_A=R_I=0}\simeq
		pH(K,-b)/d.\\
	\end{array}
\end{equation}
Thus we arrive at 
$$J_0=\begin{pmatrix}
	(1-p)H(K,a)/d & pH(K,a)/d\\
	~~(1-p)H(K,-b)/d &~~ pH(K,-b)/d
\end{pmatrix}.$$

\begin{figure}[ht]
	\centerline{\includegraphics[scale=0.400]{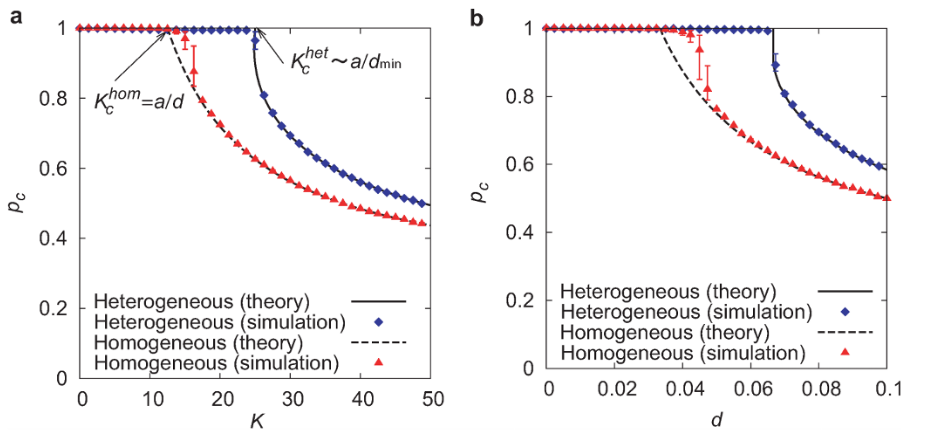}}
	\caption{ (a) Critical inactivation ratio $p_c$ as a function of the interaction strength $K$ in networked systems possessing $d \sim 0.08$ ($N=3000$ and $\langle k\rangle=240$). The dashed and solid black curves respectively stand for the analytically obtained results in Eqs.~\eqref{eq-6br} and~\eqref{sf15}. Red triangles and blue diamonds represent the numerical outcomes. The error bars correspond to the variance for $10$ network realizations. (b) The critical ratio $p_c$ with respect to the link density $d$ in networked system with $N=3000$ and $K=30$. Reprinted figure with permission from Ref.\cite{tanaka2012dynamical}.}
	\label{fig1}
\end{figure}

The equilibrium point at $R_A=R_I=0$ changes its stability near the phase transition, providing us with the following critical inactivation ratio
\begin{equation}
	\begin{array}{lcl} \label{sf15}
	{p_c}^{het} = \dfrac{H(K,a)-d}{H(K,a)-H(K,-b)} ~~~\mbox{for}~~~K > {K_c}^{het},
	\end{array}
\end{equation}
in which the definition of $H$ follows from Eq. (\ref{sf12}). The result is true if $K> a/d_{min}$ where $d_{min}=k_{min}/N$.

\par Figure~\ref{fig1} portrays a comparative analysis between the critical inactivation ratios obtained for the homogeneous and the heterogeneous networks. In Fig.~\ref{fig1}(a), we depict the critical ratio $p_c$ with respect to the interaction strength $K$, with a fixed linked density of $d \sim 0.08$ for a network of size $N=3000$ and the mean-degree $\langle k\rangle=240$. The dashed and solid black curves respectively stand for the analytically obtained outcomes in Eqs.~\eqref{sf2} and \eqref{sf15}, whereas the red triangles and blue diamonds represent the numerical results. As can be witnessed, the analytically obtained expressions are in sufficiently good agreement with the numerical results. $p_c$ values expectedly start decreasing with increasing $K$, after the respective certain critical interaction strengths ${K_c}^{hom}$ and ${K_c}^{het}$. The critical ratio $p_c$ for the homogeneous network is smaller than that for the heterogeneous one for the entire range of the coupling strength $K \in [0,50]$. Similar results are shown in Fig.~\ref{fig1}(b), but this time for varying link density $d \in [0,0.1]$ and fixed coupling strength $K=30$.  Qualitatively the same scenario remains valid here as far as the evolution of $p_c$ are concerned.  

\par This analytical study for random failures has further been generalized to targeted attacks~\cite{yuan2017robustness}. The study presents a universal formula for the critical fraction of inactive units, applicable to both random failures and targeted attacks on networked systems. It examines the impact of targeting nodes based on their degrees, starting with identical oscillators and homogeneous edge weights. The theory is subsequently extended to networks with heterogeneous edge weights and non-identical oscillators. The analytical findings are confirmed through extensive numerical simulations. On the other hand, Tanaka et al.~\cite{tanaka2014dynamical} presented a general formula for the critical inactivation ratio for interacting heterogeneous oscillators. This is done while assuming different values of the units' intrinsic parameters, instead of choosing the same fixed parameter value for all the active and all the inactive elements. The study demonstrates that increasing heterogeneity in the oscillator components of networks leads to improved dynamical robustness, as evidenced by the comparison of critical values for networks with various extents of heterogeneity. This observation is further validated for networks of Morris-Lecar neuronal systems communicating through electrical synapses.  


\subsubsection{Weighted complex networks}

The majority of the early research on dynamical robustness focused on networks without taking weightings into account. But in reality, many complex networks are indeed weighted, and the connection strength of the nodes highly influences the network dynamics~ \cite{barrat2004architecture,song2005highly,scannell1999connectional}. 
Both the topology and the strength of a network's connections have an impact on its dynamics. In particular, studies have shown that in complex networks, the presence of both degree heterogeneity and weight heterogeneity tends to impede full synchronization. However, synchronization can be notably enhanced and made unaffected by these types of heterogeneity, when the weight distribution is appropriately integrated with the degree distribution~\cite{motter2005network,motter2005weighted,motter2005enhancing}. 

\par He et al. \cite{Zhan_2013_Physica-A} first studied dynamical robustness in a weighted complex network. Their investigation highlights the roles of high and low degree nodes on dynamical robustness in weighted complex networks and generalizes the works of Tanaka et al. \cite{tanaka2012dynamical}. Study of dynamical robustness in a weighted complex network has been carried out for $N$ diffusively  coupled Stuart-Landau oscillators as expressed below,
\begin{equation}
\dot{z_j}(t)=(\alpha_j+i\omega-|z_j(t)|^2)z_j(t) +K\sum_{l=1}^N W_{jl} A_{jl}(z_l-z_j).
\label{eq-7br}
\end{equation}


\noindent Here $A = (A_{jl})$ and $W = (W_{jl})$ are the adjacency and weight matrix of the network, respectively. The weight matrix $W$ is defined as $W_{jl}=\frac{1}{k_j^\beta}$, where $k_j$ is the degree of the $j$th node and $\beta$ is a tunable parameter, whose value determines whether the network is weighted~($\beta\neq0$) or unweighted~($\beta=0$). Dynamical robustness has been studied following the  mathematical framework proposed by Daido and Nakanishi \cite{Daido_2004_PRL}.

\begin{figure}[t]
\centering
\includegraphics[width=0.8\linewidth]{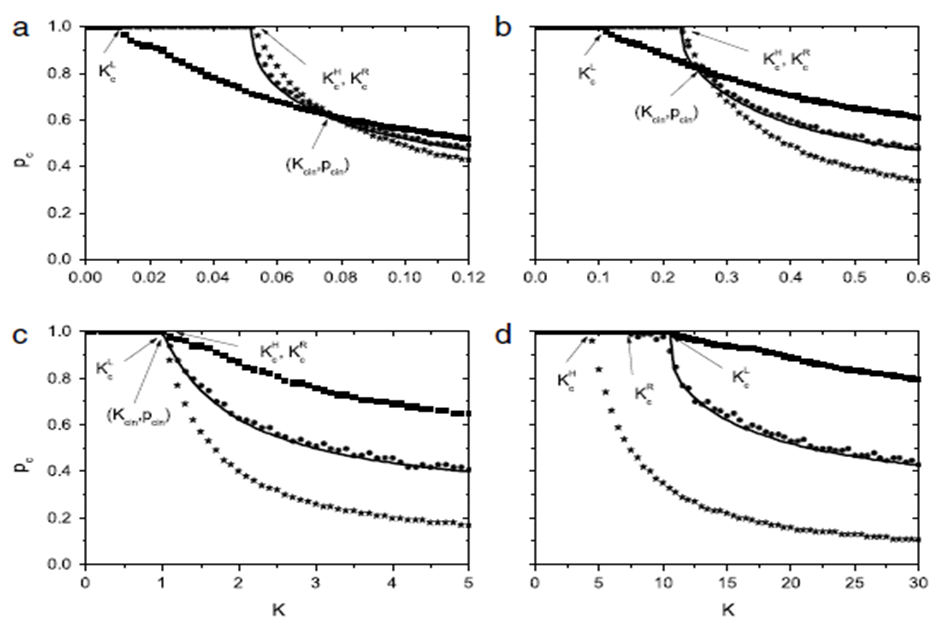}
\caption{The critical values $p_c$ (for the system~\eqref{eq-7br}) are plotted against $K$ for random inactivation (circles), targeted inactivation of high-degree nodes (stars), and that of low-degree nodes (squares) in a weighted heterogeneous networks. $K_c^R$, $K_c^H$, and $K_c^L$ represent their corresponding critical coupling strengths $K$ at $p_c = 1$. (a) $\beta = 0.2$, (b) $\beta=0.6$, (c) $\beta=1.0$, and (d) $\beta=1.4$. For $\beta<1$, an intersection point ($K_{cin}, p_{cin}$) exists, whereas for $\beta>1$, it vanishes and the curve of target on the low-degree nodes is always above that of target on the high-degree nodes. In (c), $K_c^R=K_c^H=K_c^L=K_{cin}$. Reprinted figure with permission from Ref.\cite{Zhan_2013_Physica-A}.}
\label{f-2br}
\end{figure}

\par In Fig.~\ref{f-2br}, the critical values $p_c$ are depicted against $K$ for different $\beta$ values in a weighted heterogeneous network. The graph includes data for random inactivation (circles), targeted inactivation of high-degree nodes (stars), and targeted inactivation of low-degree nodes (squares). Three distinct critical couplings at $p_c=1$ are indicated as $K_c^R$, $K_c^H$, and $K_c^L$. In Fig.~\ref{f-2br}(a-b) we observe that all the critical curves intersect at a point, denoted by $(K_{cin},p_{cin})$. Clearly for $0<\beta<1$, $p_c^L<p_c^H$, while $K<K_{cin}$ but for  $K>K_{cin}$ it becomes completely opposite.  This implies the crucial role of low-degree nodes in impacting the dynamical robustness which can only happen in a weakly weighted network and for weak coupling. For $\beta=1.0$, $p_{cin}=1$ and $K_c^R=K_c^H=K_c^L=K_{cin}$ as shown in Fig.~\ref{f-2br}(c). From Fig.~\ref{f-2br}(d) it is quite obvious that for $\beta>1$, $K_c^H<K_c^R<K_c^L$.   By applying the heterogeneous mean-field approximation, we can derive the critical value of the inactivation parameter as 

\begin{eqnarray}
p_{c}^{het}=\frac{F(K,a)-\langle k \rangle N}{F(K,a)-F(K,-b)} & \text{for $K>K_c^{hetR}$},
\label{eq-8br}
\end{eqnarray}

\noindent where $F(K,\alpha)=\sum_{j=1}^N \frac{Kk_j^{2-\beta}}{-\alpha+Kk_j^{1-\beta}}$, and $K_c^{hetR}=max_{j\in S_A}\{a k_j^{\beta-1}\}$. Here we consider $\alpha_j=a$ for an active oscillator and $\alpha_j=-b$ for an inactive oscillator.

\par The phenomenon reported in the Ref.~\cite{tanaka2012dynamical}, where heterogeneous networks are more susceptible to the failure of low-degree nodes rather than high-degree nodes, is observed specifically in weakly weighted networks and under weak coupling conditions. However, when considering the impact of weighted connections, we discover that the susceptibility to high-degree node failures is more widespread and occurs across a broader parameter range. This finding indicates a strong alignment between dynamical and structural robustness. In the case of the unweighted version of a heterogeneous network, high-degree nodes (hubs) are impacted by numerous neighbors, whereas low-degree nodes are influenced by only a small number of neighboring nodes, rendering them relatively isolated. Consequently, active low-degree nodes can sustain relatively high dynamical activity in comparison to active high-degree nodes. Consequently, carrying out targeted inactivation on low-degree nodes (rather than high-degree nodes) leads to a notable decrease in the network's dynamical activity. This aligns precisely with the observations made by Tanaka et al. in Ref.\cite{tanaka2012dynamical}. Nonetheless, as the overall coupling strength K increases, the exchange of information and dynamics among these networked nodes becomes more effective, eliminating the isolating effect previously experienced by low-degree nodes. As the weighted coupling scale $\beta$ increases, the coupling on each node (influenced by the weight matrix W) becomes more evenly distributed, diminishing the system's heterogeneity associated with the adjacency matrix $A$. A notable instance is $\beta=1$, where all nodes experience the same level of input signal intensity. Consequently, the distinctive isolating effect of low-degree nodes diminishes as well. Under these conditions, it is plausible to infer that the typical influential behavior exhibited by high-degree nodes extends across a broader parameter range, aligning with the findings of the structural robustness analysis.

Recently, Ray et al. \cite{Ray_2020_EPL} delved into aging transitions within a weighted heterogeneous network. In their investigation, weights are randomly selected from a uniform distribution $[0, w]$, where no connection exists between node degrees and the weights. Their findings reveal a direct link between weight heterogeneity and the aging transition point $p_c$. It was observed that dynamical robustness diminishes as the mean value $\bar w$ of the weight distribution increases. Heightened heterogeneity among connection weights correlates with reduced dynamical robustness. Moreover, the analytical expression of the critical value $p_c$ depends on both the mean weights and the network's average degree.


\subsubsection{Correlated networks}

\noindent Up to this point, we have explored the dynamical robustness of complex networks with structures characterized by degree distributions, which represent the probability distribution of the number of connections per node throughout the network. In particular, we have examined the dynamical robustness of these networks in relation to both homogeneously and heterogeneously connected networks featuring various degree distributions. Nonetheless, it is crucial to recognize that the degree distribution alone does not completely define the network's topology. Networks with identical degree distributions can exhibit diverse network structures. These distinctions can be quantified by examining \textit{network assortativity} concerning node degrees (degree-degree correlations), assessing the clustering coefficient, and considering various network characteristics\cite{newman2002assortative,mcgraw2005clustering}. Network assortativity, in particular, measures the correlation between a node's degree and the degrees of its neighboring nodes. In assortative networks, there is a positive correlation, meaning that nodes tend to link with others of similar degrees. Conversely, in disassortative networks, there is a negative correlation, indicating that high-degree nodes are more inclined to connect with low-degree nodes. Here, we examine the influence of network assortativity on the dynamical robustness of coupled oscillator networks, as explored in the study~\cite{sasai2015robustness}.

The assortativity coefficient $r$ is determined by computing the Pearson correlation coefficient of the degrees between pairs of connected nodes, and is calculated as follows,

\begin{eqnarray}
r = \frac{1}{\sigma^2_q}\sum_j\sum_k jk(E(j,k) - Q(j)Q(k)).
\label{cn-2}
\end{eqnarray}
Here $Q$ represents the probability distribution of the remaining degree, which quantifies the probability that a node in the end of a randomly chosen edge has $k$ edges except for the chosen one. The distribution of $Q$ is derived from the degree distribution $P(k)$ as $Q(k) = \frac{(k+1)P(k+1)}{\sum_j jP(j)}$.  The term $\sigma^2_q = \sum_k k^2Q(k) - (\sum_kQ(k))^2$ is the variance of the distribution $Q(k)$, and   $E(j,k)$ represents the joint probability distribution of the remaining degrees of two vertices. This distribution exhibits symmetry in undirected graphs and adheres to sum rules $\sum_j\sum_k E(j,k)=1$ and $\sum_j E(j, k) = Q(k)$. The assortativity coefficient $r$ in Eq (\ref{cn-2}) can be rewritten as follows,
\begin{eqnarray}
	r = \frac{4M\sum_mj_mk_m -[\sum_m(j_m+k_m)]^2}{2M\sum_m(j^2_m+k^2_m) -[\sum_m(j_m+k_m)]^2},
	\label{cn-4}
\end{eqnarray}

\noindent here $M$ represents number of edges in the network, $m\in{1,2,..,M}$ is the index of edges, and $j_m$ and $k_m$ represent the degrees of the two nodes $j$, and $k$ connected by the edge $m$. The assortativity coefficient $r$ varies within the range of $-1$ to $1$.  A value of $r$ greater than $0$ signifies assortative networks, $r=0$ indicates uncorrelated networks, and an $r<0$ implies disassortative networks.

\par
\begin{figure}[t]
\centering
\includegraphics[width=0.5\linewidth]{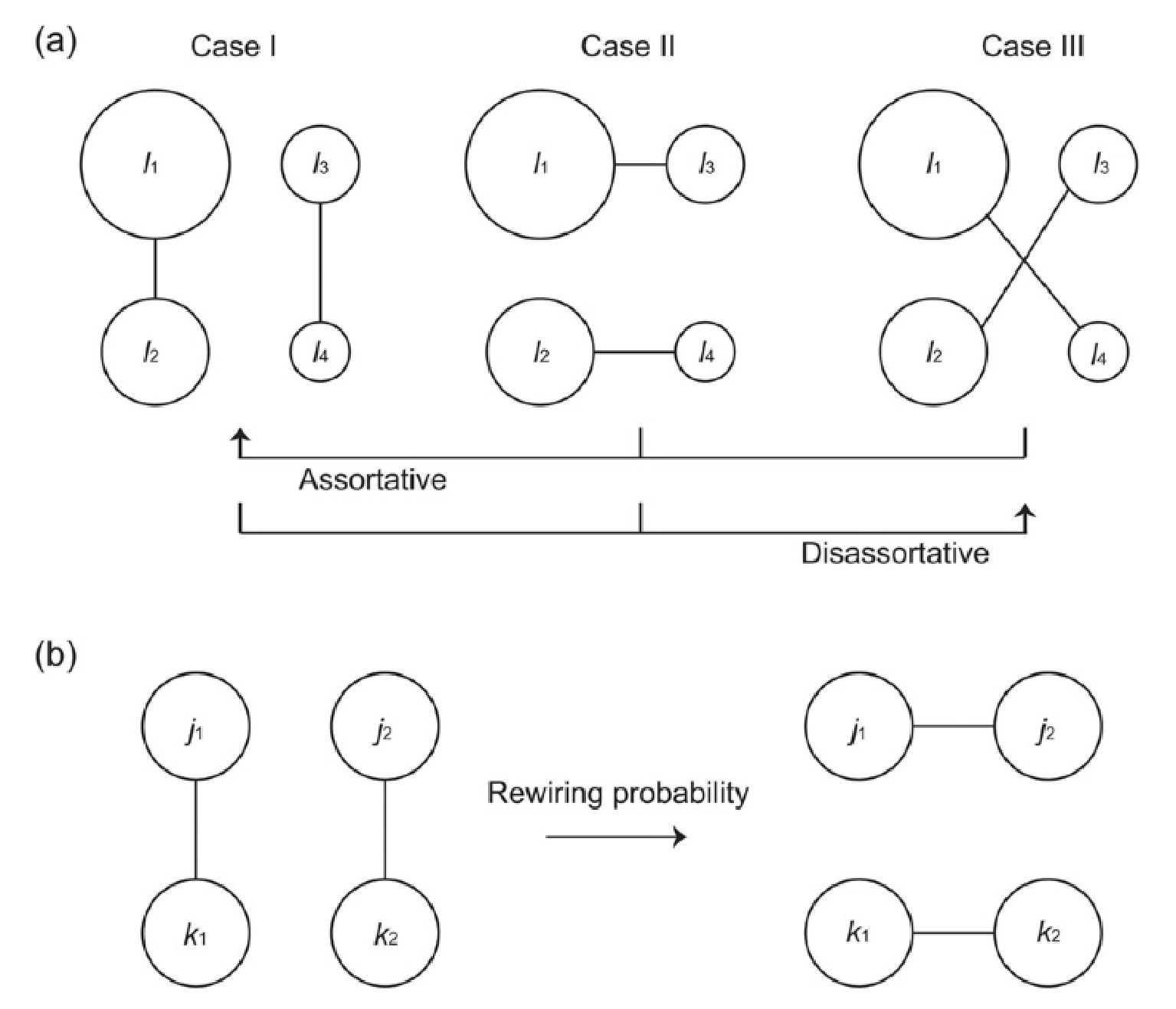}
\caption{Network rewiring  method for changing the network assortativity $r$. (a) The greedy edge-rewiring (GER) method \cite{xulvi2004reshuffling}.  (b) The stochastic edge-rewiring (SER) method. Reprinted figure with permission from Ref.\cite{sasai2015robustness}.}
\label{cn_1}
\end{figure}

 We investigate the dynamical robustness by altering $r$ through two approaches: Greedy edge rewiring (GER) \cite{xulvi2004reshuffling} and Stochastic edge rewiring (SER) \cite{newman2002assortative}. We start with an uncorrelated network, characterized by $r=0$, and then perform edge reshuffling without permitting self-loops or overlaps. We then randomly select two existing edges of the network, given by the connected node pairs $(v_1, w_1)$ and $(v_2, w_2)$. The remaining degrees of these node pairs are denoted as $(j_1, k_1)$ and $(j_2, k_2)$ respectively. i) The degree of the connected node pairs is used to guide the edge rewiring in the GER approach. We arrange the remaining degrees, namely, $j_1$, $j_2$, $k_1$, and $k_2$ in descending order and assign them new labels, $l_1$, $l_2$, $l_3$, and $l_4$, respectively, ensuring that $l_1\ge l_2\ge l_3\ge l_4$. There are  three possible ways to partition the four nodes into two pairs of connected nodes as depicted in Figure~\ref{cn_1}(a). To enhance the assortativity coefficient, we opt for Case I to establish connections between nodes with more analogous degrees if the current state is Case II or III.  Conversely, to reduce the assortativity coefficient, we employ Case III, establishing an edge between nodes with the highest and lowest degrees if  the current state is Case I or II.  With repeated  employment of edge rewiring in a greedy manner we can monotonically increase or decrease $r$  until it no longer changes. ii) In the Stochastic Edge Rewiring (SER) technique, we iteratively rewire edges in a stochastic manner, altering the network's assortativity. We aim to construct a network that adheres to a predefined joint probability distribution E(j, k) for the remaining degrees. This process is implemented using a numerical method based on~\cite{newman2002assortative}.

\begin{figure}[h]
\centering
\includegraphics[width=0.5\linewidth]{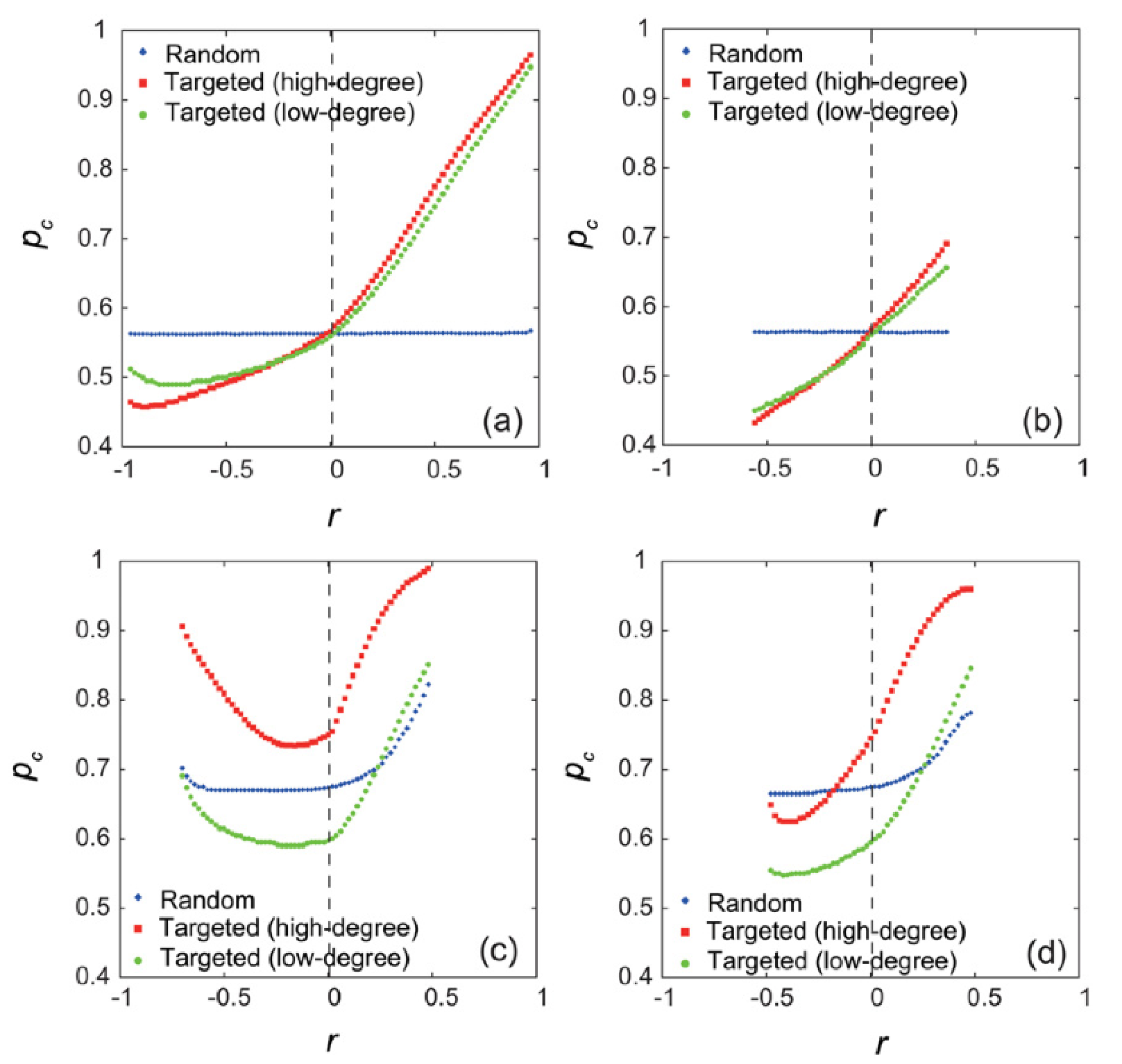}
\caption{The critical value $p_c$ is plotted  as a function of  the assortativity coefficient $r$ for random and targeted inactivation, for the system~\eqref{cn-1}. We start from a Erd\"os-R\'enyi  uncorrelated network and employ (a) GER method and (b) SER method, while altering the assortativity coefficient $r$. We start from the BA scale-free network with $r=0$ and employ (c) GER method and (d) SER method, and alter the assorttativity coefficient $r$. Reprinted figure with permission from Ref.\cite{sasai2015robustness}.}
\label{cn_2}
\end{figure}

To study the effect of network assortivity on dynamical robustness, we consider the  network model consisting of $N$ diffusively coupled Stuart-Landau oscillators which is described by,

\begin{eqnarray}
\dot{z}_j &=& (\alpha_j+i\omega - |z_j|^2)z_j + \frac{K}{N}\sum_{k=1}^{N}A_{jk}(z_k - z_j).
\label{cn-1}
\end{eqnarray}

\noindent First, we investigate the dynamical robustness of  an  uncorrelated  Erdős-Rényi random graph \cite{erdHos1960evolution}, with degrees centered around the average degree.  Subsequently, we modify the network to make it assortative or disassortative by applying the edge-rewiring algorithms outlined above. Figures~\ref{cn_2}(a) and \ref{cn_2}(b) display the critical value $p_c$ as a function of $r$ for the networks generated using the GER and SER methods, respectively. In both panels, for random inactivation, the value of $p_c$ remains almost constant, regardless of the $r$ value. This is due to the fact that the number of inactive oscillators in the vicinity of each oscillator node is not influenced by the value of $r$. The targeted inactivation of high-degree and low-degree oscillator nodes both exhibit a monotonic increase in the value of $p_c$ with $r$, as depicted in Figs.~\ref{cn_2}(a)  and \ref{cn_2}(b). This outcome can be attributed to the fact that in more assortative networks, the amplitudes of the active oscillators, which play a dominant role in determining the order parameter, are larger. Consequently, it can be concluded that network assortativity has a positive impact on the dynamical robustness of oscillator networks when facing targeted inactivation.

\par Next, we explore the impacts of assortativity on dynamical robustness of correlated networks with power-law degree distributions~\cite{barabasi1999emergence}.  The critical fraction $p_c$ plotted versus $r$ for the GER and SER techniques are shown in Figs.~\ref{cn_2}(c) and \ref{cn_2}(d), respectively. In assortative networks, it is evident from the figure that for all types of inactivation, the value of $p_c$ consistently increases as $r$ progresses from $0$. In assortative networks, connections tend to be formed between high-degree nodes and between low-degree nodes. Consequently, when targeting the inactivation of high-degree nodes, low-degree active oscillators that are connected to a few inactive oscillators can sustain large oscillation amplitudes. Likewise, when targeting the inactivation of low-degree nodes, high-degree active oscillators connected to a few inactive oscillators can also maintain large oscillation amplitudes. These nodes, which preserve substantial oscillation amplitudes, contribute significantly to the high value of $p_c$, signifying a highly robust oscillatory behavior. Thus, assortativity plays a positive role in enhancing dynamical robustness. In contrast, for disassortative network as $r$ is decreased from $0$, the dynamical robustness increases after a slight downward trend for the GER method, as shown in Fig.~\ref{cn_2}(c), but it gradually decreases until $r=-0.5$ for the SER method, as shown in Fig.~\ref{cn_2}(d). In summary, we conclude that network assortativity enhances dynamical robustness, while the impact of network disassortativity on dynamical robustness depends on the specific edge-rewiring methods employed.


\subsection{Dynamical robustness of multiplex networks}

Recent studies have further confirmed that the functions emerging within a single network can exert a notable impact on other networks. Specifically, a node within one network is often found to be a component of another network. From ecological~\cite{pilosof2017multilayer} and climate systems~\cite{donges2011investigating}, via physical and transportation systems to social networks~\cite{szell2010multirelational}, these interconnections have been identified across various contexts. This underscores the effectiveness of an interdependent~\cite{gao2012networks,buldyrev2010catastrophic}, particularly multilayer (multiplex)~\cite{boccaletti2014structure,kivela2014multilayer}, network of networks architecture in describing numerous systems and scenarios.
Thus, a multilayer network~\cite{boccaletti2014structure,kivela2014multilayer,de2013mathematical,pilosof2017multilayer,vaiana2020multilayer} is simply defined to be a network with a set of nodes, together with the concept of layers in addition
to nodes and links. These layers are often present in order to represent different types or aspects of interactions (links). In particular, a special type of multilayer network having same number of nodes in each layer, in which each node in each layer posseses exactly one connection with a node (its replica) in another layer is known to be a multiplex network~\cite{nicosia2013growing,gomez2013diffusion,verbrugge1979multiplexity,granell2013dynamical,majhi2019emergence,cardillo2013emergence}.

\par Let us here consider a multilayer network consisting of $L$ layers, each layer comprised of $N$ globally coupled nodes. Casting the dynamics of the nodes of the network by Stuart-Landau oscillatory systems, the time evolution of the entire multilayer network is then governed by the following Eq.~\cite{morino2011robustness},

\begin{equation}
	\begin{array}{lcl}\label{ml1}
		\dot {z_j}^{[l]}=\{\alpha_j + i\omega-|{z_j}^{[l]}|^2\}{z_j}^{[l]}+\dfrac{\epsilon}{N}\sum\limits_{k=1}^{N}({z_k}^{[l]}-{z_j}^{[l]})+\sigma H({z_j}^{[1]},{z_j}^{[2]},\dots, {z_j}^{[L]}),
		
	\end{array}
\end{equation}

where ${z_j}^{[l]}$ represents the state of the $j$-th node in the $l$-th layer, $j= 1,2,\dots,N;~l= 1,2,\dots,L$. Here $\epsilon$ is the intralayer coupling strength and $\sigma$ corresponds to the interlayer interaction strength with $H(\cdot)$ being the function characterizing the interlayer connections.

\par Then, three different interlayer coupling functions are assumed, namely mean-field, chain and diffusive interlayer interactions which are realized by the following forms  

\[
 H({z_j}^{[1]},{z_j}^{[2]},\dots, {z_j}^{[L]})= 
 \begin{cases}
 	\frac{1}{L}\sum\limits_{m=1}^{L}{z_j}^{[m]}& ~~~~~\text{(case I)},\\
 	{z_j}^{[l-1]}/2& ~~~~~\text{(case II)},\\
 	\frac{1}{L}\sum\limits_{m=1}^{L}({z_j}^{[m]}-{z_j}^{[l]})& ~~~~~\text{(case III)}.
 \end{cases}
 \]

Next, we modify the order parameter for the multilayer network as $R=\dfrac{1}{NL}|\sum\limits_{l=1}^{L}\sum\limits_{j=1}^{N}{z_j}^{[l]}|$ characterizes the intensity of the global oscillation in the entire multilayer networked system. We then set the value of the system parameters as  $\alpha_j=2$ for an active oscillator,  $\alpha_j=-1$ for an inactive oscillator, and $\omega=3$, while keeping the network size $N=3000$ and the coupling strength fixed at $\epsilon=8$.

\begin{figure}
	\centerline{\includegraphics[scale=0.4500]{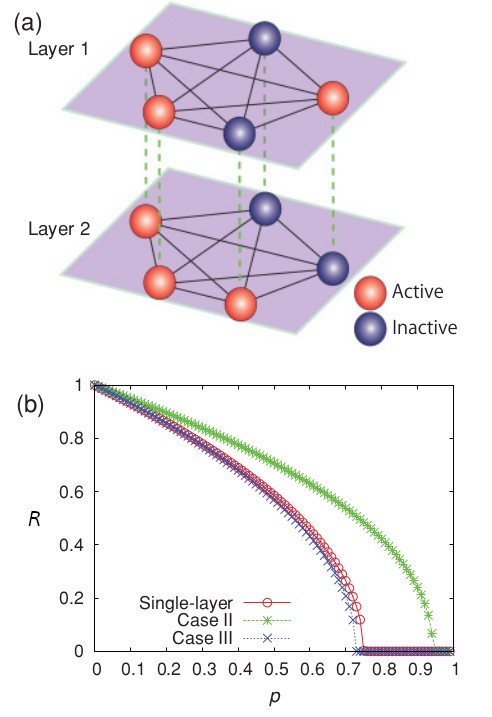}}
	\caption{(a) An exemplary multiplex network with $N=5$  nodes in each of $L=2$ layers. The red and blue circles respectively stand for the active and inactive oscillators. The ratio of the inactive units is $p=0.4$. The black solid and green dashed lines indicate all-to-all intralayer connections and local interlayer connections, respectively. (b) The order parameter $R$ with respect to the fraction of inactive elements $p$, for three different scenarios: single-layer network, multi-layer network ($L=2$, case II), and multi-layer network ($L=2$, case III). Reprinted figure with permission from Ref.\cite{morino2011robustness}.}
	\label{fig11}
\end{figure}
\par In Fig.~\ref{fig11}(a), an exemplary multi-layer network is portrayed with $N=5$ and $L=2$, in which the fraction of inactive units is chosen as $p=0.4$. Figure~\ref{fig11}(b) depicts the variation in the order parameter $R$ as a function of the inactivation ratio $p$ for the three different network setups. To be precise, the first plot (in red) corresponds to the single-layer case, i.e., when $L=1$. As witnessed, starting from the normalized unit value, $R$ monotonically decreases and eventually drops down to zero near $p \sim 0.74$ implying an aging transition around $p_c \sim 0.74$. Next, $R$ is shown (in green) for the bilayer framework ($L=2$), precisely for case II, while keeping the interlayer coupling strength $\sigma$ at $\sigma=1.5$. This time, the critical inactivation ratio $p_c \sim 0.94$ for which the aging transition occurs is much higher than that of the earlier case (single-layer network). This implies that the robustness of the bilayer network for case II is higher than that of the single-layer network. Finally, $R$ is plotted (in blue) for the multilayer network ($L=2$) under case III, for a much longer $\sigma=8$. In contrast to the previous case, now the aging transition takes place earlier than the single-layer network. Specifically, the critical ratio $p_c \sim 0.72$ of the inactive elements is smaller than that for the single-layer case. Thus, as compared to the single-layer network formulation, the robustness of the multilayer network can be higher or lower depending on the functional form of the interlayer coupling.  

 
\subsection{Dynamical robustness of long-range connectivity networks}

As far as the interactions among the constituents of a complex system are concerned, we have, thus far, confined ourselves only on the short-range direct communications. However in networked systems, interactions arise not only from the direct connections between nodes but also from indirect long-range communications facilitated by numerous other existing pathways that link the nodes. Long-range connectivity~\cite{campa2009statistical,gupta2010slow,joyce2014scaling,teles2015temperature,bountis2016mechanical,bouchet2010thermodynamics} is omnipresent in complex systems and hence has recently emerged as one of the flourishing areas of research. In particular, researchers have studied the presence of long-range interactions in various networks characterized by a power-law decay. The examples include biological networks~\cite{raghavachari1995spatially}, Rydberg atoms~\cite{saffman2010quantum}, hydrodynamic interaction~\cite{uchida2011generic}, plasmas~\cite{levin2008collisionless}, nuclear spins~\cite{alvarez2015localization}, and in climate, called teleconnections~\cite{donges2009complex}. From synchronization~\cite{marodi2002synchronization,chowdhury2010synchronization,rakshit2020synchronization,rakshit2021neuronal} to chimera state~\cite{banerjee2016chimera} and oscillation quenching~\cite{sathiyadevi2019long}, various phenomena have been examined in networks subject to long-range communications among the nodes.  

\par Here, let us first describe long-range interaction in networks through the adjacency matrices corresponding to different paths. Let $G=(V,E)$ be a network consisting of $N$ nodes in which $V$ and $E$ represent the collections of nodes and links respectively, so that $V=\{1,2,\dots,N\}$, and $E\subset V\times V$ is the set of links. Then we can characterize the network's diameter as $D=\max\big\{\mbox{dist}(j,k):j,k=1,2,\dots,N\big\}$. Here $\mbox{dist}(j,k)$ denotes the distance between the nodes $j$ and $k$.
\par The $d$-path adjacency matrix ${A}^{[d]}$ can then be written as,
\[
    {A}_{jk}^{[d]}= 
\begin{cases}
    1, & \text{if}~\mbox{dist}(j,k)=d, \\
    0,  & \text{otherwise}.
\end{cases}
\]

\par
\begin{figure}[ht]
	\centerline{
	\includegraphics[scale=0.3200]{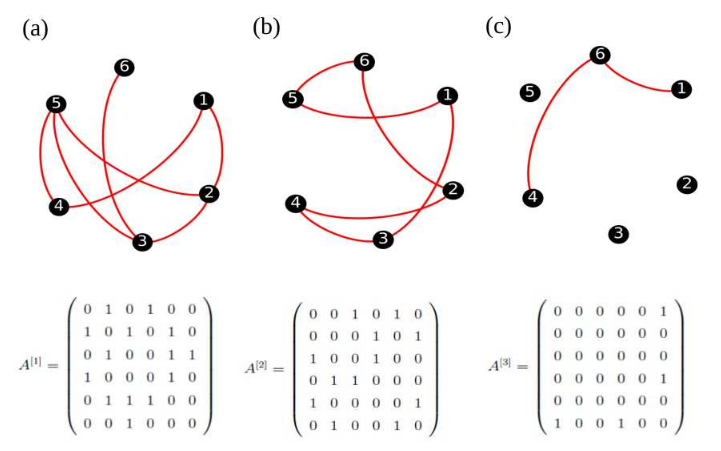}}
	\caption{A network featuring prominent long-distance connections: the $d$-path networks and their corresponding adjacency matrices are displayed in the upper and lower rows, respectively: (a) When $d=1$, it refers to the original network provided, (b) When $d=2$, it represents the network with a 2-path derived from (a). (c) When $d=3$, it indicates the network with a 3-path derived from (a). Reprinted figure with permission from Ref.\cite{majhi2022dynamical}}
	\label{fig0}
\end{figure}

 For a clear perception, we choose a small exemplary network with $N=6$ nodes with the diameter $D=3$, and in Fig. \ref{fig0}, we depict the associated $d$-path $~(d=1,2,3)$ networks along with the adjacency matrices. In Fig. \ref{fig0}(a), we display  the original given network, i.e., the direct $1$-path network, together with the associated adjacency matrix ${A}^{[1]}$ in the lower panel. Similarly, the extracted $2$- and $3$-path networks are depicted in Figs \ref{fig0}(b) and  \ref{fig0}(c) respectively, with the corresponding adjacency matrices ${A}^{[2]}$ and ${A}^{[3]}$. Then casting  each node by the dynamics of Stuart-Landau oscillators, the dynamics of the $j$-th node in the network subject to long-range communication can be described as~\cite{majhi2022dynamical}, 

\begin{equation}
\begin{array}{lcl}\label{eqq1}
\dot z_j=(\alpha_j + i\omega - |z_j|^2)z_j+\dfrac{1}{N}\sum\limits_{d=1}^{D}\sigma_d\sum\limits_{k=1}^{N}{A}^{[d]}_{jk}(z_k-z_j),j= 1,2,\cdots,N,
\end{array}
\end{equation}

In this context, $\sigma_d$ represents the interaction intensity between the $j$-th and $k$-th nodes when the distance between them is $d$, where $D$ denotes the network's diameter. Thus the strength of interaction between each pair of nodes essentially depends on the distance between them. We also choose $\alpha_j=a>0$ and $\alpha_j=-b<0$ respectively for the active and inactive set of oscillators in the system.

\begin{figure*}[ht]
	\centerline{
		\includegraphics[scale=0.400]{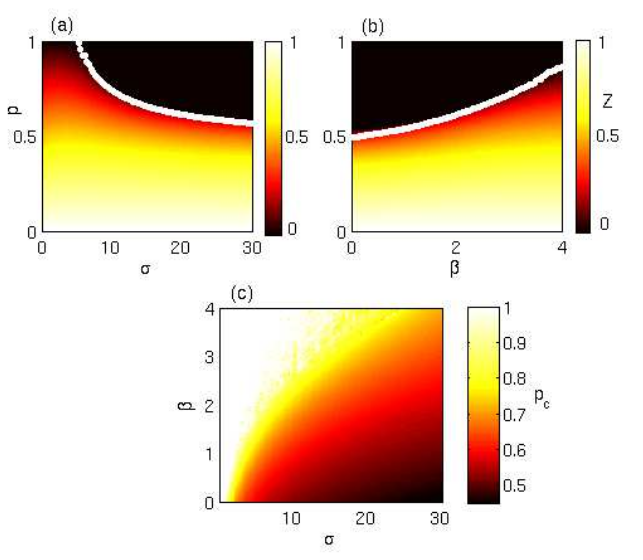}}
	\caption{ The order parameter $R$ for the system~\eqref{eqq1} with respect to the simultaneous variations of (a) the inactivation ratio $p$ and the interaction strength $\sigma$ where $\beta=2$ is fixed, (b) the rate of decay $\beta$ and the inactivation ratio $p$ where $\sigma=20$ is fixed, (c) The critical inactivation ratio $p_c$ as a function of $\beta$ and $\sigma$. The white curves in (a) and (b) correspond to the the analytical $p_c$~(Eq.~\eqref{eqq17}) . Reprinted figure with permission from Ref.\cite{majhi2022dynamical}}
	\label{fig3}
\end{figure*}

\par We specifically focus on a power-law decay of the coupling strength in relation to the distance between the nodes, which is best represented by $\sigma_d=\dfrac{\sigma}{d^{\beta}}$. Here, the exponent $\beta$ governs the rate of decay in the power law. We consider Erdos-Renyi random network architecture ($G(N,q)$ graph \cite{erdHos1960evolution}) as the underlying network with network size $N=200$ and the connection probability $q=0.05$, and we choose $a=1$, $b=1$ with $\omega=3$. Figure~\ref{fig3}(a) portrays the variation in the order parameter $R$ as a function of the interaction strength $\sigma \in [0,30]$ and the inactivation ratio $p$, in which the decay rate $\beta=2$ is kept fixed. The phase diagram explains how $R$ transits from non-null to null value and hence aging takes place in the entire range of $\sigma \in [0,30]$ as the inactivation ratio $p$ increases. The increasing values of both the interaction strength $\sigma$ and the fraction $p$ lead to faster aging and hence the network becomes more dynamically vulnerable. On this phase diagram acquired through numerical means, we depict the critical value $p_c$, which is obtained through analytical methods (discussed later), and aligns with the numerical findings. This analytical curve thus separates the aging region from the oscillatory regime. In Fig.~\ref{fig3}(b), order parameter $R$ is plotted with respect to the simultaneous variation of $p$ and the decay rate $\beta \in [0,4]$, with a fixed coupling strength $\sigma=20$. In contrast to the effect of the interaction strength $\sigma$ on the dynamical robustness, increasing values of $\beta$ inhibits aging transition and hence dynamical robustness increases. The analytically obtained $p_c$ values are plotted over the phase diagram, as before, that fits into the numerical outcome. Ultimately, in Fig.~\ref{fig3}(c), we illustrate the variations in the critical inactivation ratio $p_c$ across the parameter plane $(\sigma,\beta)$. The phase diagram describes how the robustness of the network alters due to the simultaneity of the two crucial parameters $\sigma$ and $\beta$. The critical ratio $p_c$ remains unity for sufficiently small interaction strength $\sigma$, irrespective of the value of $\beta$. However, as $\sigma$ increases aging takes place in the networked system so that $p_c$ decreases. On the contrary, $p_c$ increases for increasing values of $\beta$. This indicates that the system is dynamically vulnerable, whenever the coupling strength $\sigma$ is high and the long-range exponent $\beta$ is low.              

\par To establish a comprehensive analytical framework for evaluating the dynamic robustness of the networked system under consideration, we employ a degree-weighted mean field approximation. As a result, the system \eqref{eqq1} can be approximated as 

\begin{equation}
	\begin{array}{lcl} \label{eqq3}
		\dot{z_j} = (\alpha + i\omega - |z_j|^2)z_j + \dfrac{1}{N}\sum\limits_{d=1}^{D}\sigma_d{k_j}^{[d]}\Big[(1-p)M_A(t)+pM_I(t)-z_j\Big].\\
	\end{array}
\end{equation}

Here,
\begin{equation}
	\begin{array}{lcl} \label{eqq4}
		M_A(t) = \dfrac{\sum\limits_{d=1}^{D}\sigma_d\sum_{j\in S_A} {k_j}^{[d]}z_j(t)}{\sum\limits_{d=1}^{D}\sigma_d\sum_{j\in S_A} {k_j}^{[d]}}, 

~~M_I(t) = \dfrac{\sum\limits_{d=1}^{D}\sigma_d\sum_{j\in S_I} {k_j}^{[d]}z_j(t)}{\sum\limits_{d=1}^{D}\sigma_d\sum_{j\in S_I} {k_j}^{[d]}}
  
	\end{array}
\end{equation}
are the mean-fields with weighted degrees corresponding to the active and inactive sets of dynamical units, respectively.

where ${k_j}^{[d]}(j=1,2,...,N)$ refers to the degree of the $j$-th node related to the $d$-path network. Assuming the state variables in the form $z_j(t)=r_j(t)e^{i(\omega t+\theta)}$, Eq.~\eqref{eqq3} can be expressed as
\begin{equation}
	\small{\begin{array}{lcl}  \label{eqq6}
			\dot{r}_j = \Big(\alpha_j - \dfrac{1}{N}\sum\limits_{d=1}^{D}\sigma_d{k_j}^{[d]}-r_j^2\Big)r_j + \dfrac{1}{N}\sum\limits_{d=1}^{D}\sigma_d{k_j}^{[d]}\Big[(1-p)R_A(t)+pR_I(t)\Big],\\
	\end{array}}
\end{equation}
in which
\begin{equation}
	\begin{array}{lcl} \label{eqq7}
		R_A(t) = \dfrac{\sum\limits_{d=1}^{D}\sigma_d\sum_{j\in S_A} {k_j}^{[d]}r_j(t)}{\sum\limits_{d=1}^{D}\sigma_d\sum_{j\in S_A} {k_j}^{[d]}},
		~~R_I(t) = \dfrac{\sum\limits_{d=1}^{D}\sigma_d\sum_{j\in S_I} {k_j}^{[d]}r_j(t)}{\sum\limits_{d=1}^{D}\sigma_d\sum_{j\in S_I} {k_j}^{[d]}}.
	\end{array}
\end{equation}

Then, pursuing a similar procedure as above in Sec.~\ref{het}, we arrive at the critical inactivation ratio
\begin{equation}
	\begin{array}{lcl} \label{eqq17}
		p_c = \dfrac{H(\sigma,a)-1}{H(\sigma,a)-H(\sigma,-b)},
	\end{array}
\end{equation}
where $H$ arises from the following equation, 
\begin{equation}
\begin{array}{lcl} \label{eqq14}
		H(\sigma,\alpha)=\dfrac{\sum\limits_{d=1}^{D}\sigma_d\sum\limits_{j=1}^{N}\Big[\dfrac{{k_j}^{[d]}L_j}{L_j-N\alpha_j}\Big]}{N^2\sum\limits_{d=1}^{D}\sigma_d s^{[d]}},
	\end{array}
\end{equation}
in which $L_j=\sum\limits_{d=1}^{D}\sigma_d{k_j}^{[d]};~j=1,2,...,N$.


\section{Dynamical robustness of quantum oscillators}

So far, we have considered classical Stuart-Landau oscillators to understand different routes  leading to aging transition. We now pose the question: “Can the aging transition occur in the quantum domain, and if it does, how would it manifest?”
Motivated by this question, Bandyopadhyay et al. \cite{bandyopadhyay2023aging} investigate the occurrence of aging transitions in quantum systems, emphasizing the differences compared to classical models \cite{Daido_2004_PRL}. They examine a globally coupled quantum Stuart-Landau oscillators, which can be either active or inactive. The classification of nodes as active or inactive in the quantum context is determined by the characteristics of the dissipators in the quantum master equation. The quantum master equation for $N$ globally coupled quantum Stuart-Landau oscillators with diffusive coupling is expressed as \cite{ishibashi2017oscillation},

\begin{eqnarray}
	\label{coupled_master}
	\dot{\rho}&=&\sum_{j=1}^N\left(-i[H,\rho]+\underbrace{G_j\mathcal{D}[O_j](\rho)}+\kappa\mathcal{D}[{a_j}^2](\rho)\right)
	+ \frac{V}{N}\sum_{j=1}^N\sum_{j'=1}^N{}^{'} \mathcal{D}[a_j - a_{j'}](\rho), 
	\label{qo1}
\end{eqnarray}

\noindent where $H=\omega {a_j}^\dag a_j$. $a_j$ and ${a_j}^{\dag}$ represent bosonic annihilation and creation operators of the $j$-th oscillator, respectively. The Lindblad dissipator $\mathcal{D}[\hat{L}]$  have the form $\mathcal{D}[\hat{L}](\rho)=\hat{L}\rho \hat{L}^\dag-\frac{1}{2}\{\hat{L}^\dag \hat{L},\rho \}$, where $\hat{L}$ is an operator (we set $\hbar=1$, without any loss of generality ). The operator $O_j$ of the second term  (shown under the brace) is introduced to incorporate the concept of active and inactive elements. As system approaches the classical limit ($G_j>\kappa$), the quantum master equation becomes equivalent to the classical Stuart-Landau equation by the  relation: $\langle\dot{a}\rangle=\mbox{Tr}(\dot \rho a)$. Here $\sum_{j'}{}^{'}$ indicates that the sum does not include the condition $j'=j$. The concept of active and inactive elements can be introduced into the quantum master equation through the properties of the operator associated with the coefficient $G_j$ in the Lindblad dissipator as follows:
$$O_j = 
\begin{cases}
	{a_j}^\dag & \text{for active oscillators,} \\[10pt]
	{a_j} & \text{for inactive oscillators.}
\end{cases}$$

\noindent $O_j={a_j}^\dag$ is the dissipator $G_j\mathcal{D}[O_j](\rho)$ in Eq.~\eqref{qo1} describes a single boson, gain with a rate of $G_j$ and it lead to stable limit cycle for the jth oscillator to make it quantum active element. When  $O_j={a_j}$, the system experiences a single boson loss at a rate of $G_j$, which show the non-oscillatory or inactive quantum behaviour of jth oscillator. Similar to the case of the classical system \cite{Daido_2004_PRL}, the whole network is divided into two groups one group consists of $N_a$ active elements and the other consists of $N_i$ inactive elements and calculate fraction of inactive node $p=\frac{N_i}{N}$ in the network. 

\begin{figure}
	\centering
	\includegraphics[width=.7\textwidth]{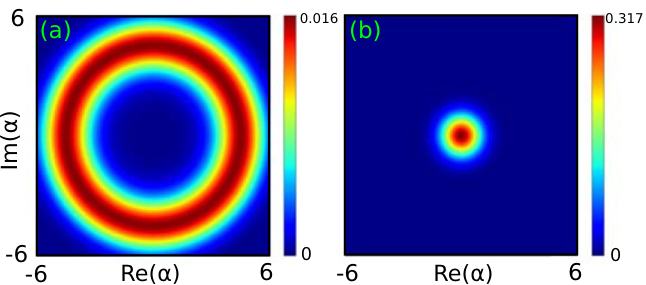}
	\caption{ (a,b) Distribution of Wigner function  in phase space for uncoupled (a) active oscillator ($G_j=4$ and $O_j={a_j}^\dag$) and (b) inactive oscillator ($G_j=2$ and $O_j=a$) at fixed parameters are $\omega=2$ and $\kappa=0.2$. Reprinted figure with permission from Ref.~\cite{bandyopadhyay2023aging}   }.
	\label{uncoupled_quantum}
\end{figure}

In the uncoupled state ($V=0$), the phase space representation of the Wigner function for active and inactive elements is depicted in Fig.\ref{uncoupled_quantum}(a) and (b), respectively. The ring-shaped Wigner function~\cite{carmichael2013statistical} indicates a quantum limit cycle at $G_j=4, O_j=a_j$ and a probability blob at 
$G_j=2, O_j=a_j\dag$ represents a non-self-oscillatory element. When dealing with a large number of oscillators, the density matrix of the many-body system can be approximately factorized as $\rho \approx \otimes_{j=1}^N \rho_j$. This approach aligns with the mean-field approximation, simplifying the master equation Eq.~\ref{qo1} into individual master equations for each oscillator, which then interact with the mean-field as follows \cite{ishibashi2017oscillation}:

\begin{align}
	\label{mf_master}
	\dot{\rho_j}&=-i[\omega {a_j}^\dag a_j,\rho_j]+G_j\mathcal{D}[O_j](\rho_j)+\kappa\mathcal{D}[{a_j}^2](\rho_j) \nonumber \\
	&+\frac{2V(N-1)}{N}\mathcal{D}[a_j](\rho)+V\left(A[{a_j}^\dag,\rho_j]-A^*[a_j,\rho_j]\right), 
\end{align}
where $A$ and $A^*$ are defined as follows: $A=\frac{1}{N}\sum_{j'=1}^{'N}\langle a_{j'}\rangle_j$ and $A^=\frac{1}{N}\sum_{j'=1}^{'N}\langle a_{j'}^\dag \rangle_j$.
Eq.~\ref{mf_master} is numerically solved by self-consistent method using QuTiP \cite{johansson2012qutip}. We take $G_j=4$ for the active elements $[j \in \{1, 2, ..., N_a\}]$, and $G_j=2$ for the inactive elements $[j \in \{N_a+1, ..., N\}]$. In the network, we differentiate between the oscillatory state and the oscillation-collapsed state by calculating the average boson number per oscillator: $Q=\frac{\bar{n}_{mf}(p)}{\bar{n}_{mf}(0)}$, where, $\bar{n}_{mf}(p)$ is the mean boson number per oscillator for a particular $p$ value.

\begin{figure}
	\centering
	\includegraphics[width=.49\textwidth]{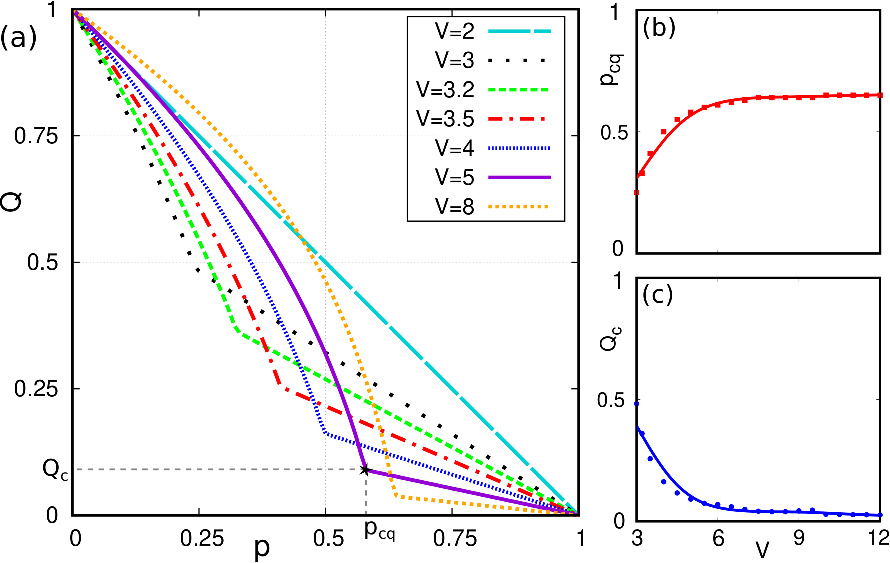}
	\caption{(a) Order parameter $Q$ is plotted against $p$ for various values of coupling strength $V$. 
		For $V=5$, the star mark indicates the knee point of the curve. The corresponding  critical values of $p$ and $Q$ are indicated as $p_{cq}$ and $Q_{c}$, respectively. (b-c) The variation of $p_{cq}$ and $Q_{c}$ with coupling strength $V$ at parameter $\omega=2$ and $\kappa=0.2$. Reprinted figure with permission from Ref.~\cite{bandyopadhyay2023aging}  }.
	\label{quantum_f2}
\end{figure}

To examine how the average mean boson number changes as the value of $p$ increases, $Q$ is plotted as function of $p$ for different coupling strengths V in Fig.~\ref{quantum_f2}(a). For $V\le 2.73$, $Q$ decreases monotonically as $p$ increases. However, once V exceeds approximately 2.73, the rate at which Q decreases exhibits two distinct phases: initially, Q declines sharply as p increases, but beyond a critical threshold $p_{cq}$, the rate of decrease becomes nearly linear. The point on the curve that marks the transition between the steep decline and the inclined linear region is referred to as the \textit{knee} point (for $V=5$ star mark in the Fig.\ref{quantum_f2}(a)). These \textit{knee} points $p_{cq}$ consider as aging transition threshold  and its associated order parameter denoted as $Q_c$. As p increases further, the curve progressively approaches Q=0 because of the growing number of inactive elements in the network. When p reaches one, all oscillators become inactive, resulting in $Q=0$, which is a straightforward scenario. The dependence of $p_{cq}$ and $Q_c$ on the coupling strength V is illustrated in Figs.~\ref{quantum_f2}(b) and (c), respectively. The results shows that $p_{cq}$ initially increases as V increases. However, for strong coupling strength, $p_{cq}$ reaches a plateau and exhibits little further change. This behavior differs from the classical case, where $p_c$ generally decreases with increasing coupling strength \cite{Daido_2004_PRL}. It's notable that $Q_c$ diminishes as the coupling strength increases, aligning with the anticipated effect of stronger coupling promoting aging. 

The key observations that distinguish the quantum aging transition from its classical counterpart are significant. Unlike in classical systems, where aging transition is marked by the complete collapse of the network, the quantum aging transition is characterized by a rapid decrease in the average mean boson number. Furthermore, the quantum aging process involves two distinct phases: initially, up to a critical ``knee" point, the order parameter decreases rapidly; beyond this point, the rate of decrease slows down. During this latter phase, all inactive elements populate the ground state through the single boson loss process. However, active oscillators never fully reach the ground state, as their relaxation is governed solely by the two-boson absorption process. This unique scenario, which has no classical counterpart, underscores the distinct nature of quantum aging processes.


\section{Dynamical robustness of biological networks}

So far, we confined ourselves to the Stuart-Landau model which is a normal form  of a nonlinear oscillating system near the Hopf bifurcation point. However, the scenario of loss of collective oscillatory activity in diverse networked systems due to the failure of their components exhibits numerous practical applications in specific biological systems as well. It is critically important to investigate how various conditions affect macroscopic activity when certain microscopic units fail and lose their self-oscillatory behavior. This is especially pertinent in biological systems, since collective functions play an integral part frequently resulting from interactions among oscillatory units that can gradually deteriorate under pathological conditions. Notable attempts have been made to further explore dynamical robustness in terms of aging transition in realistic biological networks composed of active and inactive elements, such as neuronal ensembles, and ecological systems. Motivated by these facts, in this section we highlight important results obtained for basic models in ecological and neuronal systems.

\subsection{Spatial metapopulation networks}

In recent past, the aging transition phenomenon has been studied  in the context of metapopulation survivability by Kundu et al.~\cite{Rakshit_2017_PRE}. Metapopulation dynamics, a concept used to describe the movement of spatially separated populations of one species in spatial ecology\cite{levins1969some,hanski1999metapopulation}, has shed light on the long-term dynamics of structured populations. These studies have revealed that population densities of a particular species often undergo synchronized fluctuations across extensive geographic regions \cite{ranta1995synchrony}. In this framework, a patch is typically represented as a system of differential equations that displays oscillatory solutions. Spatially structured metapopulations can be conceived as a network composed of interconnected oscillators. In this context, nodes correspond to viable habitat patches, and the links connecting these nodes signify functional pathways. This conceptual framework enables the examination of the ecological network's dynamical robustness, particularly in the context of predator-prey patches. The mathematical representation of the dynamics within a single patch is as follows,

\begin{eqnarray}\label{eq_Eco:1}
\dot x= f(x, y)=\frac{1}{\epsilon}[x(1-x)(x-\theta)-xy],\nonumber\\
\dot y=g(x, y)= xy-dy.\nonumber
\end{eqnarray}

\noindent In this context, the variables $x$ and $y$ represent the normalized prey and predator population densities, respectively. The parameter $\epsilon\in(0,1]$ signifies the time scale separation between the prey and predator populations, $\theta\in(0,1)$ represents the Allee threshold, and $d$ stands for the natural mortality rate of the predator population. The nontrivial fixed point $(d,(1-d)(d-\theta))$ exists under the condition $\theta<d<1$. This fixed point is stable when $d>\frac{1+\theta}{2}$, and a supercritical Hopf bifurcation occurs at $d=\frac{1+\theta}{2}$. When $d\leq\frac{1+\theta}{2}$, the coexistence of oscillation (stable limit cycle) and a stable extinction state (0,0) arises based on the initial population density. However, by further reducing the predator mortality rate, species extinction occurs through a boundary crisis of the limit cycle attractor.

The prey-predator model involving $N$ patches is mathematically described by the following equation,

\begin{equation}\label{eq:2}
\dot{\mathbf{X}_i} = \mathbf{F(X}_i) + \mathbf{M} \sum_{j=1}^{N}A_{ij}(\mathbf{X}_j-\mathbf{X}_i),
\end{equation}

In this framework, $\mathbf{X}i = (x_i, y_i)^T $ represents the state vector, and $\mathbf{F(X}i) = (f(x_i,y_i), g(x_i,y_i))^T$ describes the inherent dynamics of the $i$-th patch. The second term signifies diffusive coupling, illustrating interactions among species across different patches. Here, $\mathbf{M}=(\frac{ m}{deg(i)}, \frac{ m}{deg(i)})^T$ represents the dispersal matrix, where $m$ denotes the dispersal rate between patches and the term $deg(i)$ signifies the number of patches (degree) connected to the $i$-th patch. Here $A_{ij}$ stands for the adjacency matrix. In this study, an active patch indicates stable limit cycle oscillations in both populations, with $d=0.5$ set. Conversely, an inactive patch implies species extinction, with $d=0.3$.

To investigate the aging transition, we adopt the mathematical framework introduced by \cite{Daido_2004_PRL}. This approach involves considering a scenario where, in a fraction $p$ of patches, species go extinct due to the absence of dispersal between these patches. The order parameter $R$, which quantifies the level of dynamical activity in the network, is defined as follows,

\begin{equation}
R = \frac{1}{2}(R_x+R_y),
\end{equation}

where $R_x = \frac{1}{N}\sum_{i=1}^{N}(\langle x_{i,max}\rangle_t-\langle x_{i,min}\rangle_t)$, $R_y=\frac{1}{N}\sum_{i=1}^{N}(\langle y_{i,max}\rangle_t-\langle y_{i,min}\rangle_t)$, with $\langle...\rangle$ representing the long-time average. $R=0$ indicates the presence of stable steady states. To distinguish between trivial (extinction state) and non-trivial steady states, the concept of $\Delta = \Theta(\mathbf{X}_i - \delta)$ is introduced, where $\delta$ is a predefined threshold, and $\Theta(x)$ is the Heaviside step function. Non-zero values of the order parameter $R$ indicate the continued existence of the metapopulation throughout the network, whereas $R=0$ indicates  extinction  of this metapopulation.

\begin{figure}[ht]
	\begin{center}
		\includegraphics[scale=0.45]{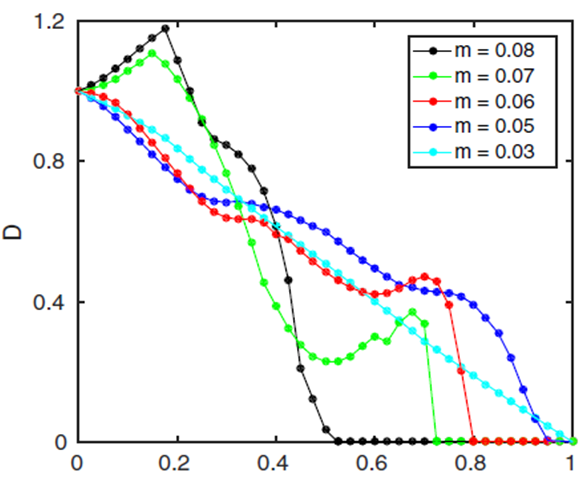}
		\caption{The normalized order parameter $D=\frac{R(p)}{R(0)}$ for the system~\eqref{eq:2} is plotted against the inactivation ratio $p$ in the globally coupled network for various dispersal values of $m$. The critical ratio $p_c$ at which $D$ reaches zero increases as the dispersal rate decreases. Reprinted figure with permission from Ref.\cite{Rakshit_2017_PRE}}
			\label{Eco_fig-1}
	\end{center}
\end{figure}


\begin{figure}[ht]
	\begin{center}
		\includegraphics[scale=0.48]{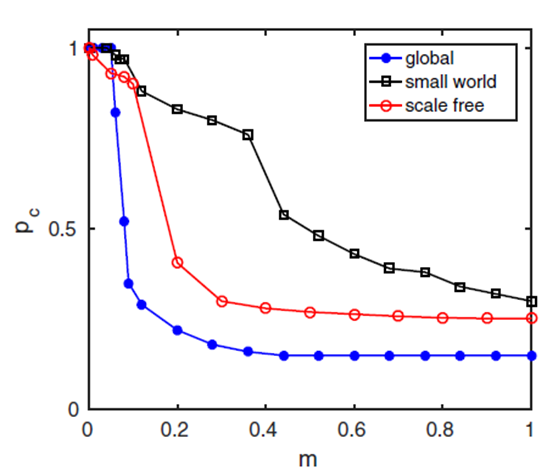}
		\caption{A comparison of the critical inactivation ratio ($p_c$) values for the system~\eqref{eq:2} in relation to the dispersal rate ($m$) is presented for three distinct network topologies: global, small-world, and scale-free. The metapopulation comprises a total of $N=500$ patches, and for the scale-free network, a random inactivation procedure has been employed. Reprinted figure with permission from Ref.\cite{Rakshit_2017_PRE}}
			\label{Eco_fig-3}
	\end{center}
\end{figure}

We now examine three distinct types of dispersal networks: global, small world, and scale-free. \\
i) In the case of an all-to-all network, as illustrated in Fig.~\ref{Eco_fig-1}, the normalized order parameter $D=\frac{R(p)}{R(0)}$ is plotted against $p$ for various dispersal rates $m$. The figure reveals that as the dispersal rate decreases, the dynamical robustness increases until it reaches a value of unity at a critical threshold of $m=0.03$, below which $p_c$ remains at unity. This implies that a lower dispersal rate supports metapopulation survivability. Additionally, we explore complex dispersal topologies, specifically small-world and scale-free networks, among patches, aiming to understand population revival in inactive patches through dispersal. \\
ii) The small-world dispersal topology exhibits aging transition behavior that is qualitatively akin to the global (all-to-all) network. \\
iii) In case of a scale-free dispersal network we  employ three distinct inactivation strategies for the patches, namely random, targeted hub (highest degree node), and targeted low-degree nodes. Analyzing the variation of $D$ in relation to $p$ across the three different dispersal topologies, it becomes apparent that in the case of random inactivation, the likelihood of metapopulation persistence is higher compared to targeted inactivation. Concurrently, it is evident that for targeted inactivation, particularly of high-degree nodes, the critical value $p_c$ is notably lower. Across all conceivable scenarios, there exists a substantial abundance of species within the metapopulation until the inactivation ratio $p$ reaches the critical value $p_c$, at which point sudden and explosive extinction occurs. In conclusion, we undertake a comparison of the aging transition across all three dispersal topologies. Figure~\ref{Eco_fig-3} illustrates the variation of the critical inactivation ratio $p_c$ with the dispersal rate $m$, clearly indicating that a small-world network exhibits the highest ecological robustness. Conversely, in the case of global dispersal, the likelihood of metapopulation extinction is notably higher compared to the more intricate complex dispersal networks.

\par As an extension to this study, multilayer framework being capable of offering an inherent structural framework to model diverse ecological systems, the authors in the Ref~\cite{kundu2021persistence} investigated the persistence in multilayer ecological network consisting of harvested patches. They considered small-world dispersal topologies in the layers for modeling the communications between the prey-predator patches.  The significant effect on the global persistence of species caused by asymmetric intralayer and interlayer dispersal strengths, along with the unique network topologies within the layers, is examined in detail.


\subsection{Neuronal networks}

\begin{figure}
	\centerline
	{\includegraphics[scale=0.360]{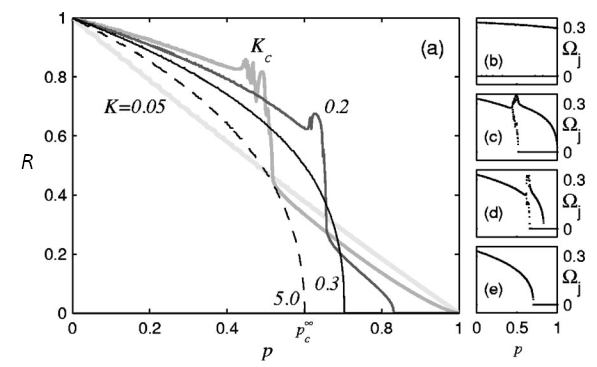}}
	\caption{ Normalized value of the ensemble's average frequency $R$ for the system~\eqref{neu2} with respect to the inactivation ratio $p$, for several	values of the interaction strength $K$. The right panels depict the individual frequencies $\Omega_j$ for (b) $K=0.05$, (c) $K=0.144$, (d) $K=0.2$ and (e) $K=0.3$. Reprinted figure with permission from Ref.\cite{pazo2006universal}}
	\label{fig19}
\end{figure}

We here present some significant results for studying neuronal ensembles. In contrast to the above discussions of dynamical robustness of networked systems in which the self-oscillatory dynamics of the inactive units is lost via inverse Hopf bifurcation, we start our discussion by focusing on the robustness of neuronal populations where the inactive neuronal systems lose their dynamism through a saddle-node bifurcation on the invariant circle (SNIC)~\cite{pazo2006universal}. To be precise, here the inactive units exhibit the regime of class-I excitability. Analogous to the earlier approach, we split the entire population of size $N$ into two sets $S_E$ and $S_A$, comprising of $pN$ excitable and $(N-Np)$ acitive systems, respectively. The dynamical evolution of the networked system reads as,   
\begin{equation}
	\begin{array}{lcl}\label{neu1}
		\dot{\bf x}_j={{\bf F}_j}({{\bf x}_j})+\dfrac{K}{N}\sum\limits_{k=1}^{N}({\bf x}_k-{\bf x}_j);~~~~~~j= 1,2,\cdots,N,
	\end{array}
\end{equation}
in which ${\bf F}_j={\bf F}_{A(E)}$ whenever $j \in S_{A(E)}$. We, specifically, choose the paradigmatic Morris-Lecar models for our analysis so that system~\eqref{neu1} becomes,
   
\begin{equation}
	\begin{array}{lcl}\label{neu2}
		C \dot{\bf V}_j=g_L(-V_j-V_L)-w_jg_K(V_j+V_K)-g_{Ca}m_{\infty}(V_j)(V_j-V_{Ca})-\phi_j(V_j-0.2)+\dfrac{K}{N}\sum\limits_{k=1}^{N}(V_k-V_j),\\
		~~~\dot{w_j}=\lambda(V_j)\{w_{\infty}(V_j)-w_j\},~~~~~~j= 1,2,\cdots,N,
	\end{array}
\end{equation}
where $m_{\infty}(V_j)=[1+\mbox{tanh}\{(V_j-v_1)/v_2\}]/2$, $\lambda(V_j)=\lambda_0[1+\mbox{cosh}\{(V_j-v_3)/v_4\}]$ and $w_{\infty}(V_j)=[1+\mbox{tanh}\{(V_j-v_3)/v_4\}]/2$. The parameters are chosen as $g_L=0.5$, $V_L=0.4$, $g_K=2$, $V_K=0.7$, $g_{Ca}=V_{Ca}=C=1$. $\lambda_0=0.33$ and $v_{1,2,3,4}=(-0.01,0.15,0.10,0.145)$. Moreover, $\phi_j=\phi_A$ assumes the value $\phi_A>\phi_*\sim 0.076$ for an active (self-oscillatory) individual dynamics whereas $\phi_j=\phi_E$ follows $\phi_E<\phi_*$ for an excitable cell. We next define the mean ensemble's frequency as a measurement of the global oscillation in the ensemble as follows,
\begin{equation}
	\begin{array}{lcl}\label{neu3}
		\Omega=\dfrac{1}{N}\sum\limits_{k=1}^{N}\Omega_k,
	\end{array}
\end{equation}
with its normalized value $R=\frac{\Omega(p)}{\Omega(0)}$, in which $\Omega_k$ is the mean frequency of the $k$-th dynamical unit. For a network of $N=500$ Morris-Lecar units, Fig.~\ref{fig19}(a) displays the normalized average frequency of the ensemble (along with the individual frequency $\Omega_j$ in the right panels) as a function of the increasing inactivation ratio $p$, for different values of the interaction strength $K$. It is reasonably discernible that for low $K$, the transition to the global quiescence state (whenever) takes place only when all the units are in the inactive (excitable) regime, i.e., for $p_c=1$. This scenario alters when we consider a coupling strength higher than $K_c \sim 0.144$ for which aging occurs even when not all the elements are excitable. For high $K$, smooth profiles of $R$ and hence of aging are observed, whereas intermediate values of $K$ leads to step-like profiles of $R$. The corresponding step-like profiles of the individual frequencies $\Omega_j$ are clearly visible in Figs.~\ref{fig19}(c) and~\ref{fig19}(d) for $K=0.144$ and $K=0.2$, respectively.  
\begin{figure}
	\centerline
	{\includegraphics[scale=0.320]{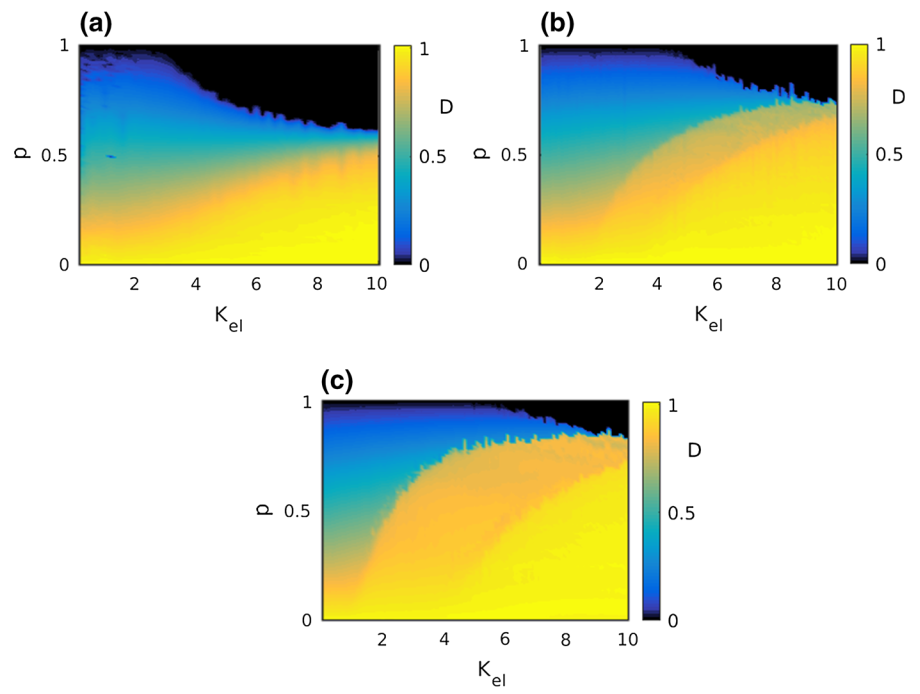}}
	\caption{ The order parameter $R$ for the system~\eqref{neuu1} with respect to the simultaneous variation in the electrical coupling strength $K_{el}$ and the inactivation ratio $p$, for	three different values of the chemical synaptic strengths (a) $K_{ch}=0$, (b) $K_{ch}=0.5$ and (c) $K_{ch}=1.5$. Reprinted figure with permission from Ref.\cite{kundu2019chemical}}
	\label{fig18}
\end{figure}

\par In contrast to the above study based upon the assumption of the presence of only linear diffusive electrical coupling, we next examine the robustness of another neuronal population which consists of neuronal systems interacting through both diffusive gap junctional and non-linear chemical synaptic communication~\cite{kundu2019chemical}. We particularly emphasize on the multi-layer framework of the neuronal ensemble and demonstrate that the chemical synapses acting through the inter-layer connections are sufficiently potential in recovering the global oscillation and hence the dynamical rhythmicity of the network. We now choose Hindmarsh-Rose neuron model in order to cast the nodes in both layers organized in a framework of bi-layer multiplex network. We presume that the neurons communicate via electrical coupling within each layer while the neurons across the layers are connected through chemical synapses. 
The intra-layer connectivities are considered to be of small-world topology, which is evident in case of brain networks. The mathematical description of the entire multi-layer network can then be given by the following equations as,
\begin{equation}
	\begin{array}{lcl}\label{neuu1}
		\dot {x}_{i,k}=a{x^2}_{i,k}-{x^3}_{i,k}-y_{i,k}-z_{i,k}+\dfrac{K_{el}}{N}\sum\limits_{j=1,~j\neq i}^{N}A_{ij}(x_{j,k}-x_{i,k})+K_{ch}(v_s-x_{i,k})\Gamma(x_{i,l}),\\
		\dot {y}_{i,k}=(a+\alpha){x^2}_{i,k}-y_{i,k},\\
		\dot {z}_{i,k}=c(bx_{i,k}-z_{i,k}+e),
	\end{array}
\end{equation}
where $N$ is the number of neurons present in each layer with $i= 1,2,\dots,N$, $k,l=1,2$ and $k \neq l$. The parameters $K_{ch}$ and $K_{el}$ account for the chemical and electrical synaptic strengths, respectively. Finally, ${(A_{ij})}_{N \times N}$ represents the adjacency matrix associated to a small-world architecture considered for each layer. Besides, the variables ${x}_{i,k}$, ${y}_{i,k}$ and ${z}_{i,k}$ correspond to the membrane potentials, ion transport across the membrane via the fast and slow channels, respectively. The synapses are assumed to be excitatory with the reversal potential $v_s> {x}_{i,k}(t)$ for all $t$. Further, the chemical synaptic function is of the form $\Gamma(x)=\frac{1}{1+e^{-10(x+0.25)}}$. With the parameters $a=2.2$, $c=0.005$, $e=5$ and $\alpha=1.6$, isolated neuronal systems display a plateau bursting for $b=9$ and exhibit a stable equipoint regime whenever $b=4$. Thus, the active units correspond to $b=b_A=9$, whereas the inactive systems are associated to $b=b_I=4$. With a similar set-up of dividing the whole ensemble into two groups of active ($(1-p)N$ number of neuronal systems) and inactive dynamical units ($pN$ number of neuronal systems), we define the following order parameter for each layer $\bar{R}_k=\sqrt{\langle ({\bf X}_{c_{k}}-\langle {\bf X}_{c_{k}} \rangle )^2 \rangle}$, where ${\bf X}_{c_{k}}=\frac{1}{N}\sum\limits_{j=1}{N}({x}_{j,k},{y}_{j,k},{z}_{j,k})$ is the centroid of the $k$-th layer ($k=1,2$) and $\langle \cdots \rangle$ stands for long-time average. The order parameter for the entire network then is defined as $\bar{R}(p)=\frac{1}{2}[{\bar{R}}_1(p)+{\bar{R}}_2(p)]$ with the normalized value $R=\bar{R}(p)/\bar{R}(0)$.

\par Assuming $N=200$ neuronal systems in each layer, and a small-world network topology (with $p_{sw}=0.05$ and average-degree $\langle k \rangle=50$) in each layer, we plot the order parameter $R$ as a function of the inactivation ratio $p$ and the gap-junctional strength $K_{el}$ for different values of the chemical synaptic strength $K_{ch}$, in Fig.~\ref{fig18}. The black regions in each of the plots correspond to $R=0$ reflecting the aging transition, that is, when the entire neuronal ensemble loses its dynamism. We start with the no-multiplexing case (i.e., with $K_{ch}=0$) in Fig.~\ref{fig18}(a), and observe that increasing $p$ leads to an aging transition, depending on the strength $K_{el}$ of the electrical coupling. Higher the gap-junctional strength, the earlier the global oscillation of the network vanishes and hence aging transition takes place. However, as we introduce multiplexing in the ensemble through the non-zero chemical synaptic strength $K_{ch}=0.5$ (cf. Fig.~\ref{fig18}(b)), we witness a significant improvement in the dynamical robustness of the networked system. This is realized in the form of a shrinked black region associated with the aging transition. We further increase the chemical synaptic strength to $K_{ch}=1.5$ and depict a similar phase diagram in Fig.~\ref{fig18}(c). With this higher $K_{ch}$, we encounter narrower black region demonstrating a further enhancement in the robustness of the system. This is how the chemical synapses are capable of enhancing the rhythmicity of the multiplexed neuronal ensemble.  \\

\par In addition to these studies, the dynamical robustness of neuronal networks is further examined in terms of the phenomenon of aging transitions in an Erdős–Rényi network of interacting Rulkov neurons based on network connectivity, connection strength, and the ratio of inactive neurons~\cite{biswas2022ageing}. Both noise-free and stochastic networks, with additive noise affecting coupling strength, are investigated. Both smooth and explosive aging transitions are witnessed in both noise-free and stochastic networks. Although, noise is found to mitigate the impact of inactive neurons and reduce the occurrence of explosive transition in the networked system.

\par Research in this area includes the illustrious study presented by Bara\'c at al.~\cite{barac2023determinants} which explores dynamical robustness in terms of the collective failures in networks of interacting heterogeneous excitable systems using the FitzHugh-Nagumo neuronal model. These networks are assumed to exhibit essential characteristics like broad-scale degree distribution, small-world feature and high modularity. The proportion of inactive excitable units, the interaction strength between them, and their proximity to the bifurcation point all influence the network failure resulting in collective aging transition. It is further demonstrated that intermediate coupling strengths prolong global network activity when high-degree nodes are inactivated first. Additionally, the most effective strategy for inducing collective failure adepends non-monotonically on coupling strength and the distance from the bifurcation point to the oscillatory dynamics of  the units.

\par Furthermore, the Ref.~\cite{liu2022analysis} studied the dynamical robustness of a multilayer neuronal network with electrical intra-layer interaction and non-synaptic ephaptic coupling between the layers. Ephaptic coupling arises due to electromagnetic induction caused by extracellular electric fields~\cite{faber2018two}, and is a form of non-synaptic interaction among neurons that plays crucial role in neuronal communication. It is worth mentioning that the inter-layer ephaptic interaction enhances the dynamical robustness of both individual layers and the entire network, contrasting with electrical coupling, which tends to weaken it. The network dynamics with such inter-layer ephaptic coupling is also validated using an analog circuit built on Multisim.


\section{Enhancement and maintenance of dynamical robustness}

In today's interconnected world, where complex systems govern various aspects of our lives, ensuring their stability and robustness has become paramount. Enhancing dynamical robustness in the form of resurrecting oscillatory activity is a crucial endeavor that safeguards the reliability and safety of critical system, ranging from power grids and transportation networks to financial systems and healthcare infrastructure. The oscillatory behavior in neurons is pivotal for processing neural information and coordinating processes related to cognitive functions and memory~\cite{varela2001brainweb,schmidt2018network}. Neurons thus demonstrate a pronounced inclination to engage in rhythmic activity both at the individual and collective level~\cite{stiefel2016neurons,singer2018neuronal}. Therefore, the interruption in the oscillatory behavior of neurons can directly impact essential neural processes. Proper functioning of cardiac and respiratory systems~\cite{jalife1998self}, as well as physiological processes like cell necrosis within organs~\cite{gurtner2007progress} relies on oscillatory dynamics. Power-grid networks necessitate stable, synchronized rhythmic activity as a requirement~\cite{motter2013spontaneous,gambuzza2017analysis}. On the other hand, the crisis of extinction of species is primarily attributed to factors such as climate change or excessive utilization of natural resources, significantly impacting the surrounding ecosystems on a broad scale~\cite{garcia2018effect,may2019stability}. In ecological networks, the extinction of patches within the meta-population can thus result in significant alterations to its overall sustainability~\cite{gilarranz2012spatial,allesina2012stability,kundu2017survivability,kundu2021persistence}. The significance and worth of investigations on the enhancement of robustness is thus evident in both natural and artificial contexts.


\subsection{Controlled diffusion method}

In the last few years, there have been a number of significant attempts in reaching an enhancement of the dynamical robustness of complex networked systems. In order to retrieve dynamism from the state of aging in damaged dynamical networks, we first describe one of the most efficient and simple procedure based upon \textit{controlled diffusion}~\cite{liu2016enhancing}. We unravel that a tiny deviation from the usual diffusive coupling among the dynamical units of a network can increase the dynamical robustness of networks quite comprehensively. Instead of easier aging, strong interaction under such controlled diffusion can support robustness of the concerned system. Let us express the dynamical evolution of the network of $N$ all-to-all coupled Stuart-Landau oscillators as,        

\begin{equation}
	\begin{array}{lcl}\label{con1}
		\dot{z_j}=(a_j+ i\omega-|{z_j}|^2)z_j+\dfrac{K}{N}\sum\limits_{k=1}^{N}(z_k-\alpha z_j);~~~~~~j= 1,2,\cdots,N,
	\end{array}
\end{equation}
where $\alpha \in [0,1]$ is a feedback parameter that controls the diffusion rate. This $\alpha$ thus differentiates this coupling form from that of the other studies which consider standard diffusive coupling without any such controlling parameter. $\alpha \in [0,1]$ controls the diffusion, while making a bridge between the direct coupling (whenever $\alpha=0$) and the usual symmetric diffusive coupling (for $\alpha=1$). This is indicative of controlled diffusion being able to represent the feature of diffusion in a large variety of real-world systems including biological and technological networks.    

\par
\begin{figure}
	\centerline{\includegraphics[scale=0.4200]{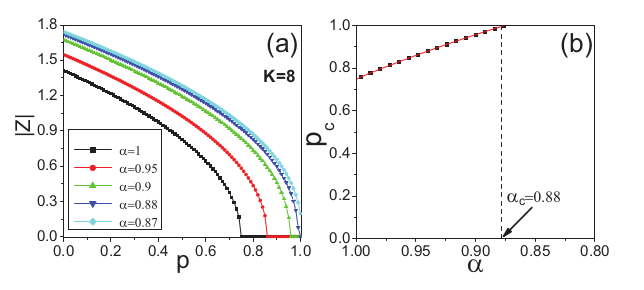}}
	\caption{ (a) The order parameter $|Z|$ for the system~\eqref{con1} as a function of the inactivation ratio $p$ for different values of the diffusion control parameter $\alpha$ for the network size $N=1000$, with the other parameters as $a=2$ for active oscillator, $b=1$ for inactive oscillator, $\omega=3$, and the interaction strength $K=8$. (b) The critical inactivation ratio $p_c$ with respect to $\alpha$ for fixed $K=8$. The red line represents the theoretical result whereas the black squares indicate the numerical outcomes. Reprinted figure with permission from Ref.\cite{liu2016enhancing}}
	\label{fig14}
\end{figure}

 Similar to the approach discussed above, setting $z_j=A$ for all the active dynamical units  $j=1, 2, ..., N-N_p$ and $z_j=I$ for all the inactive elements $j=N-N_p+1,..., N$, the system~\eqref{con1} can be reduced to the following coupled system,

\begin{equation}
	\begin{array}{lcl}\label{con2}
		\dot A=\Big(a+ i\omega-pK+K-\alpha K-|A|^2\Big)A+KpI,\\
		\dot I=\Big(-b+ i\omega+pK-\alpha K-|I|^2\Big)I+K(1-p)A.
	\end{array}
\end{equation} 
Linear stability analysis of the system~\eqref{con2} around one of the equilibrium $(A,I)\equiv (0,0)$ leads us to the following Jacobian matrix, 

$$\begin{pmatrix}
	a+ i\omega-pK+K-\alpha K & Kp\\
	~~K(1-p) &~~ -b+ i\omega+pK-\alpha K
\end{pmatrix}.$$
The critical inactivation ratio can then be determined from this Jacobian following the usual approach (explained above) as
 \begin{equation}
 	\begin{array}{lcl}\label{con3}
 		p_c=\dfrac{a(b+K)+K^2\alpha(1-\alpha)+K(b-a)(1-\alpha)}{K(a+b)}.
 	\end{array}
 \end{equation}  
It is easy to verify that this expression of $p_c$ provides the same result as in the Ref.~\cite{daido2004aging}, whenever the feedback $\alpha=1$. Let us now discuss the numerical results obtained for a network of $N=1000$ nodes and $a=2,~b=1$ with $\omega=3$. For this, we also fix the coupling strength $K$ at $K=8$. Figure~\ref{fig14}(a) depicts the variation in the order parameter $|Z|$ (as defined in Eq.~\eqref{eq-opr}) as a function of the inactivation ratio $p$ for various values of the diffusion control parameter $\alpha$. Precisely, we start with showing the results for the standard diffusion i.e., $\alpha=1$. The order parameter monotonically decreases with increasing $p$ and finally reaches to the null value characterizing the aging transition, which also confirms the observation of the Ref.~\cite{daido2004aging}. We then provide a minute deviation from the unit value of $\alpha$ and choose $\alpha=0.95$. As can be seen as a result of this tiny deviation from the usual diffusion, the order parameter $|Z|$ drops to zero for higher value of the inactivation ratio $p$ and hence the critical inactivation ratio $p_c$ increases. This essentially means that the networked system becomes more robust to the progressive inactivation of its dynamical units compared to the scenario of usual diffusive interaction. We further decrease the value of $\alpha$ to $\alpha=0.90,~0.88$ and $0.87$ respectively and witness that the $p_c$ values further increase progressively. Besides, we also observe that for $\alpha=0.87$, aging does not occur at all for any value of the inactivation ratio $p$. This is because there is a critical value $\alpha_c$ of the control parameter $\alpha$ that must be surpassed in order to have $p_c<1$. Let us now demonstrate this fact through Fig.~\ref{fig14}(b), in which we plot the critical inactivation ratio $p_c$ against the decreasing values of the diffusion control parameter $\alpha$. As can be seen, starting from a value around $p_c \sim 0.76$ whenever $\alpha=1$, the $p_c$ values increase for decreasing $\alpha$. This remains valid until $p_c$ reaches unity at around $\alpha=\alpha_c \sim 0.88$, beyond which an aging transition does not take place anymore.   

\par
\begin{figure}
	\centerline{\includegraphics[scale=0.4200]{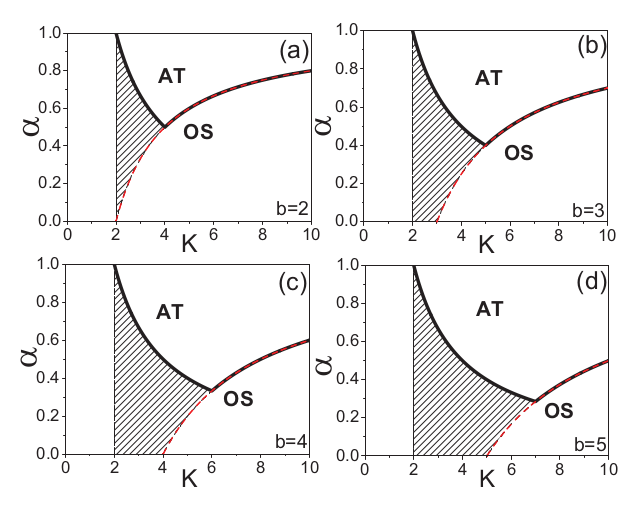}}
	\caption{Phase diagrams for the system~\eqref{con1} in the $(\alpha,K)$ parameter plane for (a) $b=2$, (b) $b=3$, (c) $b=4$, and (d) $b=5$. The other parameters remain same as Fig.~\ref{fig14}. Reprinted figure with permission from Ref.\cite{liu2016enhancing}}
	\label{fig15}
\end{figure}

 As we have already understood, for the standard diffusive coupling, aging arises for all $K>K_c=a$ and increasing $K$ results in earlier aging implying decreasing $p_c$. But one of the most interesting observations from our analysis includes the fact that for higher coupling strengths $K$, aging does not occur if the control parameter $\alpha$ is decreased from a certain value. To be specific, in our case, an aging transition takes place for the following condition on the interaction strength $K$,
\begin{eqnarray}
&
\begin{cases}
	K>a,~~~~~~~~~~~~~~~~~~\mbox{for}~~\alpha=1,\\
	\dfrac{a}{\alpha}<K<\dfrac{b}{1-\alpha},~~~~~\mbox{for}~~\alpha<1.
\end{cases}
\end{eqnarray}
Thus the interval of interaction strength monotonically decreases for decreasing $\alpha$ and eventually vanishes if $\alpha<\frac{a}{a+b}$. \\
Figs.~\ref{fig15} displays the phase diagrams in the $(\alpha,K)$ parameter plane for different values of the parameter $b$. Actually, an aging transition happens due to the competing forces between the two sets of active units possessing $A_j>0$ and inactive units with $A_j<0$, i.e., the values and magnitudes of $a$ and $b$. Through these phase diagrams, we explain how the networked system~\eqref{con1} evolves for simultaneous variation in the control parameter $\alpha$ and the interaction strength $K$, on one hand. On the other hand, we show how different magnitudes of the parameter $b$ affect the robustness of the network. 
\par Figure~\ref{fig15}(a) depicts the $(\alpha,K)$ phase diagram for $b=2$ instead of $b=1$. The upper-right region surrounded by the bold black curve represents the aging transition (AT). The lower-right area encompassed by the dashed red curve stands for the oscillatory state (OS). The shaded region describes the transition zone, where $p_c$ remains at its unit value even for decreasing $\alpha$. The aging region is clearly visible in the phase diagram, where the shaded transition zone touches the $K$-axis whenever $a<K<b$ for $b>a$. Next we progressively increase the value of $b$ and portray similar phase diagrams in the $(\alpha,K)$ plane in Figs.~\ref{fig15}(b),~\ref{fig15}(c), and \ref{fig15}(d) respectively for $b=3$, $b=4$, and $b=5$. Our observation includes that the strong interaction with controlled diffusion ($\alpha<1$) favors dynamical robustness of the network. It is also discernible that aging island expands and the oscillatory region shrinks for increasing values of $b$. Thus, the dynamism of the networked system is difficult to resurrect, whenever the attraction strength of the inactive elements becomes stronger.     

\subsection{Mean-field feedback method}

Feedback is the mechanism by which the output of a system is reintroduced into the system as input. It is widely recognized as one of the most applicable concepts across various scientific disciplines, spanning from physics and mathematics to biology and engineering~\cite{franklin2002feedback,aastrom2021feedback}. In control theory~\cite{franklin2002feedback,bechhoefer2005feedback}, dynamical systems~\cite{majhi2016restoration,chandrasekar2015feedback}, neuronal systems~\cite{huang2017positive}, physiology~\cite{vlahopoulos2017aberrant}, game theory~\cite{tilman2020evolutionary}, and also in the study of ecological science and climate science the feedback theory has made significant contributions.  

\par Instead of controlling the diffusion among the dynamical units, we now confine ourselves to the standard diffusive coupling. However, we design the networked system to be exposed to an external mean-field feedback, which we demonstrate to be capable of efficiently resuming dynamic activity in damaged networks and hence enhance the dynamical robustness of the network~\cite{kundu2018resumption}. We analyze the following mathematical form of the considered dynamical network as, 

\begin{equation}
	\begin{array}{lcl}\label{mff1}
		\dot{z_j}=(\alpha_j+ i\omega-|{z_j}|^2)z_j+\dfrac{\epsilon}{N}\sum\limits_{k=1}^{N}(z_k-z_j)+\dfrac{\eta}{N}\sum\limits_{k=1}^{N}z_k;~~~~~~j= 1,2,\cdots,N,
	\end{array}
\end{equation}
where $\epsilon$ is the diffusive coupling strength and $\eta$ accounts for the strength of the mean-field feedback. Proceeding similarly as above, assuming $z_j=A$ for the set of active elements $j=1,2,\dots,N-Np$ and $z_j=I$ for the group of inactive units $j=N-Np+1,\dots,N$, the system~\eqref{mff1} reduces to the following coupled system,
\begin{equation}
	\begin{array}{lcl}\label{mff2}
		\dot A=[a+ i\omega-p\epsilon+\eta(1-p)-|A|^2]A+(\epsilon+\eta)pI,\\
		\dot I=[-b+ i\omega+p\eta-\epsilon(1-p)-|I|^2]I+(\epsilon+\eta)(1-p)A.
	\end{array}
\end{equation}
Linear stability analysis of \eqref{mff2} around the equipoint 
 $(A,I)\equiv (0,0)$ results in the Jacobian matrix, 
$$\begin{pmatrix}
	a+ i\omega-p\epsilon+\eta(1-p) & p(\epsilon+\eta)\\
	~~(\epsilon+\eta)(1-p) &~~ -b+ i\omega+p\eta-\epsilon(1-p)
\end{pmatrix}.$$
 
Negative real parts of all the eigenvalues of this Jacobian determine the stability of the origin, leading to the critical ratio as,
 \begin{equation}
	\begin{array}{lcl}\label{mff3}
		p_c=\dfrac{(b+\epsilon)(a+\eta)}{(a+b)(\epsilon+\eta)},
	\end{array}
\end{equation}
with $\epsilon \ge \epsilon_c=a$.\\

\begin{figure}
	\centerline
	{\includegraphics[scale=0.350]{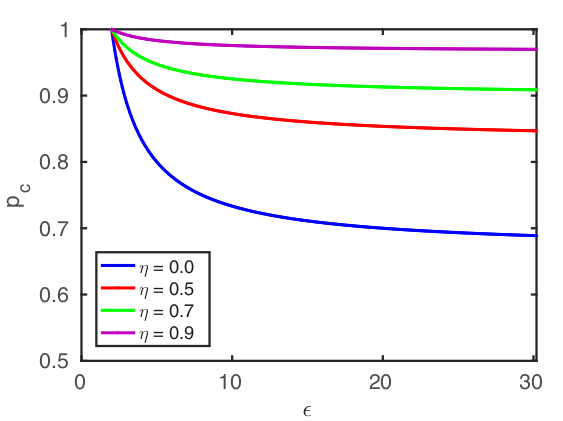}}
	{\includegraphics[scale=0.350]{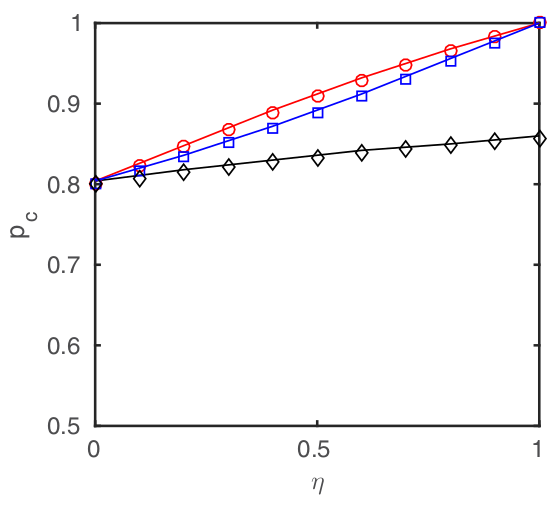}}
	\caption{ (a) The theoretically obtained critical inactivation ratio $p_c$ for the system~\eqref{mff1}, against the interaction strength $\epsilon$ for various values of $\eta$. (b) Critical ratio $p_c$ with respect to the feedback strength $\eta$ for three different mechanisms of adding feedback to the networked system, with the interaction strength fixed at $\epsilon=5$. The scenario of all the nodes, only the inactive nodes, and only the active nodes subject to feedback are in red circle, blue square and black diamond, respectively. Solid lines represent the numerical outcomes and the symbols correspond to the theoretical results. Reprinted figure with permission from Ref.\cite{kundu2018resumption}}
	\label{fig8}
\end{figure}

In Fig.~\ref{fig8}, we plot this theoretically obtained expression of the critical inactivation ratio $p_c$ as a function of the interaction strength $\epsilon$ for various values of the feedback strength $\eta$. Firstly, we present the variation in the $p_c$ values with respect to $\epsilon$ for no feedback (i.e., $\eta=0$). The initial fall of $p_c$ is quite swift for increasing $\epsilon$, however, for sufficiently high $\epsilon$ the decrement in $p_c$ becomes less sharp. When we introduce mean-field feedback with strength $\eta=0.5$, a qualitatively similar trend in the drop of $p_c$ is observed. Interestingly enough, this time, the $p_c$ values remain much higher altogether than the no-feedback case. This readily indicates that the feedback (even if it is of small strength) enhances the dynamical robustness of the networked system to a significant extent. We further increase the feedback strength to $\eta=0.7$ and $\eta=0.9$ and witness that in both the cases the $p_c$ values increase even more, whatever be the value of the interaction strength $\epsilon$, and thus makes the system~\eqref{mff1} more robust to progressive inactivation of its units. Thus, it is conspicuous that the mean-field feedback is highly effective in resuming the dynamic activity of damaged networks of active and inactive dynamical systems. 

\par We further dive deep on the possible ways of employing a mean-field feedback into the system, and study how the robustness gets affected. We portray the alteration in the $p_c$ values with respect to the feedback strength $\eta$ for feedback added to all the units, only the inactive units, and only the active units in Fig.~\ref{fig8}. There, all the solid curves stand for the numerical results, whereas the respective symbols correspond to the theoretical expressions. It is clear that the $p_c$ values increase strictly monotonically for increasing $\eta$, implying a sharp enhancement in the dynamical robustness of the system. The theoretical and numerical outcomes are in excellent agreement. We next induce a feedback to only the inactive units and following the similar procedure as above, reaching to the theoretical expression of $p_c$ as $p_c=(a(b+\epsilon))/((a+b)\epsilon+(a-\epsilon)\eta)$. We plot this expression along with the numerical results as functions of $\eta$, and observe that $p_c$ increases sharply again. Moreover, the results do not differ much from those of the previous case of feedback to all the units. This implies that even if we induce feedback to only the inactive dynamical systems, we can recover dynamism of the network. Finally, we employ feedback to only the active set of elements for which we get the theoretical $p_c=(a+\eta)(b+\epsilon)/((a+b)\epsilon+(b+\epsilon)\eta)$. Plotting this expression along with the numerical $p_c$, we observe that $p_c$ increases for increasing $\eta$, but the increment is not as sharp as in the earlier two cases. Thus, although the outcome is not as good as the other two cases, the robustness still increases even if we add feedback to only the active dynamical systems. 

\subsection{Addition of active oscillators}

Next we discuss another useful mechanism of resurrecting dynamism of damaged dynamical networks, by adding oscillatory units to the network~\cite{morino2013efficient}. We present numerical and theoretical analysis of how the additional supporting oscillators can improve the dynamical robustness of networked systems. We add at most, a single supporting oscillator to each existing dynamical units of the network. Active and inactive units are respectivley supported by $Nq_{A}$ and $Nq_{I}$ number of oscillators. The ratio of supported oscillators thus becomes $q=q_{A}+q_{I}$, and the updated size of the networked system turns out to be $N+Nq$. We denote the sets of supported and unsupported dynamical units by $S$ and $U$, respectively. Whenever a supporting oscillator is added to the $j$-th dynamical unit, the state variable of the supporting system is represented by ${z_j}^*$, the state variables for the existing units are denoted by $z_j,~j=1,2,\dots,N$. The time-evolution of the dynamical network is then described as follows,     
\begin{equation}
	\begin{array}{lcl}\label{add1}
		\dot {z_j}=(\alpha_j+ i\omega-|{z_j}|^2)z_j+\dfrac{K}{N}\sum\limits_{k=1}^{N}(z_k-z_j)+\dfrac{D}{2}({z_j}^*-z_j),~~~~\mbox{for}~j \in S,\\\\
		\dot {z_j}=(\alpha_j+ i\omega-|{z_j}|^2)z_j+\dfrac{K}{N}\sum\limits_{k=1}^{N}(z_k-z_j),~~~~~~~~~~~~~~~~~~~~~~~~\mbox{for}~j \in U,\\\\
		\dot {z_j}^*=(a+ i\omega-|{z_j}^*|^2){z_j}^*+\dfrac{D}{2}(z_j-{z_j}^*),~~~~~~~~~~~~~~~~~~~~~~~~~~\mbox{for}~j \in S,
	\end{array}
\end{equation}
where $K$ is the interaction strength between the dynamical systems in the existing network and $D$ accounts for the coupling strength between the supporting and the supported oscillators. Then the modified order parameter is $|Z|$, where $Z$ is expressed as,
\begin{equation}
	\begin{array}{lcl}\label{add2}
		Z=\dfrac{\sum\limits_{j=1}^{N}z_j+\sum\limits_{j \in S}^{}{z_j}^*}{(1+q)N}.
	\end{array}
\end{equation}
Then following the same approach as before and splitting the entire networked system into all possible sub-groups, active and inactive, the critical inactivation ratio $p_c$ is found as ~\cite{kundu2018resumption},
\begin{equation}
	\begin{array}{lcl}\label{add3}
		p_c=p_0+\dfrac{D}{J_A}\dfrac{a(b+K)}{a+b}q_A+\dfrac{D}{J_I}\dfrac{a(K-a)}{a+b}q_I,
	\end{array}
\end{equation}
in which  

\begin{equation}
	\begin{array}{lcl}\label{add4}
		p_0=\dfrac{a(b+K)}{(a+b)K},\\
		J_A=2a^2+KD-2a(K+D),\\
		J_I=(b+K)D-aD-2a(b+K).
	\end{array}
\end{equation}
Thus the term $p_0$ reflects the critical inactivation ratio in absence of any supporting oscillator added into the system ~\cite{kundu2018resumption}. 

\par
\begin{figure}
    \centerline{\includegraphics[scale=0.3500]{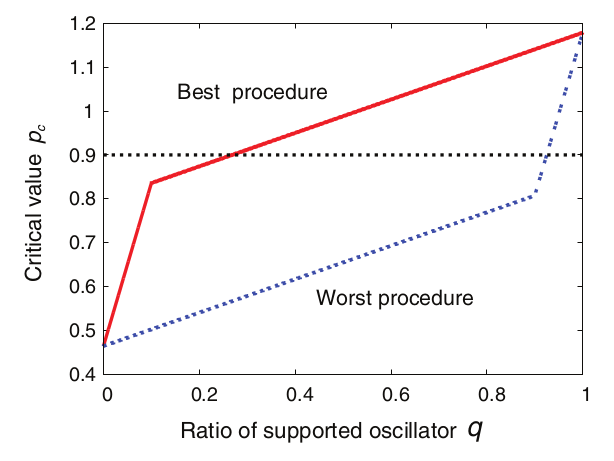}}
	\caption{Critical inactivation ratio $p_c$ (Eq.~\eqref{add3}) for the system~\eqref{add1}, with respect to the ratio of supported oscillators $q$, whenever $a=2,b=5,p=0.9$ and $D=K=8$. The red solid line stands for the best procedure when the active oscillators are preferred and supported. The blue dotted line reflects the worst procedure for which the inactive oscillators are preferred and supported. The global oscillation is resumed at the horizontal black dotted line, where $p_c$ equals $p$. Reprinted figure with permission from Ref.\cite{morino2013efficient}}
	\label{fig10a}
\end{figure}

 It can be further shown that the supporting oscillators added to the active dynamical units are much more effective in improving the dynamical robustness than the mechanism of adding supporting oscillators to the inactive units. Opposing to our intuition that the inactive systems should be supported, support to active units works more efficiently. Through Fig.~\ref{fig10a}, we delineate the variation of the critical inactivation ratio $p_c$ against the ratio of the supported oscillators $q$. We fix the fundamental parameters' values at $a=2,b=5,p=0.9$ and $D=K=8$, and depict the best procedure of preferentially choosing the active oscillators. This means $q=q_A$ for $0\le q \le 1-p$ and $q=q_I+(1-p)$ for $1-p<q \le 1$. We also plot the worst mechanism of preferentially choosing the inactive oscillators, i.e., $q=q_I$ for $0\le q \le p$ and $q=q_A+p$ for $p<q \le 1$. The global oscillation of the networked system is retrieved where $p_c$ equals $p$. As conspicuous from this portrayal, the ratio of the supporting oscillators required to develop the dynamical robustness of the network in the best mechanism is much lower than that in the worst mechanism.   


\subsection{Self-feedback delay}
Lately, researchers have explored a self-feedback delay as an additional mechanism to improve dynamical robustness within a globally coupled network in terms of mean-field interactions~\cite{sharma2021enhancement}. We exemplify the enhancement in dynamic robustness through a network of $N$ Stuart–Landau oscillators that are mutually coupled through mean-field interactions, along with self-feedback delay. The governing equation for their dynamics can be formulated as follows,
\begin{eqnarray}
\dot{z}_j(t) &=& (\alpha_j+i\omega-|z_j(t)|^2)z_j(t)+ k[\bar{z}-z_j(t-\tau)].
\label{eq75}
\end{eqnarray}
\noindent Here, $\overline{z}$ denotes the mean-field average, and $\tau$ represents the time delay in the local self-feedback component $z_j(t-\tau)$, functioning essentially as a form of negative feedback. In this context, $\omega(=5)$ denotes the intrinsic frequency of individual oscillators, while $k$ characterizes the strength of coupling. $\alpha_j$ serves as the bifurcation parameter for the oscillator indexed by $j$, as before. 
We select a network of size $N = 500$ and assign the values $\alpha_j=a=1$ to the set of active oscillators and $\alpha_j=b=-3$ to all the inactive ones. 


\par
\begin{figure}[t]
\centering
\includegraphics[width=0.5\linewidth]{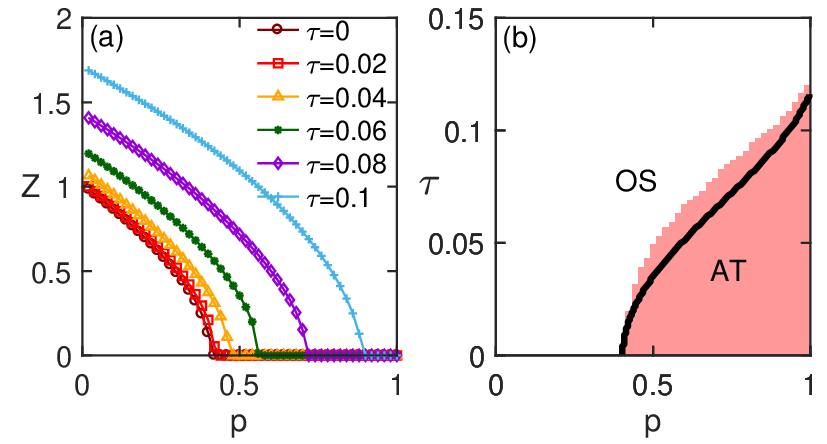}
\caption{(a) The order parameter $|Z|$ for the system~\eqref{eq75} undergoes variations based on $p$ across various $\tau$ values in coupled Stuard-Landau oscillators.
The parameters are specified as $a=1, b=-3, \omega=5$, and $k=5$. (b) Dynamical region for the coupled oscillators drawn for the ($p, \tau$) parameter plane with constant $k=5$. The white and light pink region corresponds to oscillatory state and aging transition zone denote by OS and AT respectively. Solid black line corresponding to the fit of  $p_c$, acquired from Eq.\ref{e6}. Reprinted figure with permission from Ref.\cite{sharma2021enhancement}}
\label{sfd_f1}
\end{figure}

 In Fig.~\ref{sfd_f1}(a), we illustrate the relationship between the order-parameter $|Z|$ and the inactivation ratio $p$ across various settings of the local self-feedback delay $\tau$, while keeping $k=5$. It is evident that when $\tau=0$, the order parameter $Z$ reaches zero (at a certain $p_c$), suggesting that the process of aging transition occurs at a noticeably faster rate. As the self-feedback delay $\tau$ is extended, the aging transition takes place at a higher value of the critical inactivation ratio $p_c$. This implies that adding $\tau$ has a major impact on improving dynamical robustness.
To gain deeper insight into how local self-delayed feedback affects the robustness of coupled oscillators, we have illustrated the phase transition diagram in Fig.~\ref{sfd_f1}(b) within the $\tau-p$ plane, while maintaining a constant value of $k=5$. In this illustration, region OS represents the oscillatory state, while region AT signifies island corresponding to aging transition (i.e., where $|Z|=0$). The figure clearly explores a minimal alteration in the $p_c$ value whenever very small $\tau$ is considered. However, when we elevate the value of $\tau$ towards the upper end, the critical threshold of $p_c$ also rises and ultimately reaches $p_c=1$ when $\tau \sim 0.12$. The presence of a local self-feedback delay $\tau$ is thus demonstrated to be the primary factor influencing the aging transition in the coupled oscillator system. This delay efficiently enhances the dynamical robustness of the networked system of mean-field coupled oscillators.

\par Afterward, we determine the critical value $p_c$ through an analytical method. During the aging transition, the global oscillation ceases at $p_c$, leading to the stabilization of the trivial fixed point $z_j=0,~\forall j$. We assume that the coupled system is composed of two groups, where each group consist of identical nodes. Essentially, the synchronization among the oscillators allows us to redefine the system accordingly. When assigning $z_j = A$ for the active set and $z_j =I$ for the inactive set of oscillators, the system Eq.~(\ref{eq75}) simplifies into the subsequent coupled system, as presented in \cite{daido2004aging},

\begin{eqnarray}
\dot{A}&=&(a+i\omega+kq-|A|^2)A-kA(t-\tau)+kpI,\nonumber\\
\dot{I}&=&(b+i\omega+kp-|I|^2)I-kI(t-\tau)+kqA,\nonumber\\
\label{e2}
\end{eqnarray}
\noindent where $q = 1-p$. Now, we perform a linear stability analysis to approximate Eq.(\ref{e2})  in the vicinity of the equilibrium point $A = I = 0$. The characteristic equation that arises from conducting a linear stability check around the point reads as,
\begin{align}
(a+i\omega+qk-ke^{-\lambda\tau}-\lambda)(b+i\omega+pk-ke^{-\lambda\tau}-\lambda)
-pqk^2=0, 
\label{e3}
\end{align}
\noindent Here, $\lambda=\lambda_R+i\lambda_I$, where $\lambda_R$ and $\lambda_I$ are the real and imaginary part of the eigenvalue of $\lambda$. The following equations are obtained by splitting the real and imaginary parts of Eq.~\ref{e3} and putting the real part of the eigenvalue equal to zero ($\lambda_R=0$). 
\begin{equation}
[a+qk-k cos(\lambda_I\tau)][b+pk-k cos(\lambda_I\tau)]-
(\omega-\lambda_I+k sin(\lambda_I\tau))^2+pqk^2 = 0,
\label{e4}    
\end{equation}
\begin{equation}
[a+b+k(p+q)-2k cos(\lambda_I\tau)][\omega-\lambda_I+k sin(\lambda_I\tau)] = 0,
\label{e5}
\end{equation}
\noindent where $p+q = 1$. The critical value of the inactivation ratio $p_c$ can be obtained by solving these equations. 
\begin{equation}
p_c = \frac{-ab+k(a+b)\beta+k^2\beta-kb-k^2\beta^2}{k(a-b)},
\label{e6}
\end{equation}
\noindent where $\beta = \cos(\alpha\tau)$ and $\alpha = \omega+k\sqrt{1-\left(\frac{a+b+k}{2k}\right)^2}$.
The aging transition is identified though this critical value $p_c$ of the inactivation ratio $p$. The alignment between the critical value of $p_c$ (depicted as a black solid line) and the numerical outcome (illustrated as a shaded region) for the aging transition in Fig.~\ref{sfd_f1}(b) is notably strong.
\begin{figure}[t]
\includegraphics[width=0.5\linewidth]{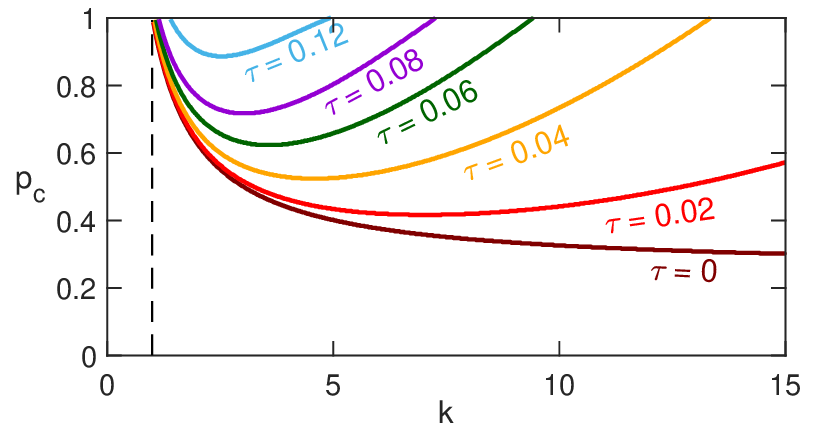}
\caption{The $p_c$ (Eq.~\eqref{e6}) varies based on the coupling strength $k$ for different local self-delay values $\tau$ in the mean-field coupled oscillators, with $p_c$ consistently decreasing as $k$ increases. Reprinted figure with permission from Ref.\cite{sharma2021enhancement}}
\label{sfd_f2}
\end{figure}
When $\tau=0$, the aging transition takes place for all values of $k>1$. As the coupling strength $k$ rises, the critical probability $p_c$ experiences a decrease.
In order to examine the influence of $k$ on $p_c$, we graph the relationship between $p_c$ and the coupling strength $k$ for various $\tau$ values, as depicted in Fig.~\ref{sfd_f2}. Remarkably, we notice that in the presence of $\tau \neq 0$, the aging transition is present solely within a limited range of $k$. In this range, the $p_c$ value gradually diminishes until it reaches its lowest point, after which it steadily rises to one. 
This observation suggests that robust network dynamics against aging is promoted by strong coupling, particularly when the feedback delay $\tau$ reaches a significant magnitude.


\subsection{Asymmetric interactions}

\noindent In the following, we report a mechanism to enhance dynamical robustness by employing asymmetric interactions between active and inactive nodes~\cite{Srilena1_2108_EPL}. We consider the following mathematical form of the coupled Stuart-Landau oscillators
\begin{equation}\label{eq_asym:1}
	\dot {z_j}=(\alpha_j+ i\omega-|{z_j}|^2)z_j+\frac{M\epsilon}{N}\sum\limits_{k=1}^{N}A_{jk}(z_k-z_j);~~~~~~j= 1,2,\cdots,N.
\end{equation}
In this context, $\epsilon$ represents the strength of diffusive coupling, and $M$ serves as an asymmetry parameter introduced into the system to differentiate the coupling strengths of active and inactive sub-populations. Specifically, we set $M=1$ for the active group and $M=m(\ge 1)$ for the inactive group of oscillators. Now based upon the assumption of $z_j = A$ for active oscillators  and $z_j = I$ for inactive oscillators, Eqs.~(\ref{eq_asym:1}) reduces to the following coupled system for a homogeneous network
 \begin{equation}
	\begin{array}{lcl} \label{eq_asym:2}
	\dot{A} &=& (a + i\omega - |A|^2)A + \epsilon pd(I-A),\\
	\dot{I} &=& (-b + i\omega - |I|^2)I + m\epsilon(1-p)d(A-I).
	\end{array}
\end{equation}
Here, $d$ is defined as $d = \langle k \rangle/(N-1)$, where $\langle k \rangle$ represents the average degree of the network(in the case of global coupling, $d = 1$). A linear stability analysis of the system (\ref{eq_asym:2}) around the origin $(A, I) =(0, 0)$ leads to the following critical inactivation ratio 
\begin{eqnarray}
 p_c &=& 1-\frac{b(\epsilon d-a)}{\epsilon d(b+am)},
 \label{eqn_asym:3}
\end{eqnarray}
\noindent for $\epsilon \geq a/d = \epsilon_c$.  From Eq.~(\ref{eqn_asym:3}), it is convincing that as one increases the  asymmetric parameter $m$, the $p_c$  value also increases. So  by merely increasing the interaction strength of the dynamical units in the inactive group compared to the active ones, it is possible to significantly enhance dynamical resilience with great efficiency.

\begin{figure}[h]
\centering
\includegraphics[width=0.5\linewidth]{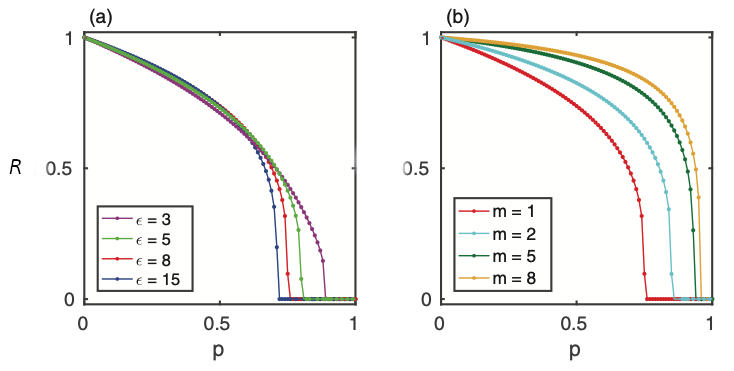}
\caption{(a) The relationship between the order parameter $Z$ and the critical ratio $p$ for the system~\eqref{eq_asym:1} is depicted for different coupling strengths, namely $\epsilon = 3, 5, 8, 15$ with a fixed value of $m = 1$. In this case, the critical inactivation ratio $p_c$ exhibits a gradual decrease as $\epsilon$ increases. (b) For various asymmetry parameter values of $m = 1, 2, 5, 8$ with a constant $\epsilon = 8$, the critical inactivation ratio $p_c$ shows a gradual increase as $m$ grows. Reprinted figure with permission from Ref.\cite{Srilena1_2108_EPL}}
\label{df_1}
\end{figure}

Next, we proceed to derive the analytical expression of $p_c$ for a scale-free network. In the case of a heterogeneous network with a large number of nodes, denoted as $N$, its behavior is primarily governed by two mean fields representing the sub-populations of active and inactive oscillators. To facilitate the application of the degree-weighted mean field approximation or the annealed network approximation to each sub-network, we assume that oscillators with the same degree within the same sub-population are indistinguishable. Based on this assumption and following the analytical methods described in the section~\ref{analytical}, we can derive the critical ratio $p_c$ as follows:
\begin{eqnarray}
p_c= \dfrac{F(\epsilon,a)-d}{F(\epsilon,a)-F(\epsilon,-b)},
 \label{eqn_asym:4}
\end{eqnarray}
for $\epsilon > \epsilon_c(=a/d_{min})$, where $d_{min} = k_{min}/N$ with the minimum degree $k_{min}=min\{k_j\}$, and $F(\epsilon,\alpha)\simeq \frac{1}{N}\sum\limits_{j=1}^N\frac{d_j^2}{d_j-\alpha/M\epsilon}.$ Here $d_j=k_j/N$ is the ratio of the degree of $j$-th oscillator and the system size.


\subsection{Low-pass filtering mechanism}

The functioning of low-pass filters has been widely utilized in diverse physical systems, including electronic and optical devices. Low-pass filters are widely present in electrical and biological networks as well~\cite{stark2012neurological}. The impact of a low-pass filter has been previously investigated within the realm of synchronization~\cite{kim2006synchronization,soriano2008synchronization}, as well as the process of oscillation suppression~\cite{zou2017revoking} and transition between limit cycles~\cite{banerjee2018transition,biswas2019effect}. 
Recently, the approach involving low-pass filtering is employed to restore the oscillatory characteristic and hence strengthen the dynamical robustness of a network.  This technique allows low-frequency signals to pass through, while potentially reducing the strength of signals with higher frequencies. The description of the mechanics of the low-pass filter for Stuart-Landau oscillators is outlined by the following set of equations 
 \cite{bera2019low},

\begin{eqnarray}
	\dot{z}_j &=& (\alpha_j+i\omega_j-|z_j|^2)z_j+\frac{\epsilon}{d_j+1}\sum_{k=1}^{N} A_{jk}(z_k - \mu_j),  \nonumber\\
	\beta\dot{\mu}_j &=& -\mu_j+z_j.
	\label{lpf_eq1}
\end{eqnarray}

\noindent For each value of $j$ from $1$ to $N$, $z_j$ represents the complex amplitude, and $\alpha_j$ indicates the intrinsic parameter of the $j$-th oscillator, signifying its distance from the Hopf bifurcation point. $\omega_j$ denotes the inherent frequency of the $j$-th oscillator, while $\epsilon$ indicates the total coupling strength, and $d_j$ denotes the degree of the $j$-th node. The second equation of (\ref{lpf_eq1}) illustrates the standard low-pass filter, and $\frac{1}{\beta}(\beta > 0)$ signifies the cutoff frequency. The utilization of a low pass filter for interaction introduces a frequency dependent impact on the dynamical behaviour. When $\beta$ approaches $0$, $\mu_j$ converges precisely to the initial $z_j$, resulting in a standard scalar diffusive type of interaction.
\begin{figure}[h]
\centering
\includegraphics[width=0.8\linewidth]{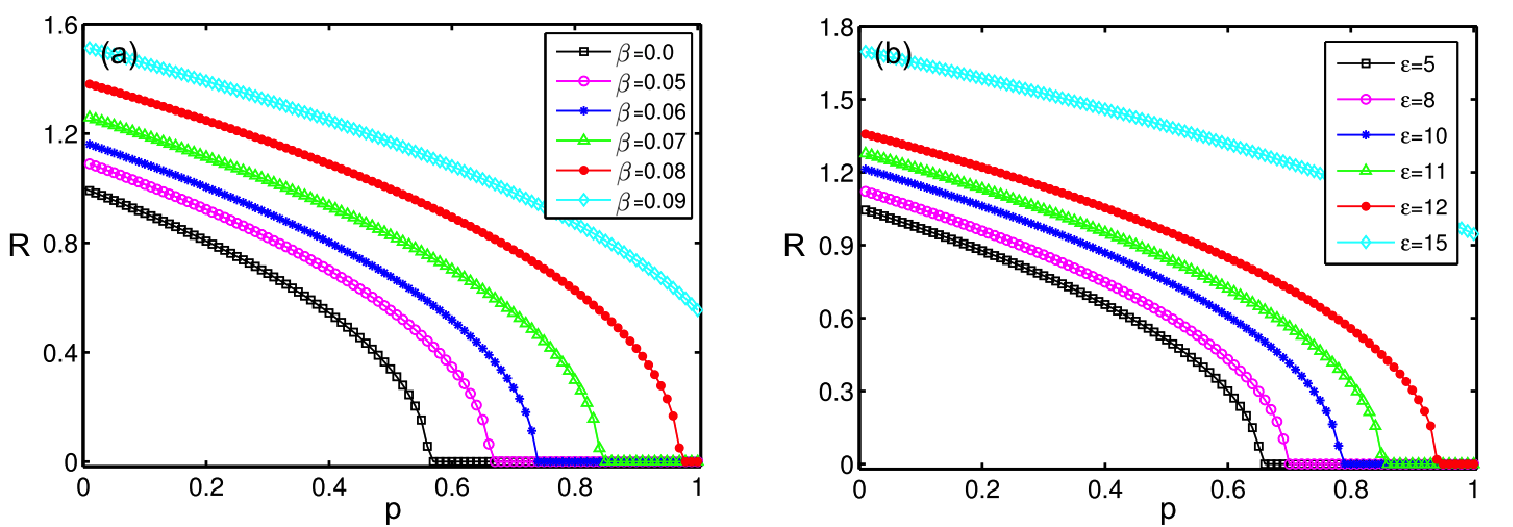}
\caption{The order parameter $R$ for the system~\eqref{lpf_eq1} as function of the inactivation ratio $p$ for (a) various $\alpha$ values at the coupling strength $\epsilon=8$, and (b) for various $\epsilon$ values at fixed value of $\beta=0.055$ for the globally coupled Stuart-Landau oscillators. Reprinted figure with permission from Ref.\cite{bera2019low}}
\label{df_1}
\end{figure}
We investigate the influence of a low-pass filter on the dynamical robustness for the globally coupled Stuart-Landau network. 
\par Our intent is to check whether for a fixed interaction strength $\epsilon$, a proper tuning of $\beta$ can enhance the dynamical robustness of the network. Choosing several values of $\beta$ for a fixed coupling strength $\epsilon=8$, we depict how the order parameter $R$ varies with the inactive ratio $p$. For $\beta=0$, the critical value of $p$ for the aging transition is $p_c=0.5625$, but as $\beta$ increases, the critical values of $p_c$ transits to a higher $p_c$ value. This scenario is depicted in Fig.~\ref{df_1}(a) by taking several exemplification increase values of $\beta$. Thus, with a proper choice of $\beta$, the coupling mechanism leads to the recovery of the oscillatory dynamics in the dynamical units from the damaged network.

\par Our next step is to further verify this result, and we plot the order parameter as a function of the inactivation ratio $p$, for a fixed value of $\beta=0.055$, and several values of $\epsilon$ (cf. Fig.~\ref{df_1}(b)). For $\epsilon=5$, the critical value of $p$ for aging transition is $p \sim 0.66$. The critical transition point keeps shifting to higher values as $\epsilon$ increases. Interestingly, with sufficiently high coupling, no AT is observed, and global oscillation of the entire network resumes even when the network contains all inactive nodes. This scenario is depicted in Fig.~\ref{df_1}(b) using the outstanding value of $\epsilon=15$. However, the improvement in dynamical robustness provided by such transition scenarios is independent of the network size. Thus the parameter $\beta$ makes the dynamical network more robust to the inactivation of the local units, and can be well-treated as an efficient mechanism for this purpose. 


\section{Machine  learning techniques to predict aging transition}
 The aging transition stands as an undesirable phenomenon in real-world systems, emphasizing the critical need to predict its occurrence while the underlying system is still in normal operation. In numerous practical applications, the governing equations of the system dynamics are often unknown, posing a challenge for the development of a mathematical model for prediction or control. As a solution, data-driven approaches, including machine learning and deep learning, have garnered substantial attention in recent years. These methodologies have the advantage of learning directly from available data or time series without the necessity of prior knowledge about the underlying system equations\cite{jaeger2001echo,tanaka2019recent, gauthier2021next,pathak2018model,hramov2021physical}.  Here we discuss 
  a machine-learning based method to explore the possibility of predicting aging transitions in a system currently experiencing a ``normal" regime with oscillations but undergoing a gradual parameter drift, potentially influenced by environmental changes\cite{rakshit2023predicting}. Specifically, we leverage a parameter-aware reservoir computing method introduced by Xiao et al\cite{xiao2021predicting}. Instead of inputting the system's intrinsic parameter values directly into the machine, we provide information regarding the fraction of inactive oscillators within the network.

The setup of Echo State Network based reservoir computing involves three components: an input layer, a reservoir network, and an output layer. In this setup, an $M$-dimensional input signal $u(t)\in R^M$ is internally fed to an $N_r$-dimensional cyclic reservoir network through an $N_r\times M$ input weight matrix $W_{in}$. The reservoir network comprises $N_r$ nodes connected in an Erdős–Rényi graph configuration, represented by an $N_r\times N_r$ weight matrix $W_{res}$. At time $t$, the state of the reservoir network is denoted by the vector $r(t)=[r_1(t),r_2(t),\ldots,r_{N_r(t)}]^T$, where $r_i(t)$ signifies the state of the $i$-th node in the reservoir at time $t$. The $N_r$-dimensional state vector of the reservoir network is then mapped to an $M-1$ dimensional output signal using the output matrix $W_{out}$ with dimensions $(M-1) \times N_r$. The updated equations governing the reservoir states are described as follows,

\begin{equation}\label{eq()}
  r(t + dt) = (1 - \alpha)r(t) + \alpha \tanh[W_{res} \cdot r(t) + W_{in} \cdot u(t)]. 
\end{equation}

\noindent Here, $\alpha\in(0,1]$ represents the leaking rate,  and $u(t)\in R^M$ the $M$-dimensional input data incorporating the system parameter value.  The input data comprises the time series data $\tilde{u}(t)$, which constitutes the first $M-1$ elements of $u(t)$, along with the corresponding parameter $p_s$, represented as the last element of $u(t)$, i.e.,

\begin{equation}\nonumber
u(t)=[\tilde{u}(t);p_s].
\end{equation}

\noindent To construct the input weight matrix $W_{in}$, we follow the methodology outlined in\cite{lu2017reservoir}. This method entails linking the $i$-th component of the $(M-1)$-dimensional input signals to $N_r/(M-1)$ reservoir nodes using the connection weights in the corresponding column of $W_{in}$. The non-zero elements of the input weight matrix are randomly selected from a uniform distribution and then scaled to fit within the interval of $[-\sigma, \sigma]$. Notably, each node of the reservoir network is connected to the parameter channel, allowing it to capture the relationship between the dynamics and the parameter value. The parameter input is determined by $(p_s-p_b)k_p$, where $p_b$ and $k_p$ serve as hyper-parameters.

We train the reservoir-computing machine separately for various parameter values and store the corresponding values of $r(t)$ successively. Following is the  conventional way\cite{pathak2018model},

\begin{align}
r_n(t) = \begin{cases} r_n(t) & \text{if n is odd} \\
r_n^2(t) & \text{ if n is even} \end{cases}
\end{align}

\noindent We organize ${r}(t)$ in the sequence of parameter values and create a single $N_r\times n_p(N_t-N_{\tau})$ matrix, forming the reservoir state matrix $R$. Here, $n_p$ is the number of training parameters, $N_t$ represents the number of training data points for each parameter value, and $N_{\tau}$ denotes the reservoir transient time. To derive an explicit expression for the output matrix $W_{out}$ through optimization, we introduce a target data matrix. This matrix encompasses all the desired outputs of the reservoir during training. Assuming the reservoir machine is trained with time series data from $n_p$ different system parameter values, the total number of training time steps becomes $n_{p}N_t$. The available data points are arranged in the sequence of parameter values and stacked into a matrix $U$ with dimensions $(M-1) \times n_p(N_t-N_{\tau})$.

The computation of the output matrix $W_{out}$ is conducted via a regression scheme with the goal of minimizing the following loss function,

\begin{equation}\label{eq()}
L = \sum_{t} \|{ U(t) - W_{out}R(t)}\| + \beta \| W_{out} \|^2  
\end{equation}
\noindent where the regularization parameter $\beta$ is employed to prevent over-fitting. The readout matrix $W_{out}$ can be determined through Ridge regression as follows:
\begin{equation}\label{Ridge-regression}
 W_{out} = UR^{T} \left(RR^T + \beta I\right)^{-1}.   
\end{equation}

In the prediction phase, the input data vector $u(t)$ is substituted with the output vector $v(t)$, creating a closed-loop, self-evolving dynamical system within the reservoir computing machine. The system updates $v(t)$ to $v(t+dt)$ following the specified rules,

\begin{gather}
r(t + dt) = (1 - \alpha)r(t) + \alpha \tanh [W_{res} \cdot r(t) + W_{in}\cdot{v}(t)],\nonumber\\
v(t) = [\tilde{v}(t), (p_{new}-p_b)k_p]^T,\nonumber\\
\tilde{v}(t + dt) = W_{out} \cdot{r}(t + dt).
\label{eq:7}
\end{gather}

\noindent We are now able to generate machine-predicted time series for a different parameter value using Eq. \ref{eq:7}. In our approach, instead of explicitly inputting the inherent parameter values of the oscillators, we train the machine by passing the information of fraction $p$ representing inactive oscillators.

\begin{figure}[h]
\centering
  \includegraphics[width=0.6\textwidth]{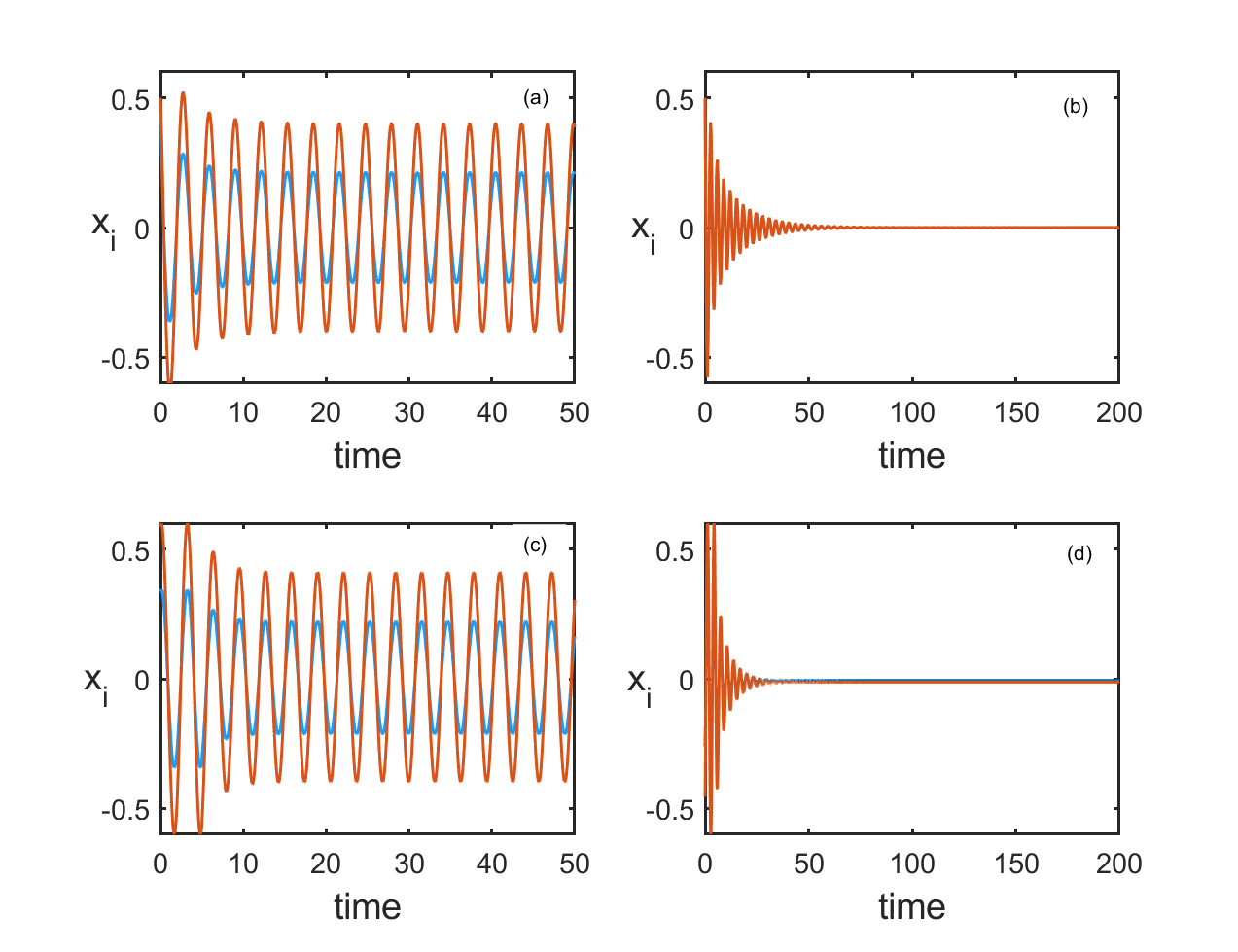}
  \caption{The observed dynamics of the network, obtained through numerical simulation of the model, are showcased alongside the forecasted dynamics generated by the machine. (a)-(b): Model simulated dynamics of active(red) and inactive(blue) units at $p=0.6$ , and $p=0.7$ respectively. We consider random initial states with model parameters are fixed at $\omega=2$, and $K=2.5$. (c)-(d): ESN predicted  dynamics of active(red) and inactive(blue) units at $p=0.6$, and $p=0.7$ respectively. ESN hyper-parameters are  fixed at $\alpha=0.31$, $\beta = 0.000001$, $\rho= 0.1893$, $\sigma=2.0$, $k_p=0.5223$, $p_b=0$, and $N_r=1200$.  We take any random point from the available trajectory as the initial condition to warm up the machine. Reprinted figure with permission from Ref.\cite{rakshit2023predicting}}
  \label{rf1}
\end{figure}

\begin{figure}[h]
\centering
  \includegraphics[width=0.5\textwidth]{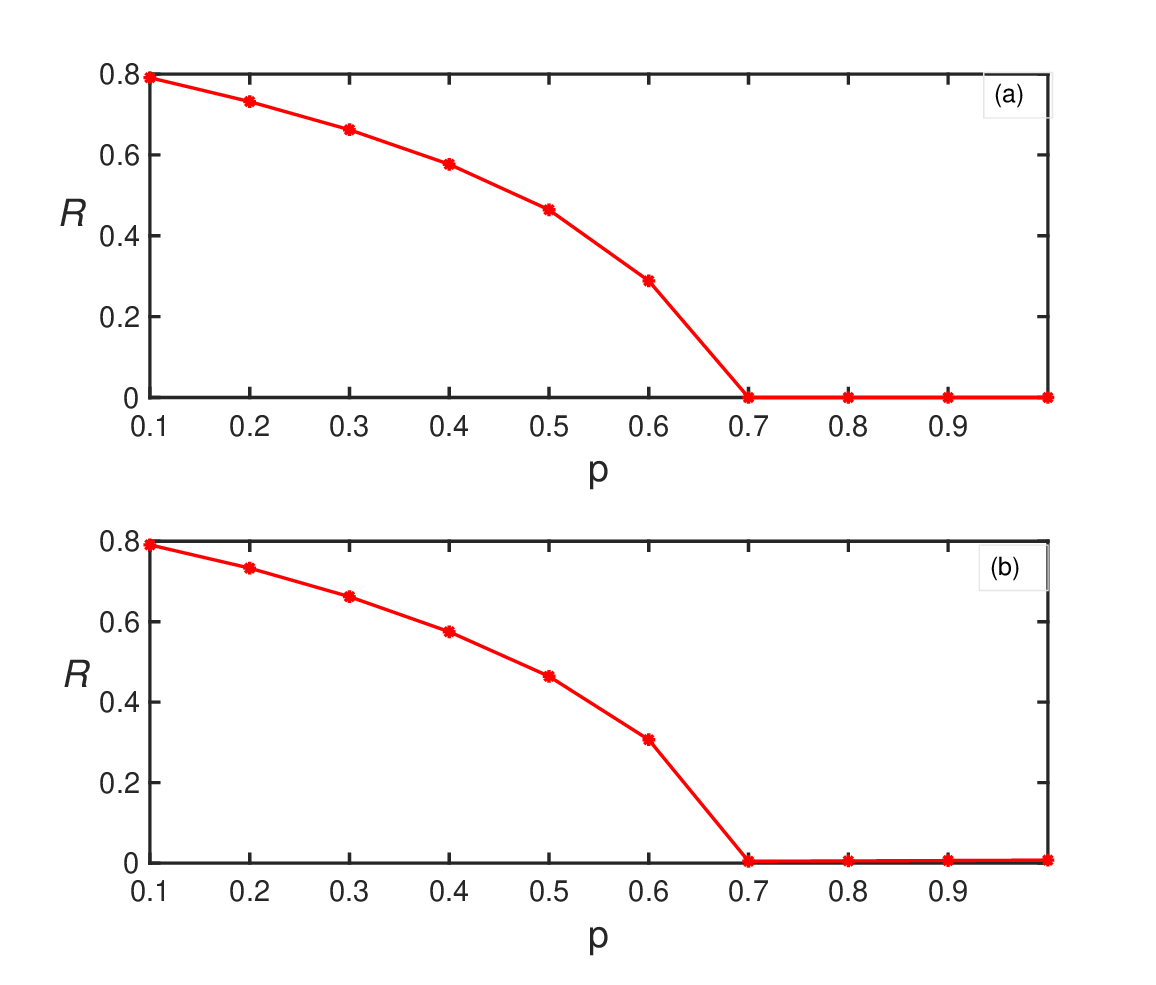}
  \caption{The order parameter $R$ is graphed as a function of the inactivation ratio $p$. (a): employing the actual network model. We consider random initial states with model parameters are fixed at $\omega=2$, and $K=2.5$. (b): machine-generated. ESN hyper-parameters are fixed at $\alpha=0.31$, $\beta = 0.001$, $\rho= 0.2193$, $\sigma=2.0$, $k_p=0.15$,  $p_b=0$, and $N_r=1200$. The machine undergoes training at three different inactivation ratios: p=0.1, p=0.3, and p=0.5. A random point from the existing trajectory is chosen as the initial condition to initialize and warm up the machine. Reprinted figure with permission from Ref.\cite{rakshit2023predicting}}
  \label{f3}
\end{figure}

We demonstrate the predictive capability of the parameter-aware reservoir computing approach in anticipating the aging transition  in $N=100$ globally  coupled Stuart-Landau oscillators, as described  by Eq.\ref{eq-1br}. To generate the { ESN} training time series, we numerically integrate Eq. \ref{eq-1br} using RK45 with a fixed step size of $\Delta t=0.01$. The training phase incorporates data for $p=0.3$, $p=0.4$, and $p=0.5$, corresponding to 30, 40, and 50 inactive oscillators in the system, respectively. This approach enables the machine to learn diverse network dynamics under varying levels of inactive oscillators. Instead of providing the ESN with time series data for all active and inactive oscillators, we simplify the input by using data from one active and one inactive oscillator for each $p$ value. Subsequently, we forecast the network dynamics for $p$ values $0.6$ and $0.7$. In Fig.~\ref{rf1}, the original dynamics of the network obtained through numerical integration are presented and compared with the predicted dynamics of the ESN. Particularly, in Figs.~\ref{rf1}(a) and \ref{rf1}(b), the time series of one active and one inactive oscillator in the network are depicted for $p=0.6$ and $p=0.7$, respectively. We observe oscillatory dynamics for both active (red) and inactive (blue) units at $p=0.6$, while at $p=0.7$, they both converge to a stable equilibrium point. The machine-predicted dynamics for $p=0.6$ and $p=0.7$ are illustrated in Figs.~\ref{rf1}(c) and \ref{rf1}(d), respectively. The ESN demonstrates accurate qualitative predictions of the dynamics and captures the quantitative behavior of the system, as evidenced by the excellent agreement between the actual and predicted time series.

 To validate our proposed method further, we train the machine using the mean-field dynamics of active ($A_z$) and inactive oscillators ($I_Z$). This aims to assess the machine's proficiency in predicting the network dynamics across the entire range of the inactivation ratio parameter $p$. Subsequently, we plot the ESN-generated order parameter $R$ (Refer Eq.~\ref{order-para}) as a function of $p$ and compare it with the corresponding curve obtained from the actual network model. Figure \ref{f3}(a) shows the order parameter $R$ against the inactivation ratio $p$ obtained from the model dynamics, while in Fig. \ref{f3}(b), we plot the same using the ESN. We observe a very good agreement between these two plots, indicating that the machine  is capable of accurately predicting the aging transition  of the original network model.

\section{Conclusions and future perspectives}

Comprehending the emerging behaviors in various scientific and technological fields relies heavily on grasping the dynamics of coupled oscillatory systems, as the presence of robust rhythmic dynamics is a prerequisite for both natural and artificial systems in these domains. It is inherent for certain oscillatory units to undergo aging, transitioning into a non-self-oscillatory state due to diverse internal and external factors. The aging of oscillatory behaviors in coupled dynamical networks has been a thriving research area, with substantial advancements achieved over the past two decades. We have comprehensively summarized studies on dynamical robustness and its enhancement in coupled dynamical networks, which is endeavoured  to amplify our understanding of aging transition in different physical and biological systems. The phenomenon of aging transition has been elaborated  by considering diverse network topologies and  coupling schemes relevant to many real-world situations.  Also, schemes proposed to enhance the dynamical robustness of networked systems which is experiencing aging transition, have been systematically reviewed. Results discussed here are believed to be constructively useful in uncovering different aging transition routes and propose suitable control mechanisms in various complex systems in science and engineering.  Here, we have endeavored to integrate and consolidate existing knowledge on dynamical robustness theory, making it more accessible to researchers in various scientific communities. This review aims to encourage more profound discussions on the aging transition and its potential reversal in complex systems. These discussions should pave the way for further research into the open problems outlined below.

\par Despite the growing body of literature on aging transitions and dynamical robustness in coupled dynamical networks, several open issues and challenges persist. Below, we highlight several open problems that may be of interest for further research and future directions.

\par First of all, it is clear that all these developments summarized in this review have specifically assumed that dyadic or pairwise interactions form the foundation for connections among the units of the system. However, for a better understanding of many complex systems, one needs to further consider more realistic different structural forms of networked systems. For instance, group interactions (of three or more entities) are prevalent in systems arising in ecology~\cite{levine2017beyond}, neuronal~\cite{santoro2023higher,stramaglia2021quantifying,yu2011higher}, and social systems~\cite{benson2018simplicial,iacopini2019simplicial}. Recent theoretical research suggests in complex systems, the inclusion of higher-order interactions, which are often represented by network generalizations like simplicial complexes or hypergraphs~\cite{battiston2020networks,majhi2022dynamics,battiston2021physics}, can have a significant impact on the system’s dynamics~\cite{skardal2019abrupt,iacopini2019simplicial,zhang2023higher,alvarez2021evolutionary}. Thus, future attention must be given to exploring aging transition in networks with higher-order interactions. So far, there have been some efforts undertaken in perceiving the phenomenon of dynamical robustness in neuronal systems, as explained above. Nevertheless, the robustness of neuronal systems subject to higher-order interactions has yet to be explored, despite their significant relevance for various processes in neuronal networks~\cite{petri2014homological}. This gap needs to be addressed in the near future.

\par Many real networks display community structures~\cite{girvan2002community,fortunato2010community,malliaros2013clustering}, where groups of nodes are highly connected to each other within their own community but have very few connections to nodes in other modules.  Examples include ecological networks, neuronal network, metabolic and regulatory networks~\cite{newman2006modularity}. It will be highly interesting to investigate dynamical robustness of networks having community structures.

\par In addition, future research should prioritize quantum oscillatory systems, as initial studies indicate that aging transitions occur in the quantum regime, albeit with different manifestations compared to classical systems. Apart from the theoretical motivation, studying aging in the quantum domain is further driven by the inevitability of system degradation or aging in real-world quantum systems, primarily due to unwanted losses such as those caused by lossy cavities~\cite{yuge2014cavity} and mechanical dissipation in optomechanical systems~\cite{aspelmeyer2014cavity, fitzgerald2021cavity}. Consequently, with the advancement of current quantum technology, we believe that the experimental realization of aging transitions is possible.

\par Further, time-varying networks~\cite{holme2012temporal,ghosh2022synchronized}, in which interactions do not persist for all the course of time, rather they arise or vanish over time, are considered to be highly capable of modeling several real-world instances. Temporal networks of static nodes and that of mobile agents are reasonably significant in this context. It also includes the crucial scenario of adaptive networks~\cite{berner2023adaptive}, which constitute a wide range of systems capable of altering their connectivity based on their dynamical state over time. This readily suggests that contemplation of temporality in the network connectivity itself is pretty essential for further grasp on dynamical robustness of the complex networked systems.

\section*{Declaration of competing interest}
The authors declare that they have no known competing financial interests or personal relationships that could have
appeared to influence the work reported in this paper.
\bibliographystyle{unsrt}       
\bibliography{dr_reference}

\end{document}